\pdfoutput=1

\documentclass[12pt,preprint]{aastex}

\usepackage{amsmath,amssymb,graphicx}
\usepackage{epsfig}
\usepackage{amssymb}
\usepackage{natbib}
\usepackage{aas_macros}
\usepackage[section] {placeins}
\usepackage{subfigure}
\usepackage{float}

\def\msun{{\rm\,M_\odot}}

\def\msun{{\rm\,M_\odot}} 

\def\zsun{{\rm\,Z_\odot}}

\newcommand{\kms}{\, {\rm km\, s}^{-1}}

\newcommand{\lya}{Ly$\alpha$ }
\newcommand{\lyaf}{Ly$\alpha$ forest}

\def\h2{${\rm\,H_2}$}

\makeatletter 

\def\kms{{\rm\,km/s}}
\def\msun{{\rm\,M_\odot}}

\def\vol#1  {{{#1}{\rm,}\ }}
\def\lya{{\rm Ly}\alpha}

\newcount\refno
\newcount\rfno
\def\eq{$^{\the\refno\ }$\advance\refno by 1}
\def\ad{\advance\rfno by 1}

\def\clock{\count0=\time \divide\count0 by 60
     \count1=\count0 \multiply\count1 by -60 \advance\count1 by \time
     \number\count0:\ifnum\count1<10{0\number\count1}\else\number\count1\fi}

\def\myputfigure#1#2#3#4#5%
{\hskip0.03\textwidth\vskip#5pt%\hfill
\makebox[0pt]{\hskip#2in
\includegraphics[width=#3\textwidth]{#1}}\vskip#4pt\hfill}
\def\civ{C~IV\ }
\def\ovi{O~VI\ }
\def\omegaovi{\Omega_{\rm OVI}}
\def\omegaciv{\Omega_{\rm CIV}}

\lefthead{Cen} 
\righthead{IGM metal enrichment}

\begin{document}

\title{Star Formation Feedback and Metal Enrichment History Of The Intergalactic Medium}

\author{
Renyue Cen$^{1}$
and 
Nora Elisa Chisari$^{2}$
} 
 
\footnotetext[1]{Princeton University Observatory, Princeton, NJ 08544;
 cen@astro.princeton.edu}
\footnotetext[2]{Princeton University Observatory, Princeton, NJ 08544;
nchisari@astro.princeton.edu}

\begin{abstract} 

Using the state-of-the-art cosmological hydrodynamic simulations of the standard cold dark matter model with star formation feedback strength normalized to match the observed star formation history of the universe at $z=0-6$, we compute the metal enrichment history of the intergalactic medium (IGM). 
Overall we show that galactic superwind (GSW) feedback from star formation can transport metals to the IGM and that the 
properties of simulated metal absorbers match current observations. 
The distance of influence of GSW from galaxies is typically limited to about $\le 0.5$Mpc
and within regions of overdensity $\delta \ge 10$.
Most \civ and \ovi absorbers are located within shocked regions of elevated temperature ($T\ge 2\times 10^4$K), 
overdensity ($\delta \ge 10$), and metallicity ($[Z/\zsun]=[-2.5,-0.5]$), enclosed by double shocks propagating outward. 
\ovi absorbers have typically higher metallicity, lower density and higher temperature than \civ absorbers. 
For \ovi absorbers collisional 
ionization dominates over the entire redshift range $z=0-6$, whereas for \civ absorbers the transition 
occurs at moderate redshift $z\sim 3$ from collisionally dominated to photoionization dominated.
We find that the observed column density distributions for \civ and \ovi 
in the range $\log N {\rm cm}^2=12-15$ are reasonably reproduced by the simulations. 
The evolution of mass densities contained in \civ and \ovi lines, $\omegaciv$ and $\omegaovi$, 
is also in good agreement with observations, which shows a near constancy at low redshifts and 
an exponential drop beyond redshift $z=3-4$. 
For both \civ and \ovi\, most absorbers are transient 
and the amount of metals probed by \civ and \ovi lines of column $\log N {\rm cm}^2=12-15$ 
is only $\sim 2\%$ of total metal density at any epoch.
While gravitational shocks from large-scale structure formation dominate the energy budget ($80-90\%$) for turning about 50\% of IGM to the warm-hot intergalactic medium (WHIM) by $z=0$, GSW feedback shocks are energetically dominant over gravitational shocks at $z\ge 1-2$.
Most of the so-called ``missing metals" at $z=2-3$ are hidden in a warm-hot ($T=10^{4.5-7}$K) gaseous phase, heated up by GSW feedback shocks. Their mass distribution is broadly peaked at $\delta=1-10$ in the IGM, outside virialized halos. Approximately $(37,46,10,7)\%$ of the total metals at $z=0$ are in (stars, WHIM, X-ray gas, cold gas); the distribution stands at $(23,57,2,18)\%$ and $(14,51,4,31)\%$ at $z=2$ and $z=4$, respectively.
%The mean metallicity of the IGM with moderate overdensities ($1-10$) that are probed by the $\lya$ forest shows a rapid increase with decreasing redshift. 
%while recent observations at $z=2-4$ have suggested otherwise. We give a possible explanation for the disagreement.

\end{abstract}
 
 
\section{Introduction}

One of the pillars of the Big Bang theory is its successful prediction of a primordial baryonic matter composition, made up of nearly one hundred percent hydrogen and helium with a trace amount of a few other light elements \citep[e.g.,][]{1998Schramm,2001Burles}. The metals, nucleosynthesized in stars later,
are found almost everywhere in the observable IGM, ranging from the metal-rich intracluster medium \citep[e.g.,][]{1997Mushotzky} to moderately enriched damped Lyman systems \citep[e.g.,][]{1997Pettini, 2003Prochaska} to low metallicity Lyman alpha clouds  \citep[e.g.,][]{2003Schaye}.
When and where were the metals made and why are they distributed as observed? We address this fundamental question in the context of the standard cold dark matter cosmological model \citep[][]{2009Komatsu} using latest simulations. Our previous simulations \citep[][]{1999bCen, 2005Cen} provided some of the earlier attempts to address this question with measured successes. In this investigation we use substantially better simulations to provide significantly more constrained treatment of the feedback processes from star formation (SF) that drive energy and metals from supernovae into the IGM through galactic winds \citep[e.g.,][]{1999bCen, 2001Aguirre, 2002bTheuns, 2003Adelberger, 2003Springel}.

Metal-line absorption systems in QSO spectra are the primary probes of the metal enrichment of the IGM as well as in the vicinities of galaxies \citep[e.g.,][]{1969Bahcall}. The most widely used metal lines include \ion{Mg}{2} $\lambda \lambda$2796, 2803 doublet \citep[e.g.,][]{1992Steidel}, \ion{C}{4} $\lambda \lambda$1548, 1550 doublet \citep[e.g.,][]{1982Young}, and \ion{O}{6} $\lambda \lambda$1032, 1038 doublet \citep[e.g.,][]{2002Simcoe}.
We here focus on the \civ and \ovi absorption lines and the global evolution of metals in the IGM. We will limit our current investigation to the observationally accessible redshift range of $z=0-6$, which in part is theoretically motivated simply because the theoretical uncertainties involving still earlier star formation are much larger.
At $z=0$ the \ovi line (together with \ion{C}{7} and \ion{O}{8} lines) provide vital information on the missing baryons 
\citep[e.g.,][]{2003Mathur, 2008Tripp, 2008Danforth, 2009Nicastro}, predicted to exist in a Warm-Hot Intergalactic Medium (WHIM) \citep[][]{1999bCen, 2001Dave}. 

For a well understood sample of QSO absorption lines, one could derive the cosmological density contained in them \citep[e.g.,][]{2009Cooksey}.
%Many studies have measured the \civ cosmological mass density
%from observations of \civ absorbers in the sightlines of QSOs.  
%Samples vary in size and completeness.
Early investigations indicate that $\omegaciv$ remains approximately constant in the redshift interval $z \sim 1.5 - 4$ \citep[][]{2001Songaila,2005Songaila,2003Boksenberg}. There have been recent efforts to extend the measurements of  $\omegaciv$ to $z<1.5$ \citep[][]{2009Cooksey} and to $z>5$ \citep[][]{2006Simcoe,2006Ryanweber, 2009Ryanweber,2009Dodorico, 2009Becker}.  Observations in these redshift ranges have been difficult to carry out because \civ transition moves to the UV at low redshift and to the IR band at high redshift. \citet[][]{2009Dodorico} find evidence of a rise in the \civ mass density for $z<2.5$. \citet[][]{2006Simcoe} and \citet[][]{2006Ryanweber} found evidence of \civ density at $z\sim 6$ being consistent with estimations at $z \sim 2-4.5$. More recently, however, \citet[][]{2009Becker} set upper limits for $\omegaciv$ at $z\sim 5.3$ and \citet[][]{2009Ryanweber} observe a decline in intergalactic \civ approaching $z=6$, which we will show are in good agreement with our simulations.

The ionization potential of \ovi and the relatively high oxygen abundance are very favorable for production of \ovi absorbers in the IGM \citep[e.g.,][]{1983Norris, 1986Chaffee}. The rest wavelength of OVI ($1032,1037$\AA) places it within the Ly-$\alpha$ forest, which makes the identifications
of these lines more complicated,
although being a doublet helps significantly.
At $z\ge 2 $, however, \ovi absorption can probe the metal content of the IGM in ways complementary to what is provided by \civ lines. For example, the \ovi lines can probe IGM that is hotter than that probed by the \civ lines and can reach lower densities thank to higher abundance. There are now several observational studies at redshifts $z=2-3$ that describe the properties of \ovi absorbers and attempt to estimate
the \ovi mass density, $\omegaovi$ 
\citep[][]{2002Carswell,2002Bergeron, 2004Simcoe, 2006Simcoe, 2008Frank, 2008Danforth, 2008Tripp, 2008Thomb}.

At $z\sim 2-3$ there is a missing metals problem: only 10-20\% of the metals produced by all stars formed earlier have been identified 
in stars of Lyman break galaxies (LBG), in damped Lyman alpha systems (DLAs) and $\lya$ forest.
%i.e., cold-warm gas and stars.
The vast majority of the produced metals appear to be missing \citep[e.g.,][]{1999bPettini}. The missing metals could be in hot gaseous halos of star-forming galaxies \citep[][]{1999bPettini, 2005Ferrara}. We will show that most of the missing metals are in a warm-hot ($T=10^{4.5-7}$K)
but diffuse IGM at $z=2-3$ of overdensities of $\sim 10$ that are outside of halos.

%Is the low density IGM pristine? Recent studies of the metallicity of the low density IGM, based on the novel method of pixel optical depth (POD) 
%\citep[e.g.,][]{1998Cowie, 2002Aguirre}, have yielded intriguing results. Specifically, they suggest that the metallicity of IGM at densities that are close to the mean density of the universe increases with increasing redshift \citep{2003Schaye}. Our simulations do not produce such a trend and we provide reasons that may reconcile the disagreement.

The outline of this paper is as follows. In \S 2 we detail our simulations and the procedure of normalizing the uncertain feedback processes from star formation. Results on the metal enrichment of the IGM are presented in \S 3. In \S 3.1 we give a full description of the properties of the \civ and \ovi lines at $z=0-6$, followed \S 3.2 discussing \civ and \ovi absorbers as metals reservoirs. We devote \S 3.3 to a general discussion of global distribution
of metals, addressing several specific topics, including the metallicity of the moderate overdense regions at moderate redshift, the missing metals at $z\sim 3$. Conclusions are given in \S 4.

\section{Simulations}\label{sec: sims}

\subsection{The Hydrocode}

Numerical methods of the cosmological hydrodynamic code and input physical ingredients have been described in detail in an earlier paper \citep[][]{2005Cen}. The simulation integrates five sets of equations simultaneously: the Euler equations for gas dynamics in comoving coordinates,
time dependent rate equations for hydrogen and helium species, the Newtonian equations of motion for dynamics of collisionless (dark matter) particles,
the Poisson equation for the gravitational potential field and the equation governing the evolution of the intergalactic ionizing radiation field, all in cosmological comoving coordinates. The gasdynamic equations are solved using a new, improved hydrodynamics code, ``COSMO" \citep[][]{2008Li} 
on a uniform mesh. The rate equations are treated using sub-cycles within a hydrodynamic time step due to the much shorter ionization time-scales
(i.e., the rate equations are very ``stiff"). Dark matter particles are advanced in time using the standard particle-mesh (PM) with a leapfrog integrator.
The Poisson equation is solved using the Fast Fourier Transform (FFT) method on the uniform mesh. The initial conditions adopted are those
for Gaussian processes with the phases of the different waves being random and uncorrelated. The initial condition is generated by the
COSMICS software package kindly provided by E. Bertschinger (2001).

Cooling and heating processes due to all the principal line and continuum atomic processes for a plasma of primordial composition with
additional metals ejected from star formation. Compton cooling due to the microwave background radiation field and Compton cooling/heating due to
the X-ray and high energy background are computed.
The cooling/heating due to metals is computed using a code based on the Raymond-Smith code assuming ionization equilibrium that takes into account the presence of a time-dependent UV/X-ray radiation background, which we have included in our simulations since \citet[][]{1995Cen} and has now been performed by other investigators \citep[e.g.,][]{2010Shen}.

We follow star formation using a well defined, Schmidt-Kennicutt-law-like prescription used by us in our previous work and similar to that of other investigators \citep[e.g.,][]{1996Katz, 1996Steinmetz, 1997Gnedin}. A stellar particle of mass $m_{*}=c_{*} m_{\rm gas} \Delta t/t_{*}$ is created
(the same amount is removed from the gas mass in the cell), if the gas in a cell at any time meets the following three conditions simultaneously:
(i) contracting flow, (ii) cooling time less than dynamic time, and  (iii) Jeans unstable, where $\Delta t$ is the time step, $t_{*}={\rm max}(t_{\rm dyn}, 10^7$yrs), $t_{dyn}=\sqrt{3\pi/(32G\rho_{tot})}$ is the dynamical time of the cell, $m_{\rm gas}$ is the baryonic gas mass in the cell and $c_*=0.03$ is star formation efficiency \citep[e.g.,][]{2007Krumholz}. Each stellar particle is given a number of other attributes at birth, including formation time $t_i$, initial gas metallicity and the free-fall time in the birth cell $t_{dyn}$. The typical mass of a stellar particle in the simulation is about $10^6M_{\odot}$; in other words, these stellar particles are like coeval globular clusters. All variations of this commonly adopted star-formation algorithm essentially achieve the same goal: in any region where gas density exceeds the stellar density, gas is transformed to stars on a timescale longer than the local dynamical time and shorter than the Hubble time. Since these two time scales are widely separated, the effects, on the longer time scale, of changing the dimensionless numbers (here $c_*$) are minimal. Since nature does not provide us with examples of systems which violate this condition (systems which persist over many dynamical and cooling time scales in having more gas than stars), this commonly adopted algorithm should be adequate even though our understanding of star formation remains crude.

Stellar particles are treated dynamically as collisionless particles subsequent to their birth. Feedback from star formation, the effects of the cumulative SN explosions %and AGN output 
known as Galactic Superwinds (GSW) and metal-enriched gas, will be described in more detail in the next subsection. While the code can self-consistently compute the ionizing UV-X-ray background using sources and sinks in the simulation, here we use the \citet{1996Haardt} spectra for all runs such that we do not introduce additional variations due to otherwise varying UV backgrounds in the different runs.
However, a local optical depth approximation is adopted to crudely mimic the local shielding effects: each cubic cell is flagged with six hydrogen ``optical depths" on the six faces, each equal to the product of neutral hydrogen density, hydrogen ionization cross section and scale height,
and the appropriate mean from the six values is then calculated; analogous ones are computed for neutral helium and singly-ionized helium. In computing the local ionization and cooling/heating balance for each cell, 
self-shielding is taken into account to attenuate the external HM ionizing radiation field.
Both these two shielding effects are essential in order to obtain self-consistent radiation background evolution and neutral hydrogen evolution.

%%%%%%%%%%%%%%%%%%%%%%%%%%%%%%%%%%%%%%%%%%%%%%%%%%%%%%%%%%%%%%%%%%%%%%%%%%%%%%%%%%%%%%%%%%%%%%
\begin{deluxetable}{lccll}
\tablecolumns{5}
\tablewidth{0pc}
\tablecaption{Simulations}
\tablehead{
\colhead{Run} & \colhead{Box (Mpc/h)}
& \colhead{Res (kpc/h)}  & \colhead{DM ($\msun$)} & \colhead{$e_{GSW}$} }
\startdata
N & 50 & 24 & $1.1\times 10^7$ & $0$   \\ 
L & 50 & 24 & $1.1\times 10^7$ & $3\times 10^{-6}$   \\ 
{\bf M} & {\bf 50} & {\bf 24} & {\bf $1.1\times 10^7$} & {\bf $7\times 10^{-6}$}   \\ 
H & 50 & 24 & $1.1\times 10^7$ & $1\times 10^{-5}$   \\
{\bf MR} & 50 & 48 & $8.8\times 10^7$ & $7\times 10^{-6}$
\enddata
\tablecaption{
The first gives a letter label for each run.
The second, third and fourth columns give
the comoving box size, comoving spatial resolution
and dark matter particle mass.
The last column indicates the GSW strength. 
\label{table1}}
\end{deluxetable}

\subsection{Cosmological and Physical Parameters of the Simulations} 

We have run a set of four new simulations of a WMAP5-normalized \citep[][]{2009Komatsu} cold dark matter model with a cosmological constant: 
$\Omega_M=0.28$, $\Omega_b=0.046$, $\Omega_{\Lambda}=0.72$, $\sigma_8=0.82$, $H_0=100 h {\rm km s}^{-1} {\rm Mpc}^{-1} = 70 {\rm km} s^{-1} {\rm Mpc}^{-1}$ 
and $n=0.96$. The adopted box size is $50$Mpc/h comoving and with $2048^3$ cells of size $24$kpc/h comoving; the dark matter particle mass and mean baryonic mass in a cell are equal to $1.1\times 10^7\msun$ and $2.6\times 10^5\msun$, respectively.
Some of the key parameters for the four simulations are summarized in Table 1. The only difference among the four main runs is the strength of the GSW feedback:
(N) no GSW, 
(L) low GSW feedback, 
(M) moderate GSW feedback and (H) high GSW feedback.
In the next subsection we will determine which feedback strength produces the star formation rate history that matches observations. 
We run an additional lower resolution simulation with $1024$ cells a side, each of size $48$kpc/h (run ``MR'') 
to test convergence of results.
When computing results using run ``MR", we multiply the metallicity
of each cell in run ``MR" by a constant factor such that its mean metallicity 
at any epoch match that of run ``M".
We obtain an additional set of results by changing the amplitude of the UV background, run ``M2'', 
where it is reduced to one half of that in ``M''.

\placetable{table1}

\subsection{Mechanical Feedback from Star Formation}

It is well known that without impeding processes to counter the cooling and subsequent condensation of baryons, the stellar mass in the universe would be overproduced -- the ``overcooling" problem \citep[e.g.,][]{1991White, 1991Cole, 1992Blanchard}. Feedback from star formation is believed to play the essential role to prevent gas from overcooling. The key question is: Where does the feedback from SF throttle gas cooling and condensation?

%REPHRASE this paragrpah.
We consider three independent lines of evidence to address this question. First, while metals from supernovae ejecta can be accelerated to velocities
exceeding the escape velocity,
the whole interstellar gas is very difficult to be blown away, even in starburst galaxies,
based on simulations \citep[e.g.,][]{1999MacLow},
although their adopted feedback strength may be on the low side.
Second, observed normal galaxies in the local universe tend to be relatively gas poor
\citep[e.g.,][]{2009Zhang}.
Their progeniors or their building blocks were presumably gas rich in the past when
most of the star formation occurred.
This implies that, once gas has collapsed, it would turn into stars on a time scale
that is shorter than the Hubble time.
Finally, if gas were able to collapse inside halos without hinderance,
the observed soft X-ray background would be overproduced
by more than an order of magnitude \citep[][]{1999Pen, 2001Wu}.
These three lines of evidence together suggest that 
feedback from star formation likely exerts its effect outside
normal stellar disks, probably in regions
that are tens to hundreds of kiloparsecs from halo centers,
before too much gas has either been collected inside the virial radius or cooled and condensed onto the disk.

It is currently difficult to fully model GSW in a cosmological simulation, although significant progress has been made to provide a better treatment of the multi-phase interstellar medium \citep[e.g.,][]{1997Yepes, 2003Springel}. It is likely that a combination of both high resolution and detailed multi-phase medium treatment (perhaps with the inclusion of magnetic fields and cosmic rays) is a requisite for reproducing observations.

Here we do not attempt to model the {\it causes} and generation of GSW, but, instead, to simply assume an input level of mass, energy and metals, and carefully compute the {\it consequences} of GSW on the surrounding medium and on subsequent galaxy formation. 
%For this purpose our code is very well designed.
Our simulations have a resolution of $24$kpc/h comoving (see Table 1), which may provide an adequate resolution for this purpose, given the aforementioned lines of evidence that feedback from star formation likely exerts most of its effects in regions on scales larger than tens of kiloparsecs.
In our simulations, GSW energy and ejected metals are distributed into 27 local gas cells centered at the stellar particle in question, weighted by the specific volume of each cell \citep[][]{2005Cen}. 
The temporal release of the feedback at time $t$ has the following form, all being proportional to the local star formation rate:
$f(t,t_i,t_{dyn}) $~$\equiv (1/ t_{dyn})
[(t-t_i)/t_{dyn}]$~$\exp[-(t-t_i)/t_{dyn}]$. 
Within a time step $dt$, the released GSW energy and mass to the IGM from stars are
$e_{GSW} f(t,t_i,t_{dyn}) m_* c^2 dt$ and 
$e_{mass} f(t,t_i,t_{dyn}) m_* dt$, respectively.
We fix $e_{mass}=0.25$, i.e., 25\% of the stellar mass is recycled with the ejecta metallicity of $5\zsun$. Metals, collectively having the observed solar abundance pattern, are followed as a separate hydro variable (analogous to the total gas density or nuetral hydrogen, HeI density, HeII density)
with the same hydrocode. We do not introduce any additional ``diffusion" process for the metals. We note that cooling process is never turned off, before or after the deposition of thermal energy, and hydrodynamic coupling between ejected baryons and surrounding gas is not turned off either,
a departure from some of the previous simulations \citep[e.g.,][]{2002Theuns, 2005Aguirre, 2006Oppenheimer, 2008DallaVecchia, 2010Shen}.
This is physically made possible in part due to a deposition of energy at scales that are comparable or larger than 
the Sedov radius in our current simulations, thanks to our limited spatial resolution.

%*******************************************************************
\begin{figure}[h]
\centering
\vskip -0.5in
\epsfig{file=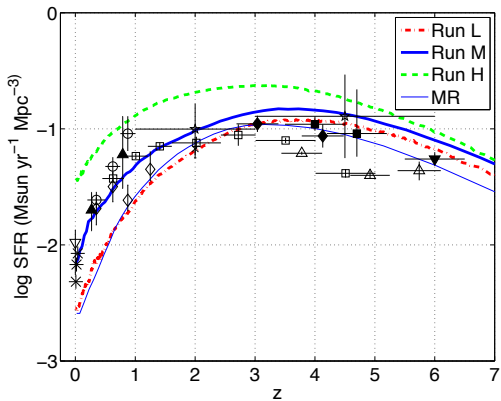,angle=0,width=3.0in} %,height=2.5in}
\vskip -1.0in
\caption{
Star formation rate density as a function of redshift
for three models with differing feedback coefficients
$e_{GSW}=7\times 10^{-6}$ (run ``M", thick solid curve),
$e_{GSW}=1\times 10^{-5}$ (run ``H", dot-dashed curve),
$e_{GSW}=3\times 10^{-6}$ (run ``L", dashed curve),
and run ``M2" (thin solid curve),
compared with observational data taken from (from low to high redshift):
\citet[][3 asterisks at $z\sim 0$]{Heavens04},
\citet[][open inverted triangle at $z=0$]{Nakamura04},
\citet[][open circles]{Lilly96},
\citet[][filled triangles]{Norman04},
\citet[][open diamonds]{Cowie99},
\citet[][open squares]{Gabasch04},
\citet[][cross at $z=2$]{Reddy05},
\citet[][open stars at $z=2$ and 4.5]{Barger00},
\citet[][filled diamonds at $z=3,4$]{Steidel99},
\citet[][filled squares at $z=4,4.7$]{Ouchi04a},
\citet[][open triangles at $z=3-6$]{Giavalisco04},
%\citet[][filled circle at $z\sim 6$]{Bunker04},
and
\citet[][filled inverted triangle at $z=6$]{Bouwens05}.
%and
%\citet[][open stars without error bars]{Thompson06}. %did not plot them
The data are converted to the values with the Chabrier IMF and
common values are assumed for dust extinction for the UV data.
}
\label{fig:sfr}
\end{figure}
%***************************************************************************

The GSW strength is therefore controlled by one single adjustable parameter, $e_{GSW}$. We normalize $e_{GSW}$ by the requirement that the computed
star formation rate (SFR) history matches, as closely as possible, the observations
over the redshift range $z=0$ to $z=6$ where comparisons can be made.
Figure~\ref{fig:sfr} shows the SFR history for the three runs with non-zero $e_{GSW}$, (L,M,H).
What is immediately evident is that the mechanical feedback strength from star formation 
has a dramatic effect on the overall SFR history, especially at low redshift ($z\le 3$).
At the resolution of the simulation,
run ``M" provides the best and excellent match to observations,
where run ``L" and ``H", respectively, over- and under-estimate
the SFR at $z<2$.
At the time of this writing we prefer to avoid introducing additional ad hoc physics
to remedy this and are instead content with the ballpark agreement 
at $z>3$ between simulations and observations, given the large uncertainties
in the observational data as evidenced by the large dispersion among different observations.
At redshift zero we find that the stellar densities in the three models 
(L,M,H) are $\Omega_*=(0.011, 0.0048, 0.0030)$, which should be compared to 
the observed value of $\Omega_{*,obs}=0.0041\pm 0.0006$
\citep[][]{Cole01}.
%%%%%%%%%%%%%%%%%%%%%%We need here Omega_* for MR and M2.
Our experiments indicate that, had we set $e_{GSW}=0$,
the amount of stellar density $\Omega_*$ at $z=0$ would exceed $0.015$, in serious disagreement with observations.
In this respect model ``M" also agrees better with observations.
Our findings are in agreement with 
\citet[][]{2003Springel} and \citet[][]{2006Oppenheimer} in that 
star formation rate history depends sensitively on the stellar feedback,
but in disagreement with \citet[][]{2010Shen} who find otherwise.
All the subsequent results presented are based on run ``M". 
There is some indication that a model between ``M" and ``L" might
provide a better match to the observations at low redshift ($z<1$) 
if the compilation of \citet{Hopk06a} is used. But we note that such a model may run into a worse agreement with observations with respect to $\Omega_*$ at $z=0$. Currently, it is difficult to reconcile the observations of star formation rate history and $\Omega_*$ at $z=0$. One might appeal to an evolving IMF to provide an attractive reconcilation between the possible discrepancy \citep[][]{2008Dave}. This is well beyond the scope of this investigation. In any case, a slight varied simulation, say, using an $e_{GSW}$ value between the ``M" and ``L" would give qualitatively comparable results. 
In order to test for numerical convergence 
we run one additional simulation, ``MR'', 
which has the same parameters as run ``M'' but have half the resolution.
To test the dependence of results on the extragalactic UV background 
we run our software pipeline through run ``M"
but with halving the amplitude of the UV background, called run ``M2".

%**********************************************************************
\begin{figure}[ht]
\centering
\vskip -0.6in
\epsfig{file=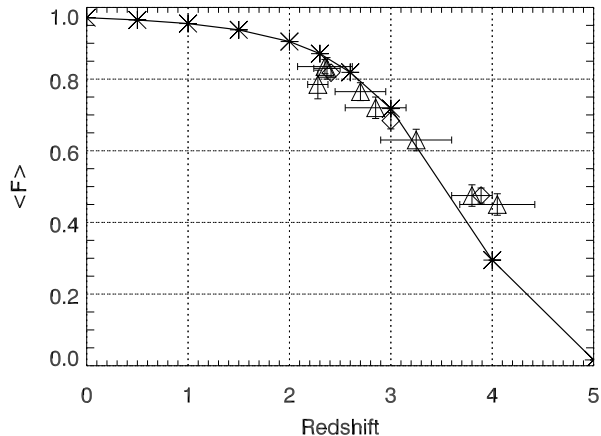,angle=0,width=4.5in}
\vskip -2.7in
\caption{
shows the mean flux for $\lya$ forest as a function of redshift.
Our computed results are shown in asterisks. Diamonds
correspond to mean transmitted flux
values for each quasar in the sample of 
\citet[][]{2000McDonald},
and triangles correspond to the mean flux for the same
observational data but binned in redshift
intervals: $[3.39,4.43]$, $[2.67,3.39]$ and $[2.09,2.67]$.
 }
\label{fig:meanF}
\end{figure}
%****************************************************************************

It is prudent to make a self-consistency check for the value of $e_{GSW}$ that is empirically determined. The total amount of explosion kinetic energy from Type II supernovae with a Chabrier IMF translates to $e_{GSW}=6.6\times 10^{-6}$.
Observations of local sturburst galaxies indicate that nearly all of the star formation produced kinetic energy (due to Type II supernovae)
is used to power GSW \citep[e.g.,][]{2001Heckman}. Given the uncertainties on the evolution of IMF with redshift the fact that newly discovered prompt Type I supernovae contribute a comparable amount of energy compared to Type II supernovae, we argue that our adopted ``best" value of 
$e_{GSW}=7\times 10^{-6}$ is consistent with observations and entirely within physical plausibility.

\subsection{Mock Spectra and Identification of Absorption Lines} 

The photoionization code CLOUDY \citep[][]{1998Ferland}  is used post-simulation to compute the abundance of \civ and \ovi,
adopting the UV background calculated by \citet{1996Haardt}. For $\lya$ absorption lines we use the computed neutral hydrogen density distribution directly from the simulation that was already using the \citet{1996Haardt} UV background in the rate equations for hydrogen and helium species.
We have checked that the radiation field is consistent with observations by comparing the simulated mean transmitted flux as a function of redshift with observations.
Figure~\ref{fig:meanF} shows the mean transmitted $\lya$ flux as a function of redshift from the simulation in comparison with observations. We see the $\lya$ forest produced in the LCDM model using the adopted UV background provides an adequate match to observations over most of the redshift range compared, $z=0-4$. 
At $z\gtrsim 4$, our results do not seem to coincide with observations. We attribute this to the UV background used: we have only considered a quasar background, while at these high redshifts the UV radiation coming from galaxies should have a significant effect on the $\lya$ forest. Nevertheless, we do not expect this to be an issue on the metal species considered in the following sections. 
These correspond to much higher energies than 1~ryd 
that are not affected by the UV contribution from galaxies to the ionizing radiation.

We generate random synthetic absorption spectra for each of the three absorption lines by producing optical depth distribution along lines of sight parallel
to one of the three axes of the simulation box, based on density, temperature and velocity distributions in the simulation (i.e., our calculations include redshift effects due to peculiar velocities and thermal broadening). 
The code used is similar to that used in our earlier papers \citep[][]{1994Cen, 2001Cen}. We identify each absorption line as a contiguous region 
in the flux spectrum between a down-crossing point and an up-crossing point, both at a flux equal to $0.85$. Note that flux equal to $1$ corresponds to no absorption. For each identified line we compute its equivalent width (EW), Doppler width ($b$), mean temperature ($T$), mean metallicity ($Z$) 
and mean gas overdensity ($\delta$), weighted by optically depth of each pixel. 
We do not attempt to perform Voigt profile fitting, a procedure often used to analyze observed spectra.
Because of this, we tend to not generate some of the very low column lines that are purely an effect of profile fitting process.
Also, precise comparison between our mock absorbers and observed ones is not possible for some quantities, such as Doppler width distributions.

\section{Results}

\subsection{\ion{C}{4} $\lambda \lambda$1548, 1550 and 
\ion{O}{6} $\lambda \lambda$1032, 1038 absorption lines}

\begin{figure*}%[h] %h
\centering
%\vskip -0.9in
\vskip -0.7in
\begin{tabular}{cc}
\epsfig{file=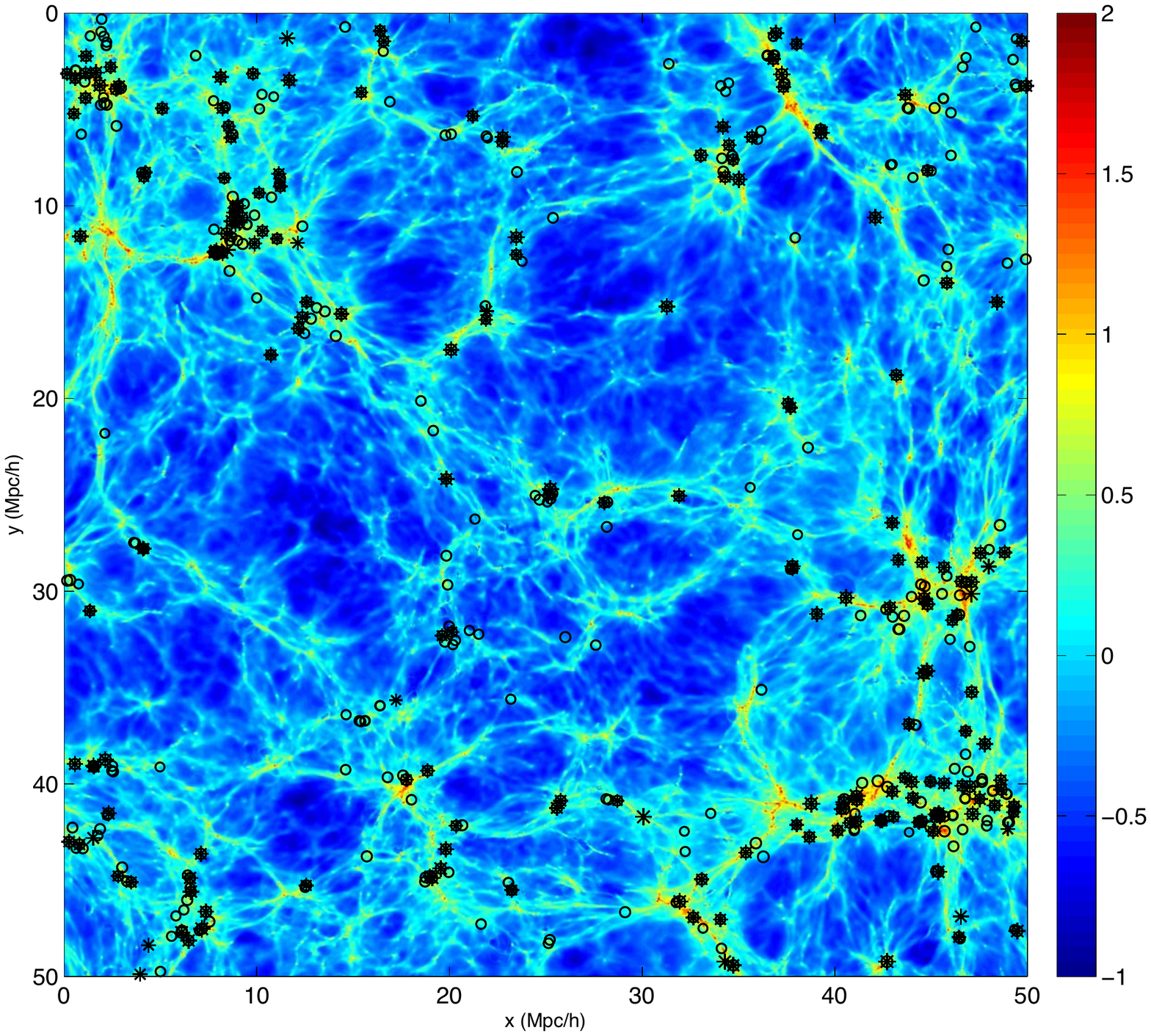,angle=0,width=3.4in} &
\hskip -0.4in
\epsfig{file=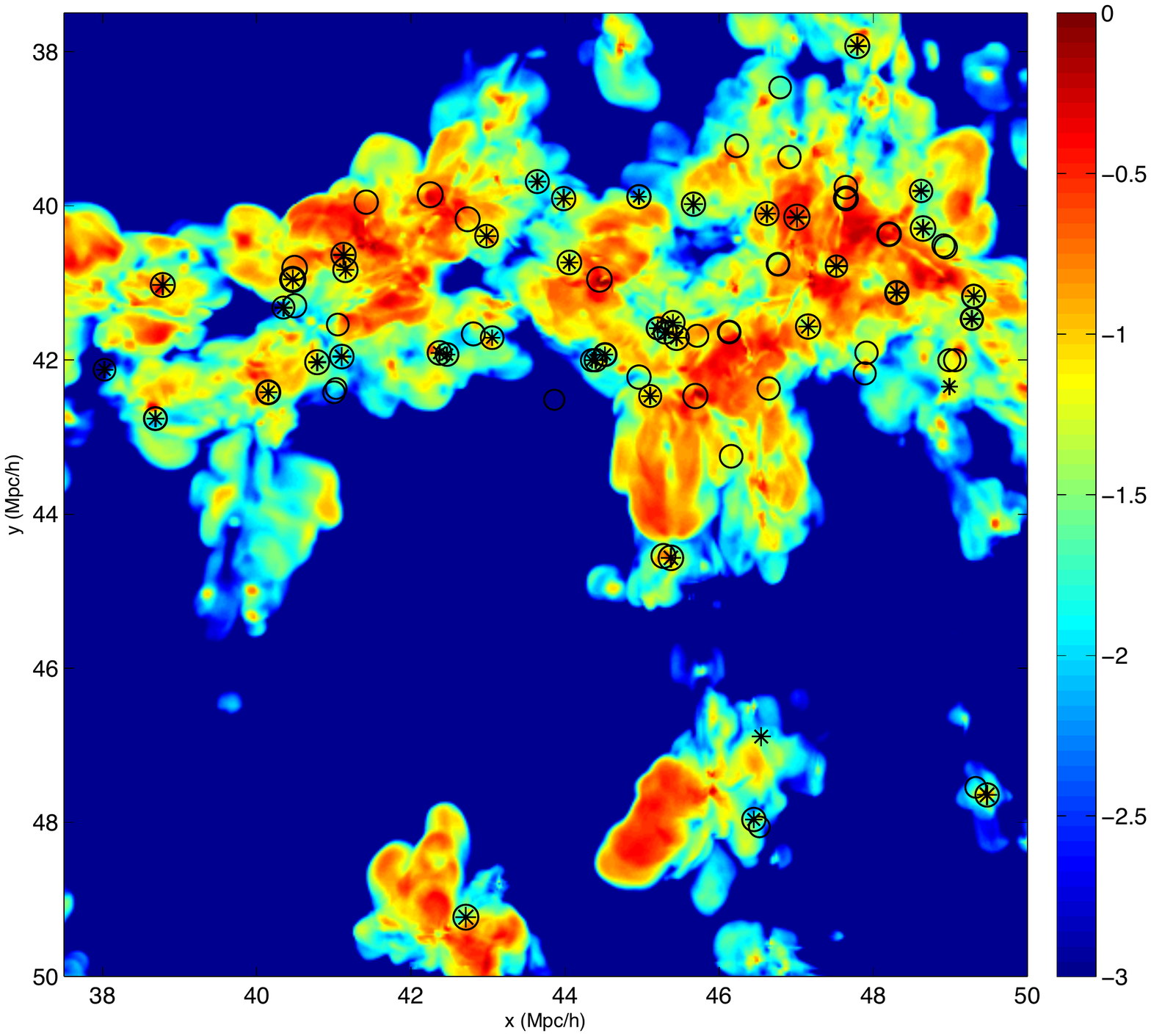,angle=0,width=3.4in} \\
\epsfig{file=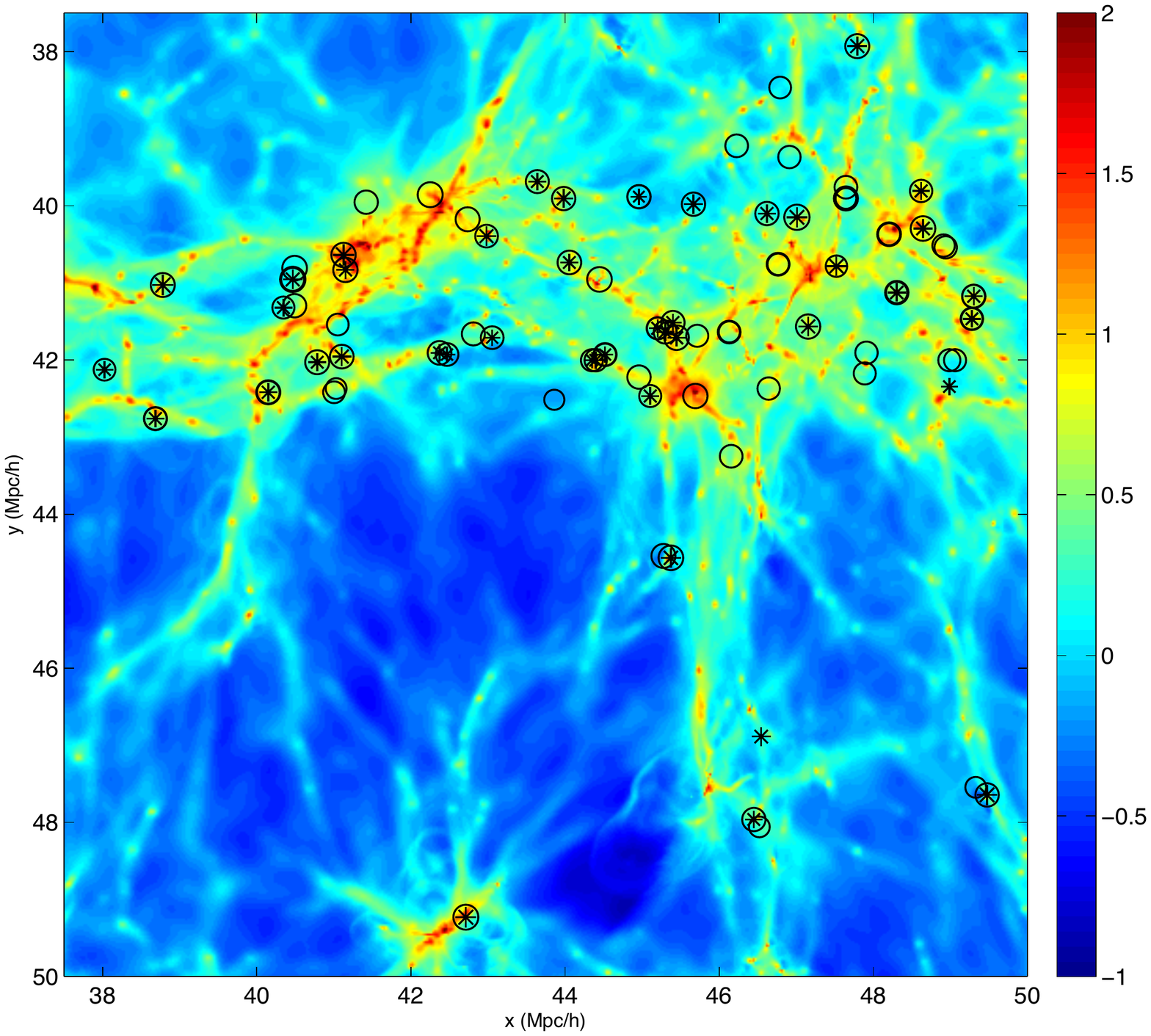,angle=0,width=3.4in} &
\hskip -0.4in
\epsfig{file=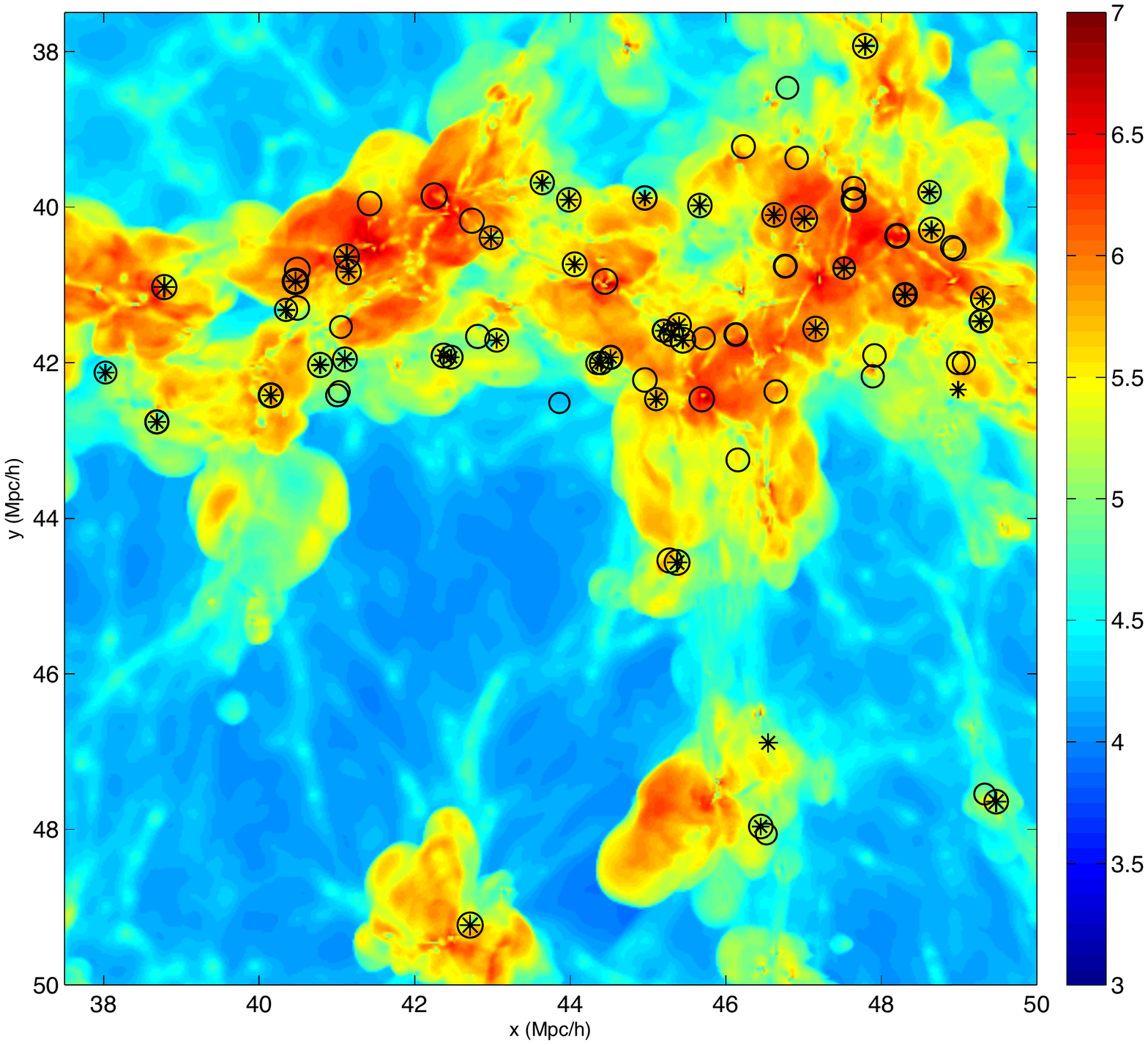,angle=0,width=3.4in} \\
\epsfig{file=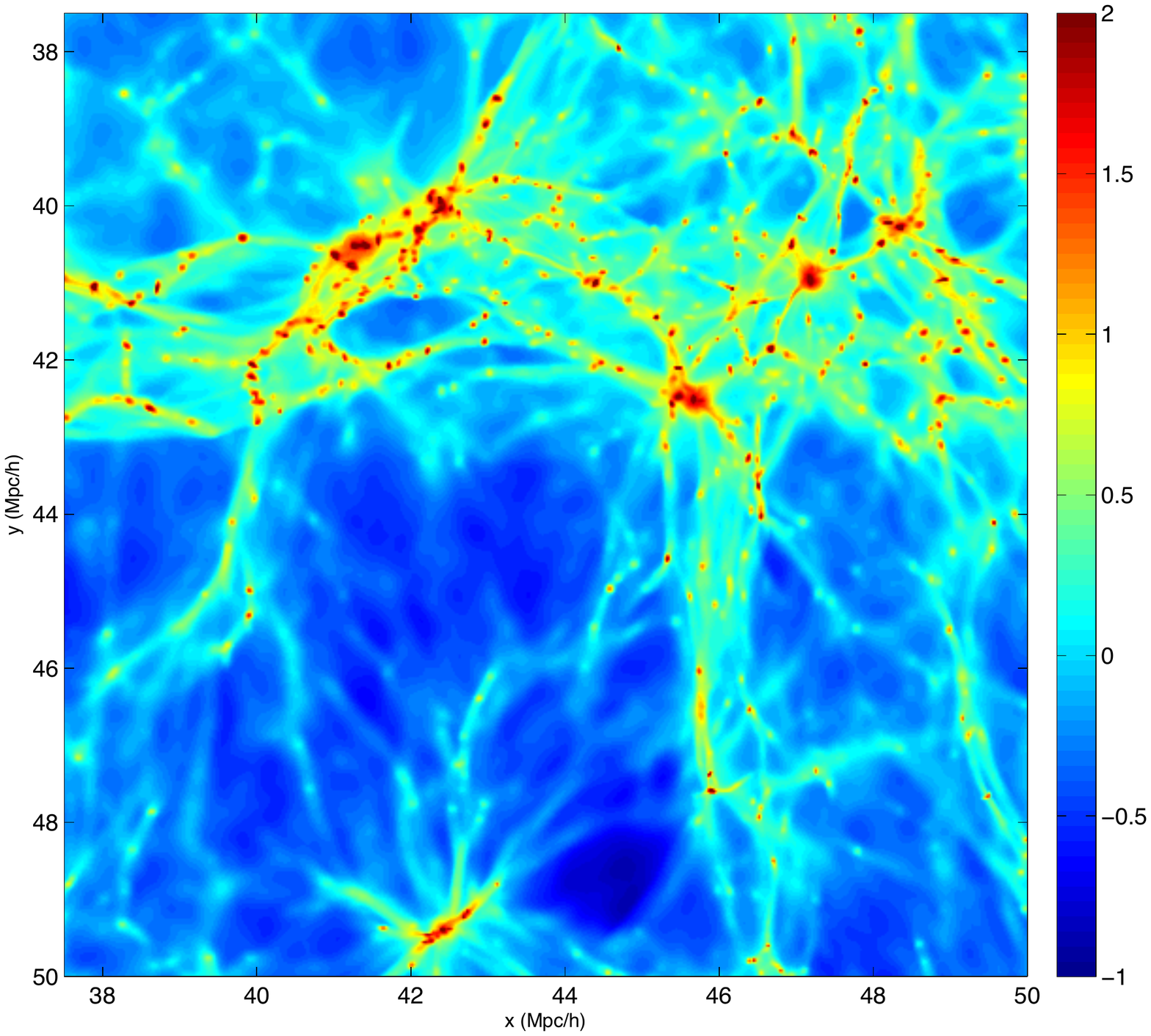,angle=0,width=3.4in} &
\hskip -0.4in
\epsfig{file=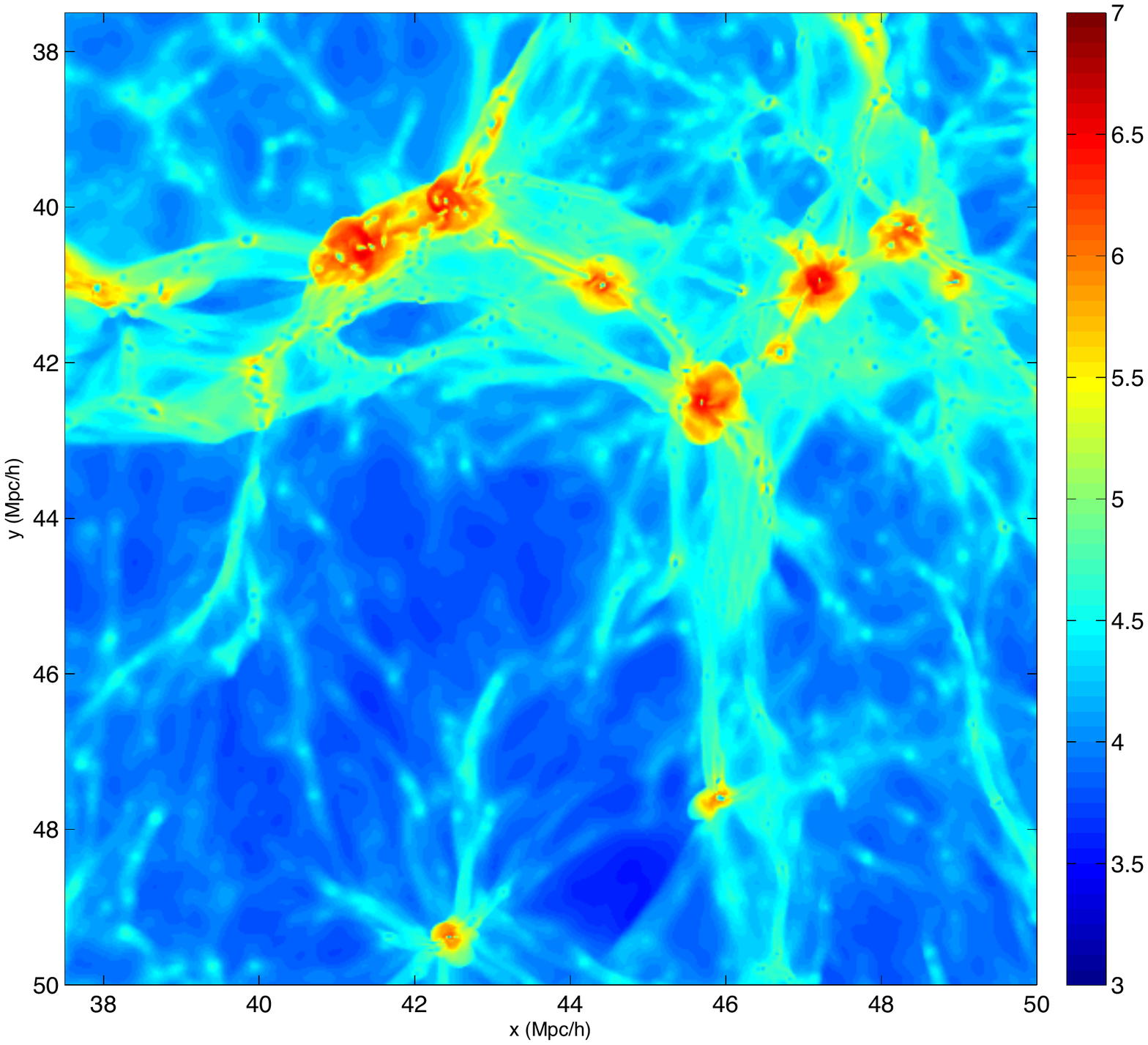,angle=0,width=3.4in} \\
\end{tabular}
\caption{
The top-left panel shows a slice of gas surface density in units of the mean gas surface
density at $z=2.6$ of size $50\times 50$(Mpc/h)$^2$ comoving and a depth of $3.125$Mpc/h comoving.
The \civ absorption lines are indicated by black asterisks, produced by sampling 
the slice using $8000$ random lines of sight.
The \ovi absorption lines are indicated by black circles. 
The top-right panel shows a zoom-in slice of the gas density of 
size $12.5\times 12.5$(Mpc/h)$^2$ comoving and a depth of $3.125$Mpc/h comoving,
corresponding to the lower right corner of the top-left panel,
while the bottom two panels show the corresponding gas temperature in Kelvin
and gas metallicity in solar units.
}
\label{fig:CIVslice}
\end{figure*}

We begin with a visual examination of density, temperature and metallicity distribution
of IGM at $z=2.6$ and compare cases with and without star formation feedback,
shown in Figure~\ref{fig:CIVslice}.
Comparing the density structures in runs with (middle left panel)
and without (bottom left panel) GSW 
we see that
the effect of GSW on the overall appearance of large-scale density structure is visually 
non-striking and the filamentary skeleton of the large-scale density distribution
remains intact.
An important and visually discernible effect of GSW is to ``smooth" out density concentrations in 
the dense (red) knots: the high density peaks ($>10^2$; red regions) in the run without GSW are 
substantially higher than those with GSW; 
examples include the knots at $(47,41)$Mpc/h, $(45.5,42.5)$Mpc/h and $(42,40)$Mpc/h.
This effect is of course reflective of the sensitivity on GSW of the SFR history, 
which in turn allowed the observations of SFR history to provide a powerful constraint on GSW,
as shown earlier in Figure~\ref{fig:sfr}.

The effect of GSW on low density (blue) regions seems small,
likely because GSW do not reach there and/or become weak even if reaching there.
The effect of the GSW on intermediate regions, a.k.a, filaments,
is most easily seen by comparing the temperature distributions
of the run with (middle right) and without (bottom right) GSW. 
We see that large-scale gravitational collapse induced 
shocks at this redshift tend to center on dense regions with a spatial extent that
is not larger than about $100-300$kpc/h;
these are virialization and infall shocks due to gravitational collapse 
of high density peaks.
Some of the larger peaks are seen to be enclosed by shocks of temperature
reaching or in excess of $10^7$~K (note that the displayed picture is inevitably
subject to smoothing by projection thus the higher temperature regions 
have their temperatures somewhat underestimated).
Galaxies form in the center of the filamentary structures
where collapse of pancake structures occurs.
Most of the shock heated volume from green ($10^5$K)
to red ($10^7$K) are clearly caused by GSW, 
because they appear prominent only in the simulation with GSW.
The GSW shock heated IGM seems to extend as far as $\sim 0.5$Mpc/h from galaxies.
The temperature of this shock heated gas falls in the WHIM temperature range
of $10^5-10^7$~K; we will discuss this more quantitatively in \S 3.2.
Inspecting the temperature (middle right) and metal density (top right) 
distribution with GSW reveals that metal enriched regions, ``metal bubbles",
coincide with temperature bubbles.
This indicates that GSW energy and metal deposition are tightly coupled. 
Most of the affected regions have a size of a few hundred kiloparsecs to about 
one megaparsec, suggesting that this is the range of influence of GSW
in transporting most of the metals to the IGM.
%The predicted effect has perhaps received 
%dramatic confirmation in the \ovi absorption measurements
%of Stocke \etal (2005),
%where they find that \ovi absorbers 
%are invariably found within $800$~kpc of galaxies.

We now inspect visually typical physical locations of \civ and \ovi absorption lines,
shown as asterisks (\civ) and circles (\ovi) in the top two rows in Figure~\ref{fig:CIVslice}.
The interesting feature is that 
\civ and \ovi absorbers tend to avoid ``voids" and are almost exclusively
located around filamentary structures with most of them seemingly residing in regions of
an overdensity of $\sim 3-30$;
however, limited resolution of our simulation prevent us from reaching firm conclusion on this at this time.
For every \civ absorber that is produced, there is
almost always an \ovi absorber along the same line of sight.
As we will see, all these paired-up \civ and \ovi in fact arise from around the same regions in space.
The converse is not necessarily true; a lower fraction of \ovi absorbers
do not have \civ counterparts within the depth of the projected slice of $3.125$Mpc/h comoving
and they tend to be
located in regions that are slightly
further away from high density peaks than those occupied
by \ovi lines with associated \civ lines.
The vast majority of both \civ and \ovi absorbers 
appear to be located in regions that have been swept by feedback shocks, as evidenced by
the similarly looking shock heated temperature bubbles 
(middle right panel of Figure~\ref{fig:CIVslice})
and metal enriched bubbles emanating from collective supernovae in star-forming galaxies
(upper right panel of Figure~\ref{fig:CIVslice}).
The \civ and \ovi lines, either collisionally ionized or photoionized,
unequivocally stem from regions that are shock heated and metal enriched by feedback from 
star formation; this conclusion will be 
confirmed quantitatively later.

The typical metallicity and temperature of the \civ and \ovi absorbers appear to be around
 $[C/H]\sim -2$ and $T\sim 10^{4.5-5.5}$K.
Typical $\lya$ forest clouds have comparable densities
but are at a significantly lower temperature, $T\sim 10^4$K and a lower metallicity $[C/H]\sim -3$. 
These properties indicate that, while most of the \civ and \ovi 
absorption lines may have comparable overdensity
compared to typical hydrogen \lyaf absorption lines
($N_{\rm HI}\sim 10^{13}-10^{15}$), the former are located in
somewhat hotter regions with somewhat higher metallicity than the latter.
Moreover, while many \civ and \ovi lines often coincide along the same line of sight
within a short distance, it will be shown that the actual gas properties of 
regions that produce them are significantly different.

\begin{figure*} %[h]
\centering
\vskip -0.7in
\begin{tabular}{cc}
\hskip -0.5in
\epsfig{file=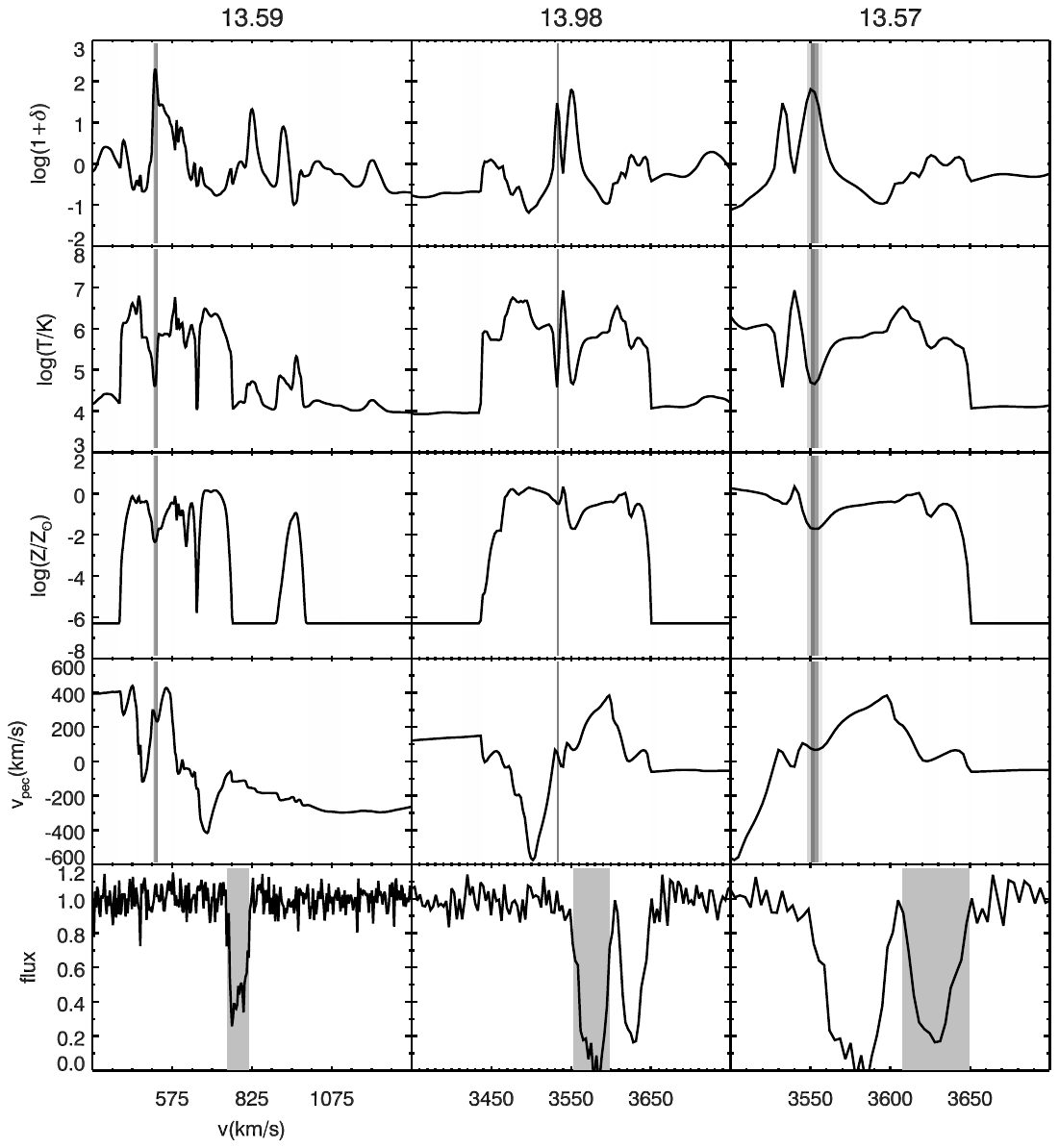,angle=0,width=5.35in,height=10in} &
\hskip -2.3in
\epsfig{file=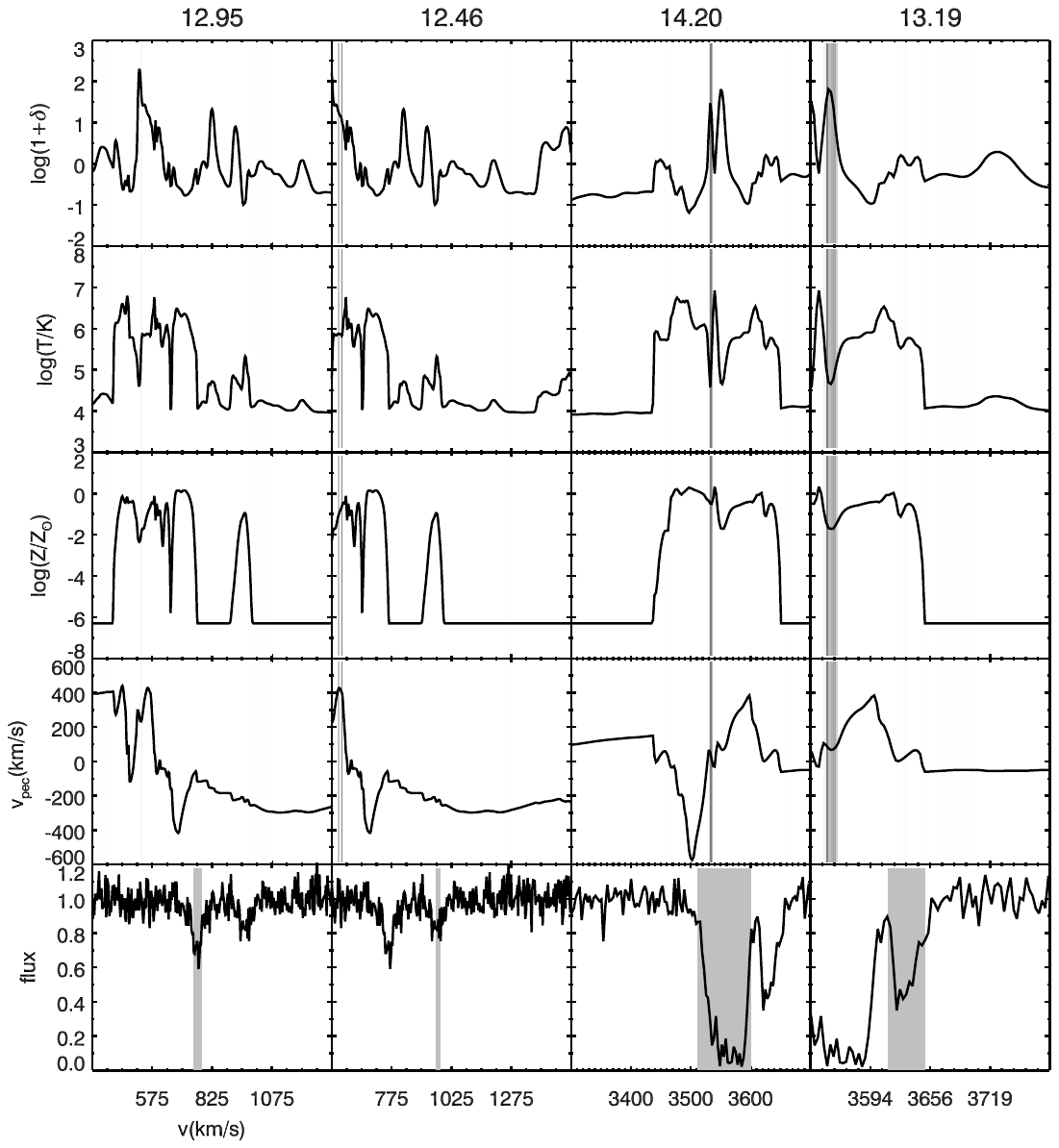,angle=0,width=5.35in,height=10in} 
\end{tabular}
\vskip -4.0in
\caption{
shows the physical properties of all \civ absorption lines (left)
and \ovi absorption lines (right) with column greater $10^{12}$cm$^{-2}$
along a random line of sight of length equal to the simulation boxsize of $50^{-1}$Mpc at $z=2.6$.
Small regions around of all identified \civ lines along each sightline are shown in separate columns.
Aside from the flux distribution shown at the bottom panel in velocity (Hubble) space,
all other panels of physical variables are shown in real space.
Each identified \civ absorption line in the bottom panel is indicated by a shaded region 
with the value of the log of its column density.
The corresponding physical location that produces the line is shown by
a shaded vertical line with dark shades indicating larger contributions to the column of the line.}
\label{fig:spectrum1}
\end{figure*}

\begin{figure} %[h]
\centering
\vskip -0.7in
\begin{tabular}{cc}
\hskip -0.5in
\epsfig{file=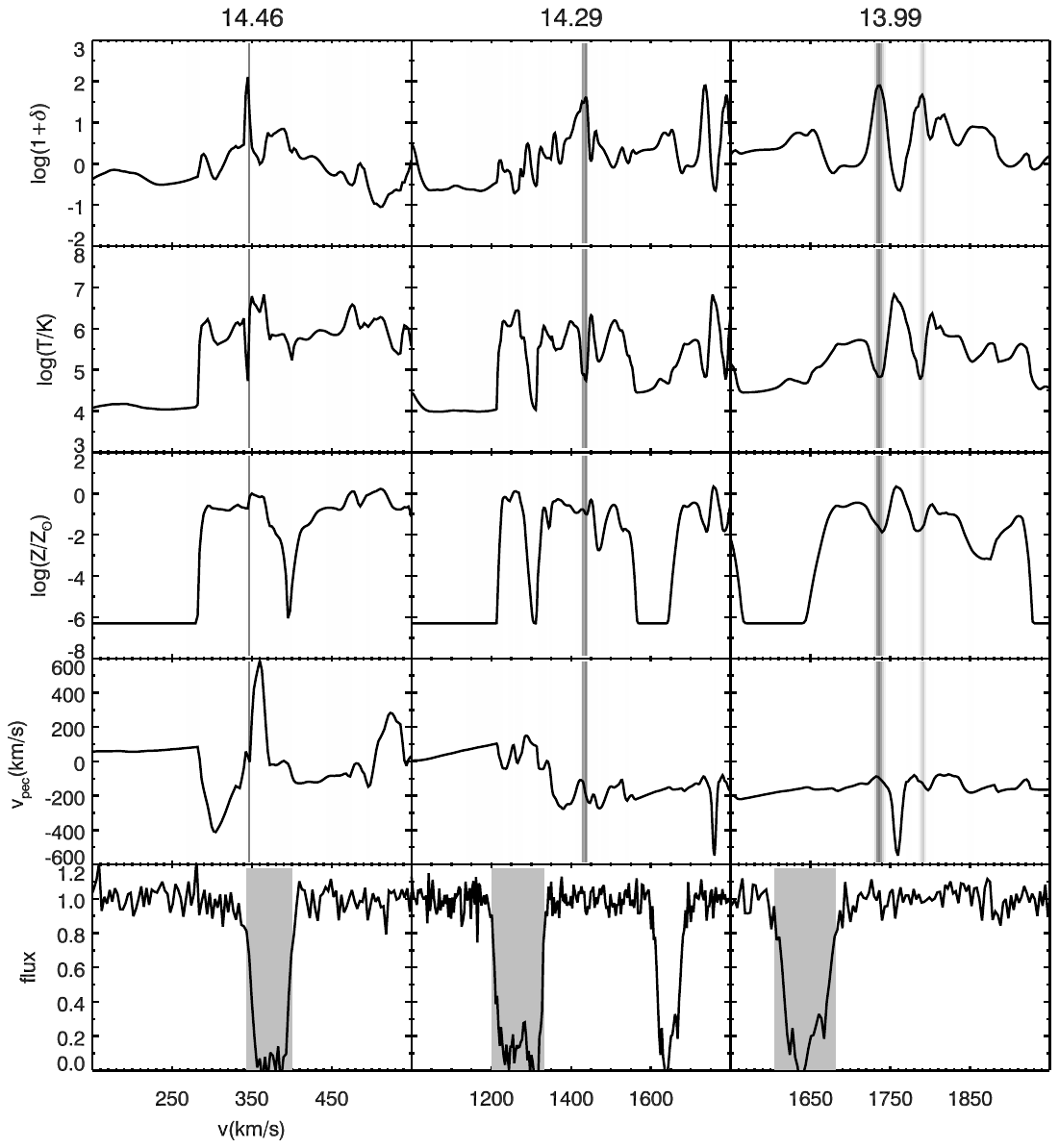,angle=0,width=5.35in,height=10in} &
\hskip -2.3in
\epsfig{file=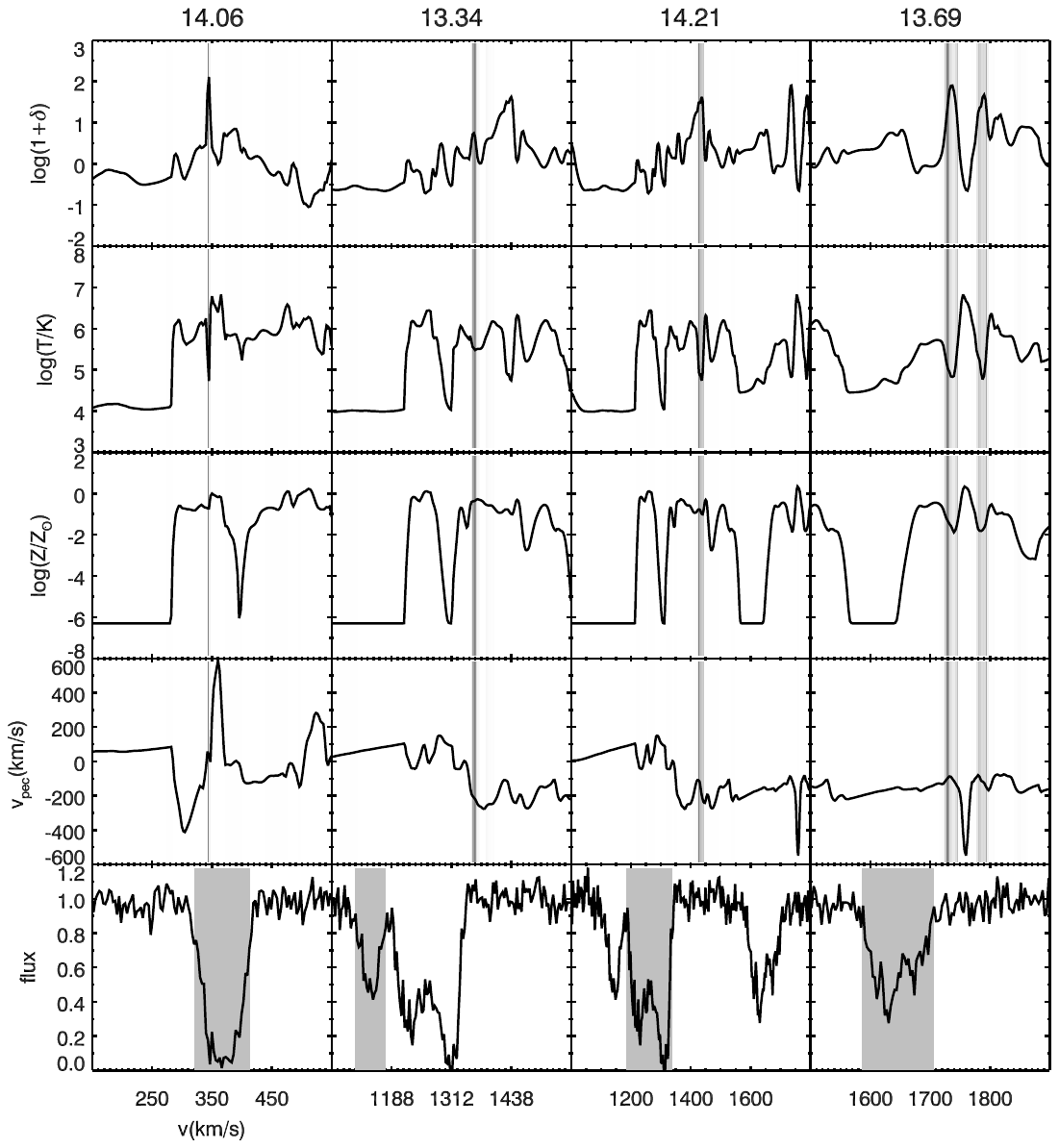,angle=0,width=5.35in,height=10in} 
\end{tabular}
\vskip -4.0in
\caption{
this is similar to 
Figure~\ref{fig:spectrum1} but for another random line of sight.
}
\label{fig:spectrum2}
\end{figure}

\begin{figure} %[h]
\centering
\vskip -0.7in
\begin{tabular}{cc}
\hskip -0.5in
\epsfig{file=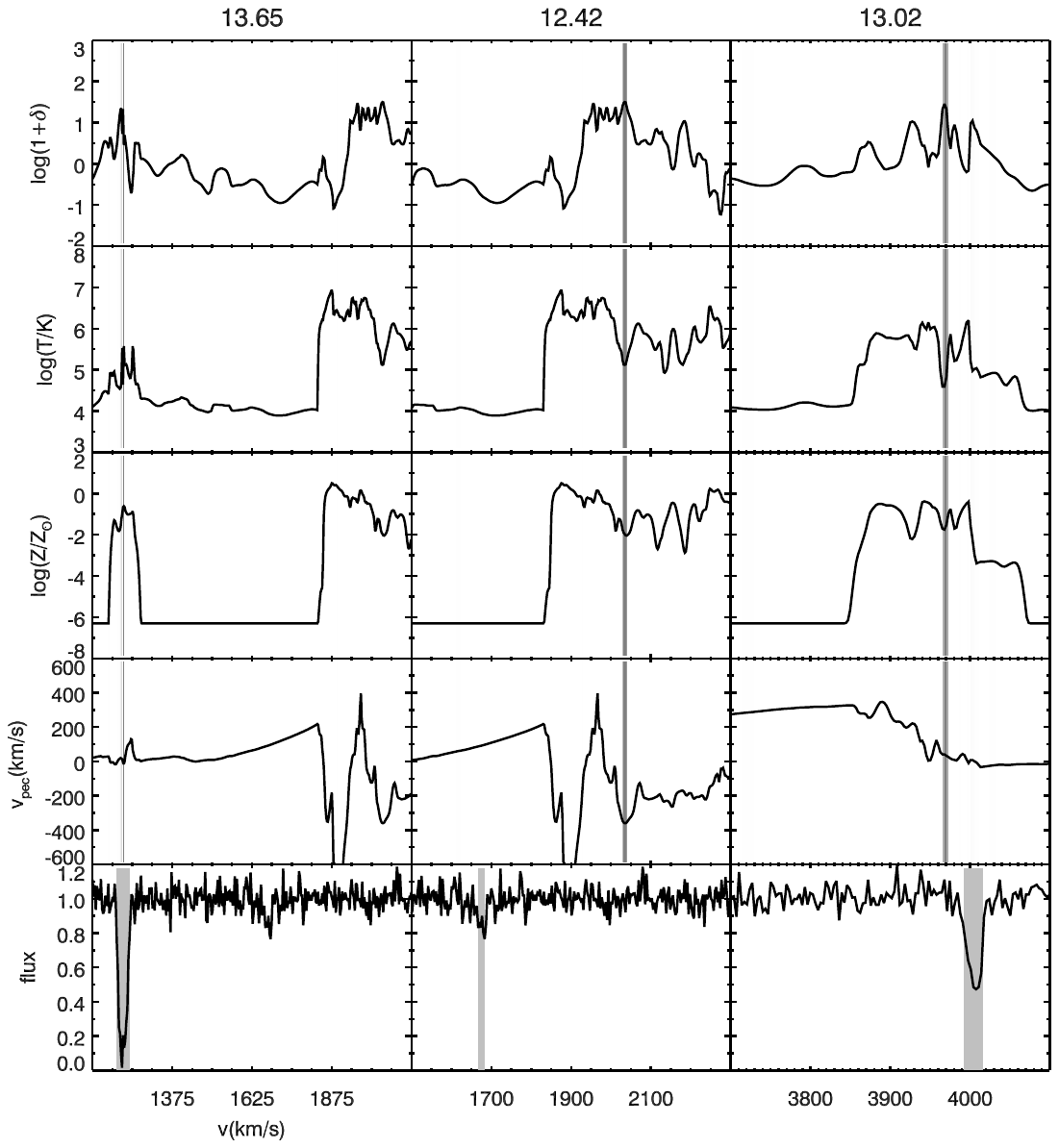,angle=0,width=5.35in,height=10in} &
\hskip -2.3in
\epsfig{file=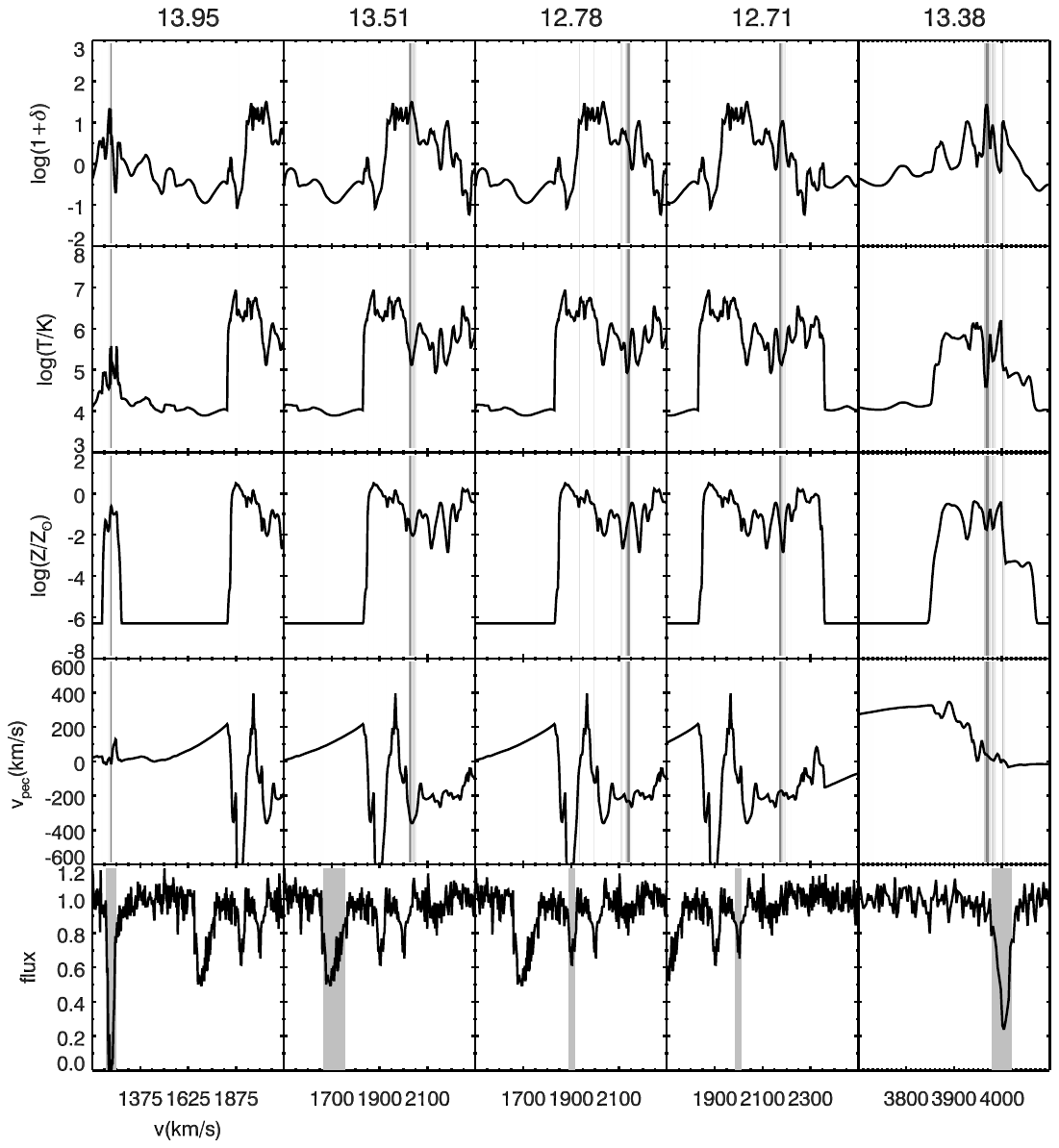,angle=0,width=5.35in,height=10in} 
\end{tabular}
\vskip -4.0in
\caption{
this is similar to 
Figure~\ref{fig:spectrum1} but for another random line of sight.
\vskip 2.0in
}
\label{fig:spectrum3}
\end{figure}

Let us now examine the physical properties of \civ and \ovi absorbers in greater detail.
Figures~\ref{fig:spectrum1},\ref{fig:spectrum2},\ref{fig:spectrum3}  
show three random sightlines through the simulation box.
In order to better see details we have concatenated all the zoomed-in regions around identified 
\civ and \ovi lines for each sightline
to one panel, separated into columns.
The left panels are for \civ lines and right for \ovi lines.
Several interesting properties of \civ and \ovi absorbers may be gleaned.
First, both \civ and \ovi absorbers sit in regions 
with significantly elevated temperature 
(i.e., $>2\times 10^4$K)
of widths of $\sim 100$km/s or larger, i.e., a few hundred physical kiloparsecs or larger,
which are then connected with the general photo-ionized IGM of
lower temperature of $\sim 10^4$K
(2nd row from top in Figures~\ref{fig:spectrum1},\ref{fig:spectrum2},\ref{fig:spectrum3}).
The density structures
(top row in Figures~\ref{fig:spectrum1},\ref{fig:spectrum2},\ref{fig:spectrum3})
show that the densities in the regions of allevated temperatures
span a wide range from $\delta\sim 0$ to $\sim 100$
and there is no clear positive correlation between density and temperature 
(although there is a strong anti-correlation between them near density peaks).
This suggests that the elevated temperatures in these regions    
are not caused by gravitational compression.
It is also clearly seen that 
at the two locations demarcating each high temperature region, there is a shock-like density jump (of a factor of a few).
A closer examination of the peculiar velocity structures
(2nd panel from bottom in Figures~\ref{fig:spectrum1},\ref{fig:spectrum2},\ref{fig:spectrum3})
shows evidence of a double shock propagating outward, with the shock fronts coincidental with
the temperature and density jump.

Second, there is a tight correlation between gas temperature and gas metallicity
(middle row in Figures~\ref{fig:spectrum1},\ref{fig:spectrum2},\ref{fig:spectrum3})
in the sense that higher temperatures have higher metallicity and 
each region with elevated temperature is bordered by a synchronous drop in both
temperature and metallicity on two sides.
This is a strong indication that the elevated temperature is caused by a double shock
originating from a alaxy or small group of galaxies due to GSW,
which plays the double role of both shock heating the surrounding IGM and metal-enriching it.
To reiterate this important point, \civ and \ovi absorbers are located in regions that have been swept through by
metal-enriched feedback shocks, which are still propagating outward and ``separate"
the \civ and \ovi absorbers from the general IGM of temperature $T\sim 10^4$K by about $100$km/s or more.
Because of the high temperatures probed by \civ and \ovi lines, 
they are not in general correlated with $\lya$ lines on scales $\le 100$km/s.
The latter probe 
typically lower temperatures. 
Overall, the locations of \civ and \ovi lines are closely correlated.
The overall spatial extent of \ovi lines, in terms of their distance from galaxies,
are somewhat larger than that of \civ lines, as seen in 
Figure~\ref{fig:CIVslice} and Figures~\ref{fig:spectrum1},\ref{fig:spectrum2},\ref{fig:spectrum3}
and will be verified by their origin being in somewhat lower density gas than \civ lines
(see Figure~\ref{fig:CIVdnddelta} below). 

%This paragraph is hereby modified.
%Third, a closer look at the exact locations of \civ and \ovi absorbers
%reveals that each region that produces a \civ absorption line of column density $10^{12}-10^{15}$cm$^{-2}$
%mostly often, not always, reside within $\sim 50$km/s or $\sim 0.5$Mpc/h comoving 
%from density peaks of height $\sim 100$ times the mean density.
%This is strongly suggestive that these moderate column density \civ absorptions lines tend to be
%located within $\sim 0.5$Mpc/h comoving from a galaxy, in good agreement with
%observations \citep[e.g.,][]{2001Chen, 2005bAdelberger}. 
%This is also consistent with the maximum gas temperatures in the vicinities of the density peaks being 
%below $10^7$K, 
%which indicates that the associated galaxies are likely $\le L_*$. 

Third, many \civ absorbers appear to be paired up with \ovi absorbers.
For brevity, our convention is that we count absorption lines from left to right in each panel.
For example, 
the first and fourth \ovi lines  in the right panel 
can be respectively paired up with the first and third \civ lines in the left panel 
of Figure~\ref{fig:spectrum1};
the first, third and fourth \ovi lines in the right panel
can be respectively paired up with the first, second and third \civ lines in 
the left panel of Figure~\ref{fig:spectrum2};
the second and fifth \ovi lines in the right panel 
can be respectively paired up with the second and third \civ lines in 
the left panel of Figure~\ref{fig:spectrum3}.
The \ovi lines that appear together with \civ lines 
seem to have relatively low temperature ($T\sim 10^{4.5}-10^5$K), probably
with a significant photoionization component.
Note that collisional ionization makes maximum contribution to \ovi production
at $T=10^{5.5}$K, whereas for \civ this happens at $T=10^{5.0}$K.
Thus, it appears that relatively low-temperature \ovi lines 
are often paired up with a \civ line, for which both photoionization and collisional ionization
may be relevant.
The excess of \ovi lines compared to the number of \civ lines 
is likely due to the difference in the number of collisionally ionized cases
for the two lines, given the difference in the optimal temperatures 
for collisional ionization for \civ and \ovi lines.
Note that with collisional ionization alone, the abundance of each species
drops when the temperature moves away from the optimal temperature to either side (lower or higher) - 
a factor of $\sim 10$ drop when temperature differs from the optical temperature by a factor of two.
Roughly speaking, while the probability of an associated \ovi line for a given \civ line is close to unity,
the probability of an associated \civ line for a given \ovi 
is somewhat lower.  A more detailed study of this issue will be performed in
sections to come.

%*****************************************************************************************
\begin{figure} %[h]
\centering
\vskip -0.7in
\hskip -2.5in
\epsfig{file=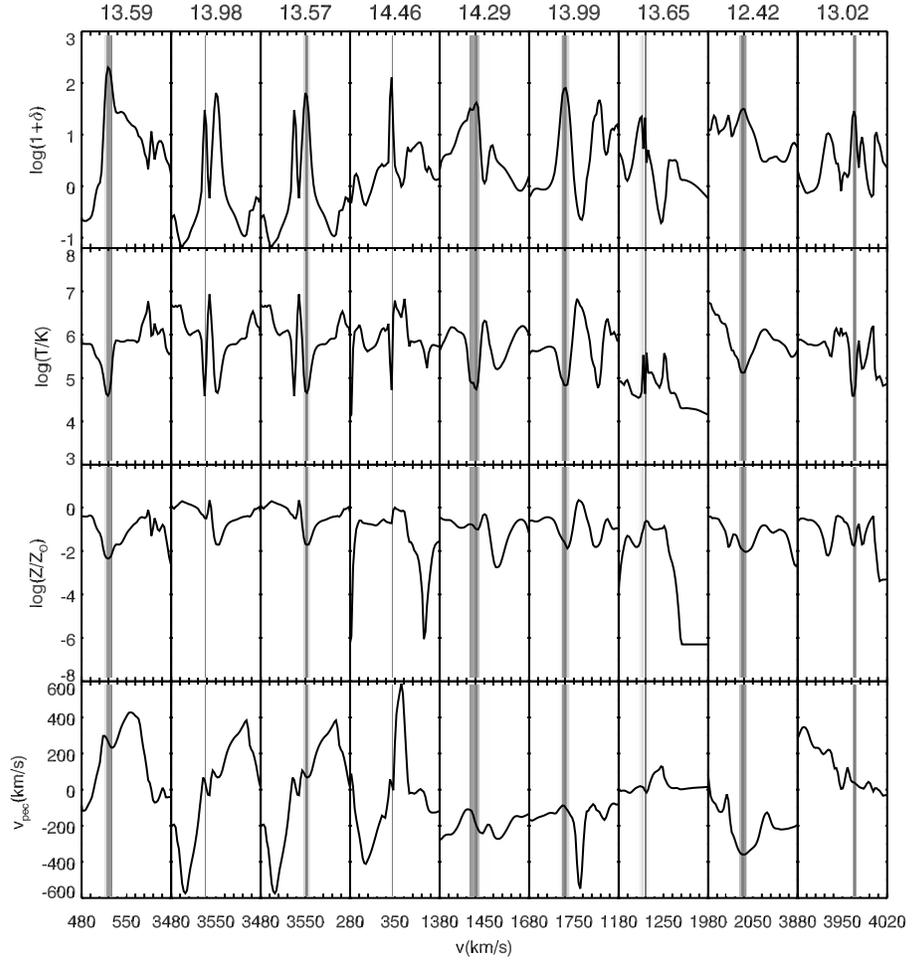,angle=0,width=9.00in}%,height=10in} 
\vskip -5.0in
\caption{
shows a close-up view of the region around each \civ line in real space,
where the physical size along the line of sight has been translated to
velocity using $\Delta v=H(z) \Delta x$.
Each tickmark is $10$\kms.
}
\vskip 2.0in
\label{fig:zoomCIVlines}
\end{figure}

%*******************************************************************************************
\begin{figure} %[h]
\centering
\vskip -0.7in
\hskip -2.5in
\epsfig{file=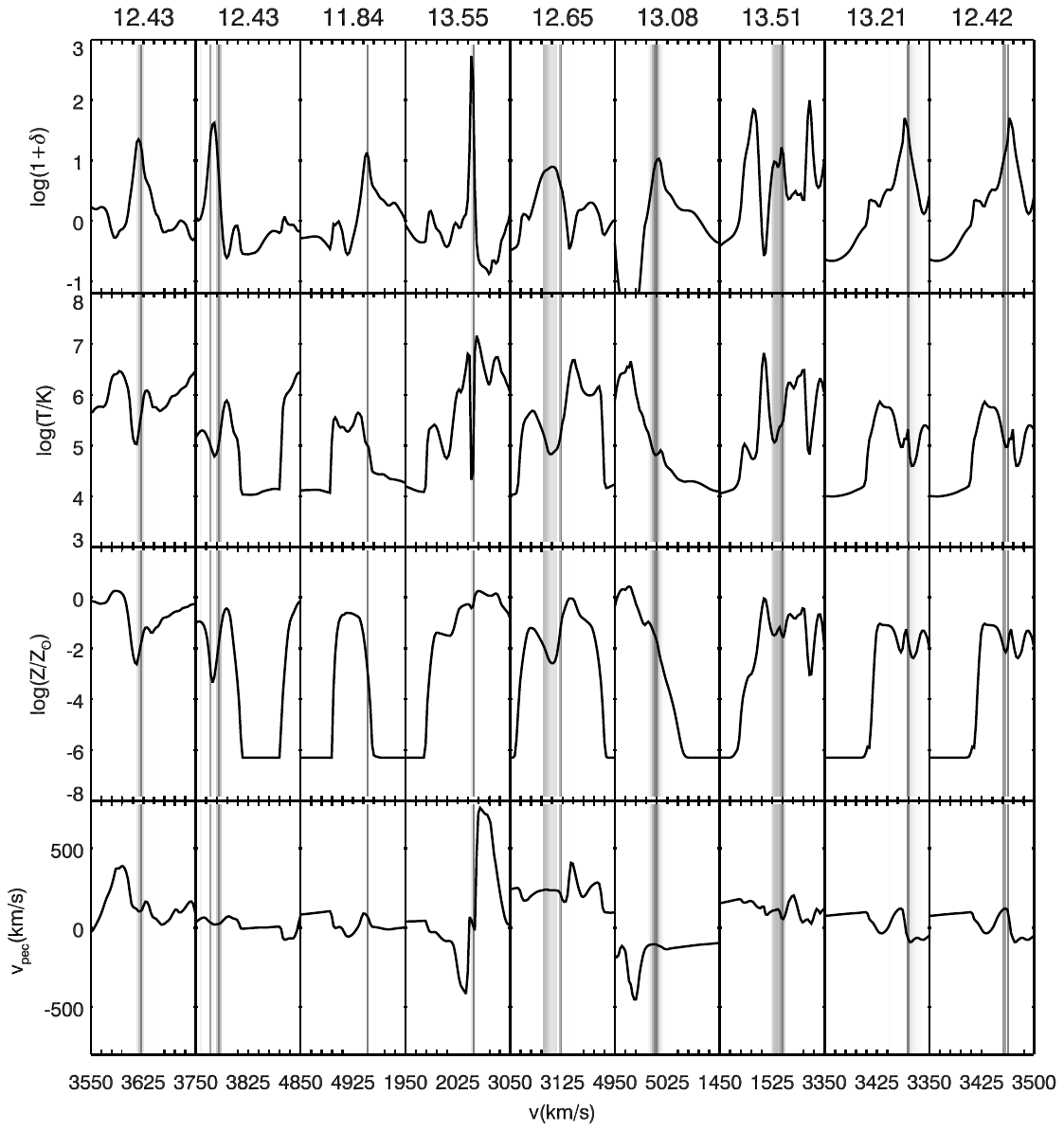,angle=0,width=9.00in}%,height=10in} 
\vskip -5.0in
\caption{
shows a close-up view of the region around each \ovi line in real space,
where the physical size along the line of sight has been translated to
velocity using $\Delta v=H(z) \Delta x$.
Each tickmark is $10$\kms.
}
\vskip 2.0in
\label{fig:zoomOVIlines}
\end{figure}
%**********************************************************************************************

Finally, in Figure~\ref{fig:zoomCIVlines} we show a close-up view of several randomly chosen \civ lines.
It is clear that the regions contributing to a \civ line
tend to be centered or nearly centered on a local density peak along the line of sight,
which almost always corresponds to a trough in temperature.
It is also evident that the spatial extent of the \civ producing region is limited
to about up to $10$\kms, corresponding to about comoving $100$kpc/h,
with some regions much narrower than that.
As a consequence, even though the velocity gradients in the intermediate vicinities
(i.e., the whole surrounding region of elevated temperature)
of \civ-producing regions are often large (with $dv/dr\sim$ a few $100\kms$ per comoving Mpc),
the velocity gradients in the actual \civ-producing regions is smaller,
which, in conjunction with the narrowness of the \civ-producing region,
limits the velocity contribution to the Doppler width, as will be shown quantitatively later.
Physically, this tells us that
each \civ absorber tends to arise primarily
from a narrow region in real space
that have previously thermalized through feedback shocks, have cooled and 
are presently relatively quiescent.
There does not appear to be a visible correlation between the LOS size of \civ lines and
the column density; some of the high column \civ lines shown (the second and fourth panel
from left) appear to come from very narrow regions of size $\ll 100$kpc comoving which
appear to have very steep velocity gradients (for example,
the fourth from left line with log of column equal 14.46).
%%FIX THIS
%Our findings are in agreement with observations of \citet{1996Rauch} that
%indicate a quiescent environment for \civ systems,
%but in disagreement with their interpretation of them being ensembles of dwarf galaxies.
The \civ lines are mostly intergalactic in origin, not from inside galaxies.

We next examine several randomly chosen \ovi lines in close-up
shown in Figure~\ref{fig:zoomOVIlines} and make
 detailed comparisons of the physical properties with \civ lines, when possible.
We note three points.
First, in Figures~\ref{fig:spectrum1},\ref{fig:spectrum2},\ref{fig:spectrum3}  
we noted that most \civ lines ($\ge 10^{13.5}$) 
have associated \ovi lines that have comparable column densities.
This indicates that both \civ and \ovi lines of relatively high column ($\ge 10^{13.5}$) 
tend to arise in regions in or near density peaks and temperature troughs.
Second, a typical \ovi line tends to have a lower column density due to 
a steeper column density distribution of \ovi lines (see Figure~\ref{fig:CIVdndN} below).
Third, \ovi lines often lie in regions that are offset from density peaks
by $\sim 10-100$\kms, and often these density peaks 
do not have corresponding temperature troughs.
This is clear evidence that many, lower column \ovi lines arise from regions that 
are not physically bound and instead they are mostly transient, stemming from density 
and temperature fluctuations in shock heated regions in the neighborhood of galaxies.
It may be that the steeper column density distribution for \ovi lines
has its origin in the abundance of these more transient structures.
The low density, shock heated regions may have 
temperatures that are too high to produce equally abundant \civ lines in conjunction with 
a lower abundance of carbon than oxygen.

%***************************************************
\begin{figure*}[h]
\centering
\vskip -0.5in
\begin{tabular}{cc}
\hskip -0.75in
\epsfig{file=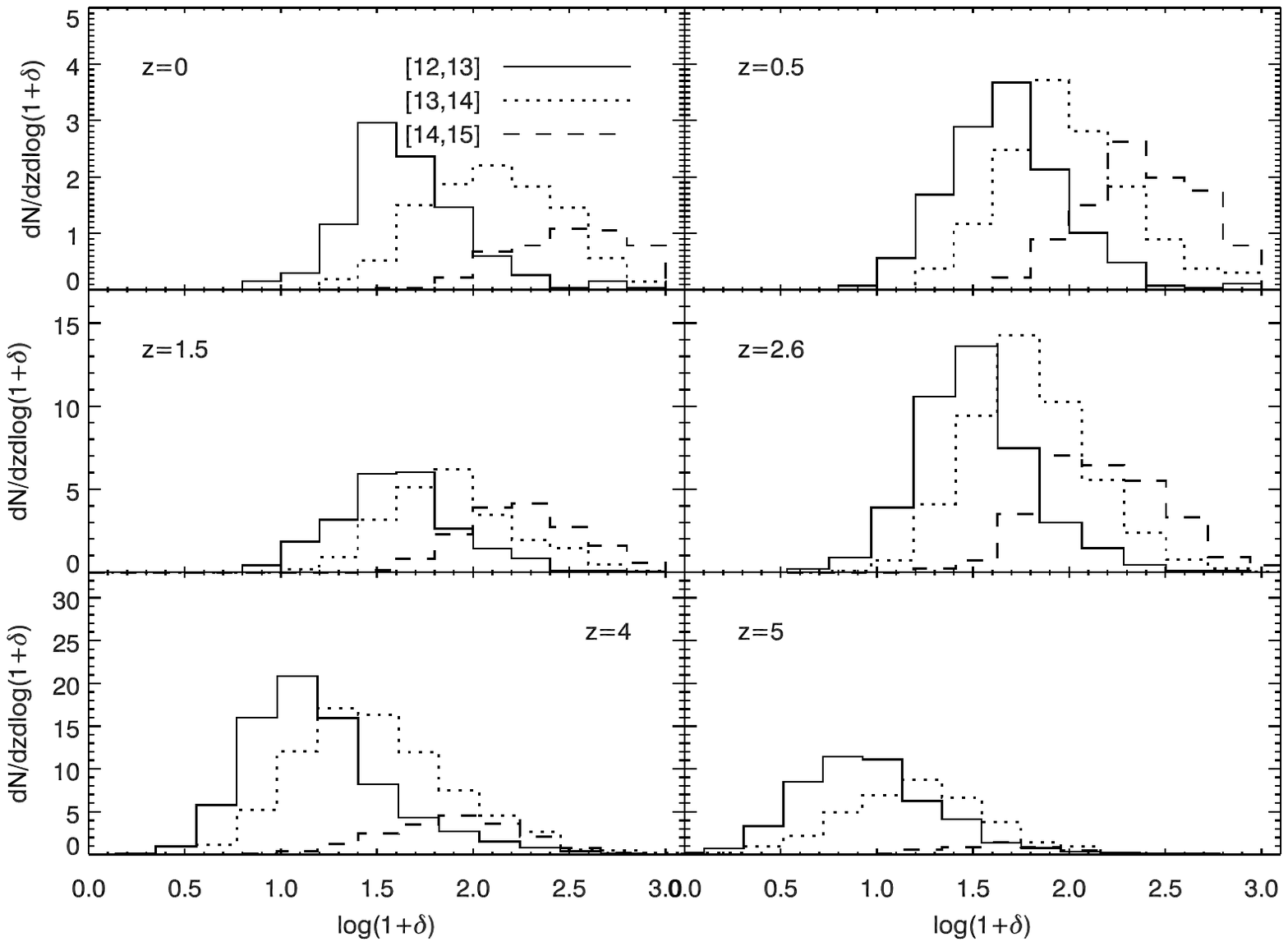,angle=0,width=4.3in} &
\hskip -1.1in
\epsfig{file=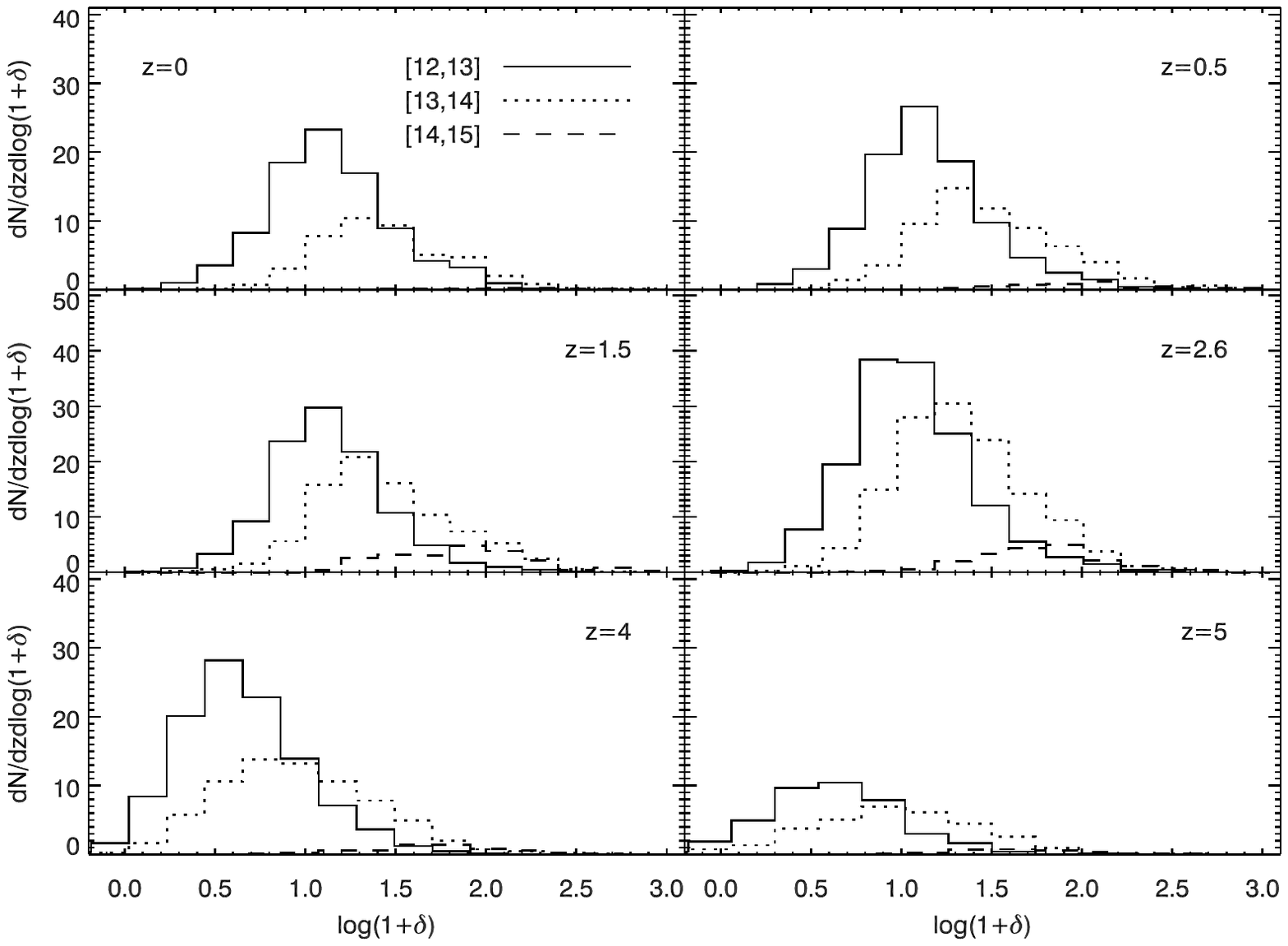,angle=0,width=4.3in} \\
\end{tabular}
\vskip -2.5in
\caption{
Left panel shows the distribution of gas overdensity of regions that produce the CIV absorption lines
at six different redshifts, $z=0,0.5,1.5,2.6,4,5$,
separately for three subsets of lines of column density in the range
of $\log$N$_{\civ} {\rm cm}^2$=[12,13],[13,14],[14,15], respectively.
Right panel shows the counterpart for \ovi absorption lines.
}
\label{fig:CIVdnddelta}
\end{figure*}
%**********************************************************

We now quantify the properties of \civ  and \ovi absorbers by different projections through the
multi-dimensional parameter space spanned by several fundamental physical variables.
Figure~\ref{fig:CIVdnddelta}  
shows the distribution of gas overdensity for \civ  (left) and 
\ovi absorbers (right) at six different redshifts, $z=(0,0.5,1.5,2.6,4,5)$.
First, a comparison of the three histograms for three subsets of \civ and \ovi absorbers
in each panel indicates that higher column \civ and \ovi absorbers are produced, 
on average, by higher density gas.
Second, there is a clear trend that 
\civ absorbers trace increasingly more overdense regions
with decreasing redshift.
For example, while the location of the vast majority of \civ absorbers with $\log($N$_{\civ}{\rm cm}^{2})=[12,13]$
appears to be outside virialized regions (i.e., overdensity less than about $100$)
at $z>2.6$, a significant fraction of them reside in 
virialized regions at $z<1.5$;
the same is true for higher column \ovi absorbers.
A comparison to \ovi absorbers reveals a striking contrast:
the vast majority of \ovi absorbers with $\log($N$_{\civ}{\rm cm}^{2})\le 14$ are located 
outside virialized regions {\it at all redshifts}.
In addition, typical \ovi lines arise
from somewhat lower density regions than \civ lines.
%I would change this:
For example, for 
\ovi absorbers of $\log($N$_{\civ}{\rm cm}^{2})=[12,13]$, the typical overdensity
peaks at $\delta\sim 5$ for \ovi absorbers versus $\sim 10$ for \civ lines at $z=2.6-5$, 
which jumps to $\delta\sim 10$ for \ovi absorbers versus $\sim 50$ for \civ absorbers at $z=1.5$.
A more quantitative analysis of the cross correlation between \civ and \ovi absorption lines
and galaxies will be presented in a later paper.

%*********************************************************
\begin{figure*}[h]
\centering
\vskip -0.5in
\begin{tabular}{cc}
\hskip -0.75in
\epsfig{file=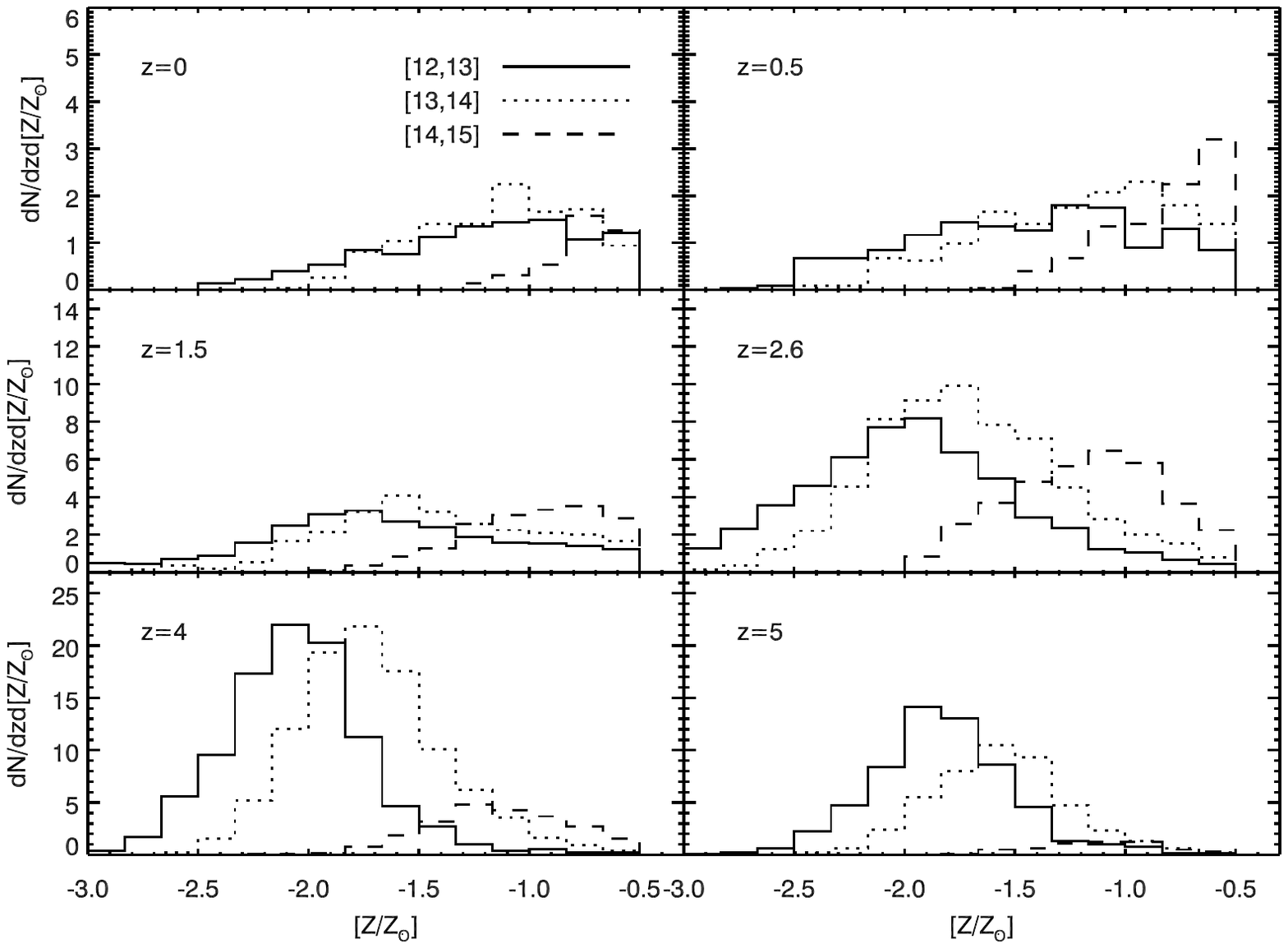,angle=0,width=4.3in} &
\hskip -1.1in
\epsfig{file=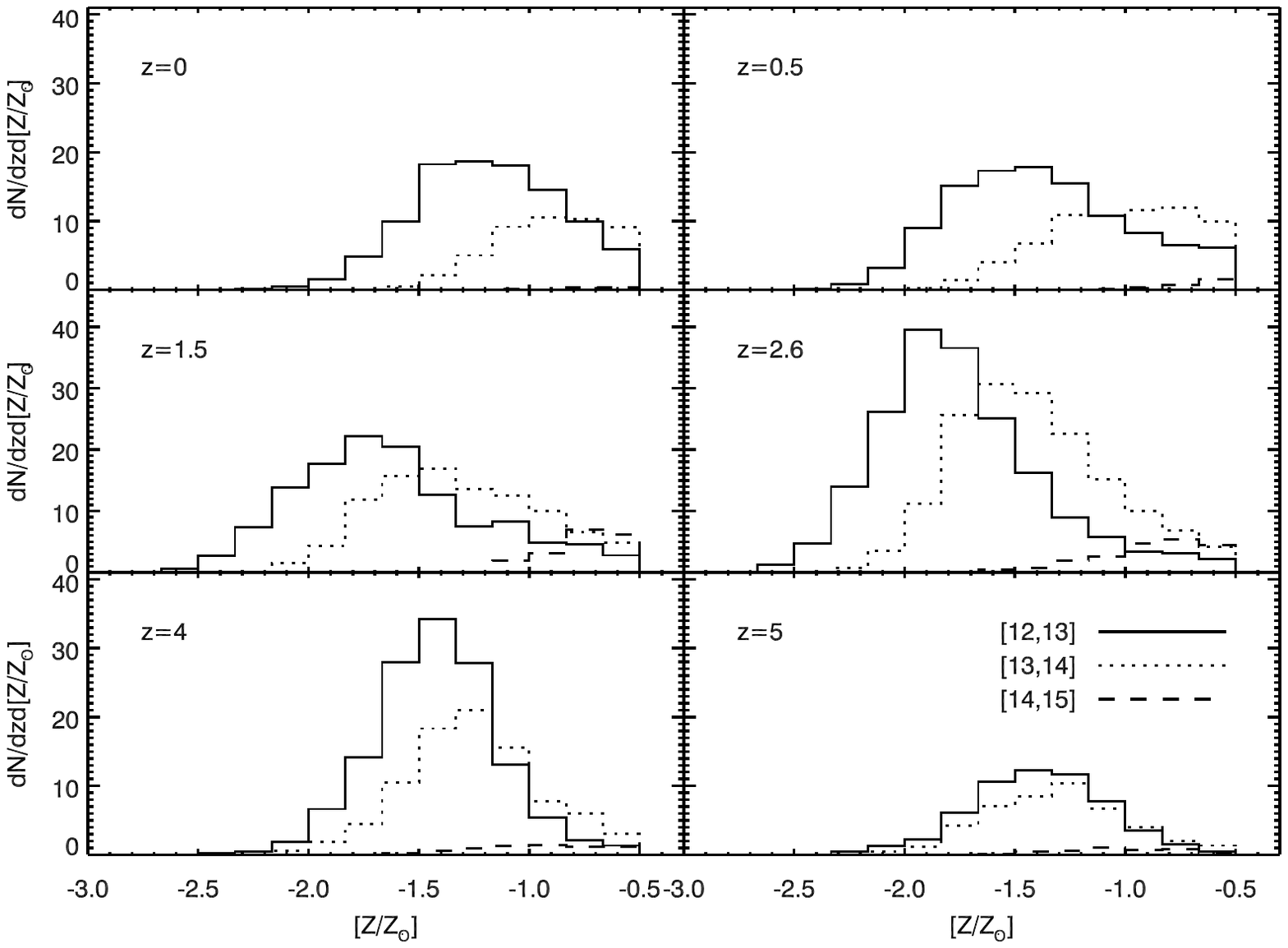,angle=0,width=4.3in}  \\
\end{tabular}
\vskip -2.5in
\caption{
The left panel shows the distribution of gas metallicity in solar units 
of regions that produce the CIV absorption lines
at six different redshifts, $z=0,0.5,1.5,2.6,4,5$,
separately for three subsets of lines of column density in the range
of $\log$N$_{\civ} {\rm cm}^2$=[12,13],[13,14],[14,15], respectively.
Right panel shows the counterpart for \ovi absorption lines.}
\label{fig:CIVdndZ}
\end{figure*}
%**********************************************************

Figure~\ref{fig:CIVdndZ}  
shows the distribution of gas metallicity for \civ  (left) and 
\ovi absorbers (right) at six different redshifts, $z=(0,0.5,1.5,2.6,4,5)$.
We see that \civ absorption lines arise from gas with a wide range of
metallicity from [C/H]=-3 to -0.5, peaked approximately around -2.5 to -1.5
at $z>0.5$.
%The relative distribution of \ovi and \civ absorbers shifts with time.
At $z>2.6$ the distribution for \ovi lines is roughly like taking the left end 
of each corresponding \civ distribution and squeezing the whole distribution rightward
by an amount of $\sim 0.5-1.0$.
So the metallicity distributions for \ovi absorbers are generally cut off at a higher
metallicity than those for \civ absorbers at the low end by about $0.5-1.0$ and
peak at a metallicity that is higher by this factor. 
The situation appears to start reversing at $z=1.5$ such that
at $z<0.5$ the fraction of high metallicity \civ absorbers
exceeds that of \ovi absorbers.
What is also interesting is that 
the typical metallicity of \civ and \ovi lines 
displays a non-monotonic trend at a fixed column density.
For \ovi absorbers, at $z=4-5$ the metallicity of \ovi lines with 
$\log($N$_{\ovi}{\rm cm}^{2})=[12,14]$ peaks at $[Z/\zsun]=-1.5$ to $-1.0$,
which moves to a lower value of $[Z/\zsun]=-2.0$ to $-1.5$ at $z=2.6$,
then slightly moves back up to $[Z/\zsun]\sim -1.5$ at $z=(1.5,0.5,0)$.
For comparison, the overall behavior for \civ lines is as follows:
the metallicity of \civ lines with 
$\log($N$_{\civ}{\rm cm}^{2})=[12,14]$ peaks at $Z=-2.0$ to $-1.5$ at $z=5$,
at $[Z/\zsun]\sim -2$ at $z=4$,
followed by a very broad distribution peaking at $Z=-2$ to $-1$ at $z=1.5$ to $z=2.6$
with a larger fraction reaching a relatively high metallicity gas with $[Z/\zsun]>-1$.

The overall trend in metallicity evolution with redshift for the \civ and \ovi absorbers
could be understood as follows. Let us first note
that the ionizing radiation background at $z=4,5$ is about ($1/3,1/30$) of that $z=2.6$, which in turn is larger than that at $z=(1.5,0.5,0)$ by a factor of $\sim (2,7,30)$.
At $z=4-5$ both \civ and \ovi absorbers
are predominantly collisionally ionized with the temperatures
peaking at $10^5$K and $10^{5.5}$K, respectively, as shown below
in Figure~\ref{fig:CIVdndT}. 
These regions are relatively closer to galaxies, from which metal-carrying
shocks originate and have relatively high metallicities.
At lower redshift $z=2.6$ larger regions around galaxies have been enriched with metals
and the rise of the ionizing radiation background produces a large population
of photoionized \civ and \ovi lines at lower temperature and lower metallicity.
%changed to < rather than =
Towards still {\bf lower redshift $z<1.5$}, the decrease of the mean gas density in the universe
demands a rise in overdensity of the \ovi-bearing gas in order to produce a comparable
column density, causing a shift of these regions to be closer to galaxies where both
metallicity and density are higher, seen in Figure~\ref{fig:CIVdnddelta}. 

%As a side note, since damped Lyman alpha systems (DLAs) are believed to arise within galaxies
%(at least at redshift $z\le 2$) and have a typical (very weakly evolving) metallicity of $\sim -1$ 
%\citep[e.g.,][]{1999aPettini, 2000Prochaska},
%this trend of \civ and \ovi metallicity is consistent with the fact that 
%a significant fraction of the absorbers
%come from regions inside virial radii of galaxies 
%at lower redshift, as seen in Figure~\ref{fig:CIVdnddelta}. 
%Clearly, the temperatures of DLAs ($T\le 10^4$K) and \civ and \ovi absorbers ($T\le 10^{4.5}-10^{5.5}$K) 
%are very different.
%The spatial coexistence of DLAs systems and higher 
%column density \civ and \ovi absorbers ($\log($N$_{\civ}{\rm cm}^2)\ge 14-15$)  
%inside galaxies at widely disparate temperatures, densities and pressures, but at comparable metallicity,
%is indicative of efficient metal mixing inside galaxies.

The combination of lower density (Figure~\ref{fig:CIVdnddelta}) 
and higher metallicity (Figure~\ref{fig:CIVdndZ})  
for the typical (low) column density \ovi absorbers compared to \civ absorbers
is reminiscent of metal-carrying shocks propagating through inhomogeneous
medium, exactly the situation one would expect of the feedback shocks 
from galaxies entering the highly inhomogeneous IGM.
Given the widespread steep density gradients (steeper than $-2$) in regions
just outside the virial radius of galaxies,
these shocks could not only heat up lower density regions to higher temperatures but
also enrich them to higher metallicity.
The feedback shocks generically propagate in a direction that 
has the least resistance and 
is roughly perpendicular to the orientation of a local filament where a galaxy sits,
as seen clearly in Figure~\ref{fig:CIVslice} and shown
previously \citep[e.g.,][]{2002bTheuns, 2005Cen}.
While higher density regions, on average, tend to have higher metallicity
(as we will show later), the dispersion is sufficiently large that 
the reverse and other complex situations often occur in some local regions.
This appears to be what is happening here, at least for some regions that
manifest in \civ and \ovi lines.

%*********************************************************
\begin{figure*}[h]
\centering
\vskip -0.5in
\begin{tabular}{cc}
\hskip -0.75in
\epsfig{file=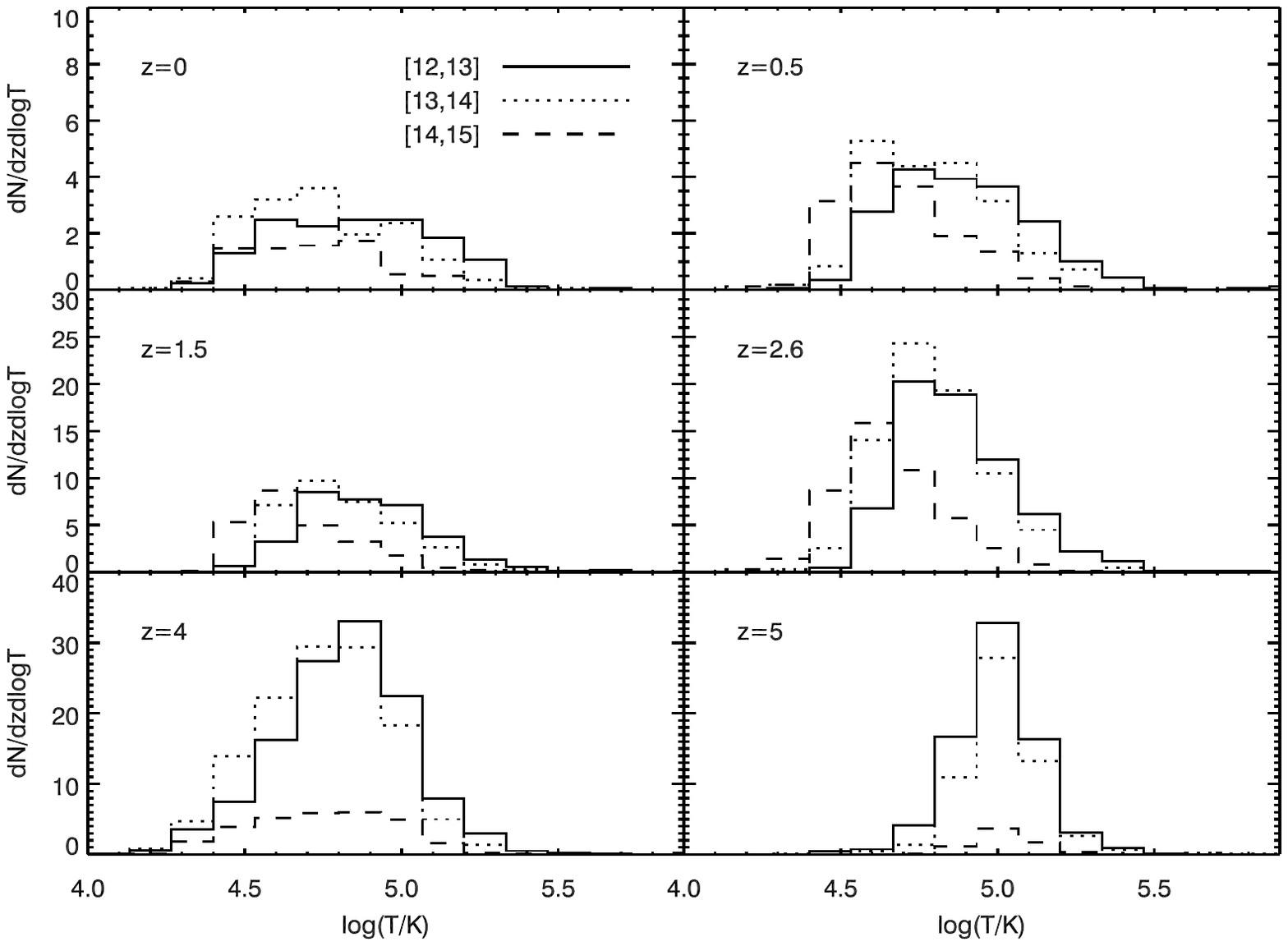,angle=0,width=4.3in} &
\hskip -1.1in
\epsfig{file=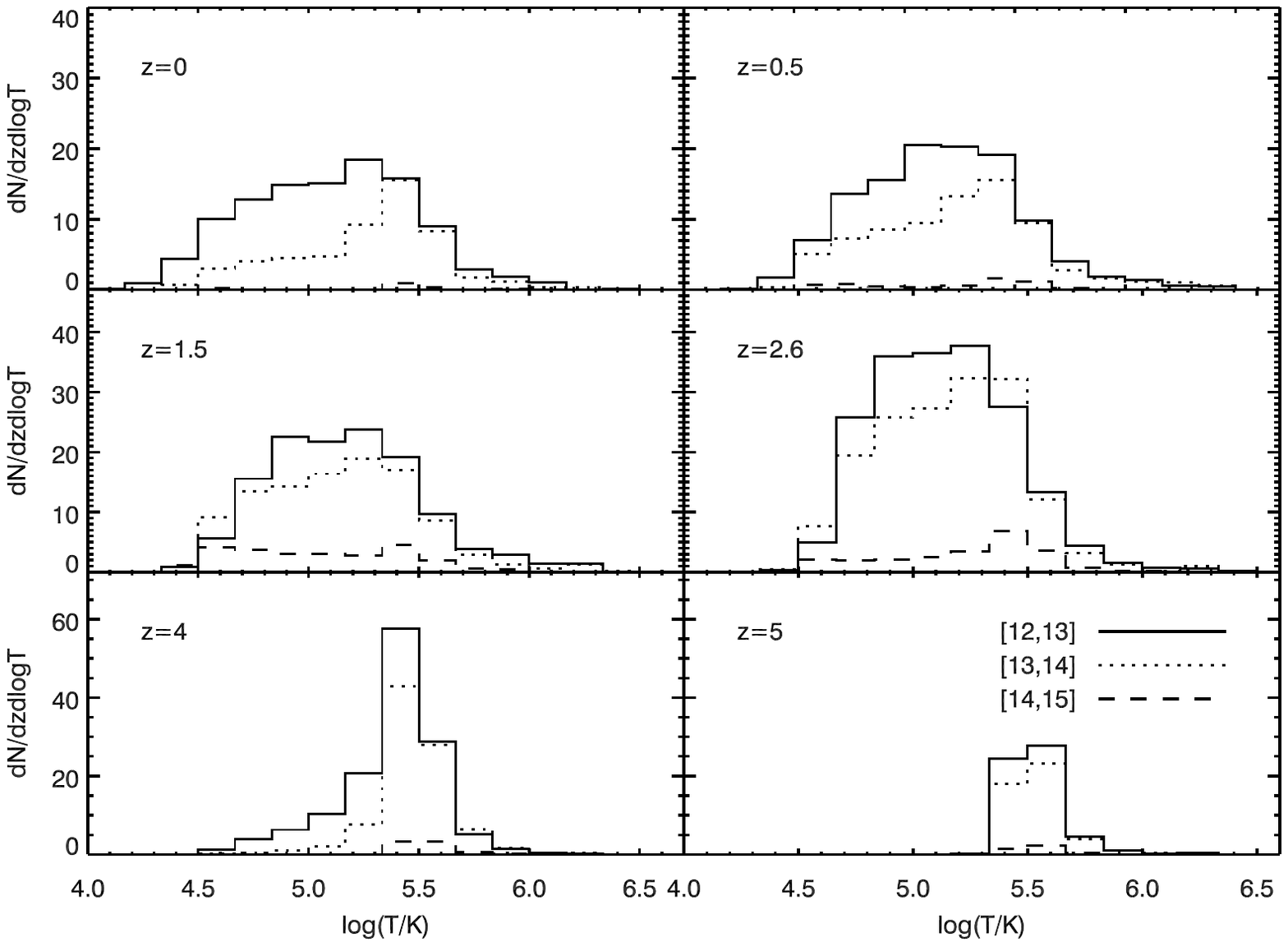,angle=0,width=4.3in}  \\
\end{tabular}
\vskip -2.5in
\caption{
Shows the distribution of gas temperature of regions that produce the \civ absorption lines
at six different redshifts, $z=0,0.5,1.5,2.6,4,5$,
separately for three subsets of lines of column density in the range
of $\log$N$_{\civ} {\rm cm}^2$=[12,13],[13,14],[14,15], respectively.
Right panel shows the counterpart for \ovi absorption lines.
}
\label{fig:CIVdndT}
\end{figure*}
%**********************************************************

Figure~\ref{fig:CIVdndT}  
shows the distribution of gas temperature for \civ (left) and \ovi (right) absorbers.
We see that the temperatures of \civ absorbers at $z=5$ and \ovi absorbers at $z=4-5$
narrowly peak at $10^{5}$K and $10^{5.5}$K, respectively,
suggesting that collisional ionization makes the dominant contribution
to both species and the two types of absorbers arise from different regions.
The rapid drop in the amplitude of the UV radiation background beyond $z=3$ 
and increase in gas density with $(z+1)^3$ is the primary reason for diminished
component of photoionized \civ and \ovi absorbers at these high redshifts.
At redshift $z<2.6$ the distributions for the two absorbers become
progressively broader ranging from $10^{4.3}$K to $10^{5.5}$K for \civ absorbers,
and from $10^{4.3}$K to $10^{6}$K for \ovi absorbers. 
Thus, at $z<2.6$
both \civ and \ovi absorbers are a mixture of photoionized and collisionally ionized ones.
For both  \civ and \ovi lines, 
while the temperature distributions of \ovi lines at $z<2.6$ 
are broad, there is no significant segregation in temperature
of lines of different column densities.
Recall that there is a noticeable correlation between column density
and overdensity for both \ovi lines and \civ lines (Figure~\ref{fig:CIVdnddelta}).
This is likely indicative of complex, inhomogeneous nature of metal enrichment process around galaxies.

%***********************************************
\begin{figure*}[h]
\centering
\vskip -0.5in
\begin{tabular}{cc}
\hskip -0.75in
\epsfig{file=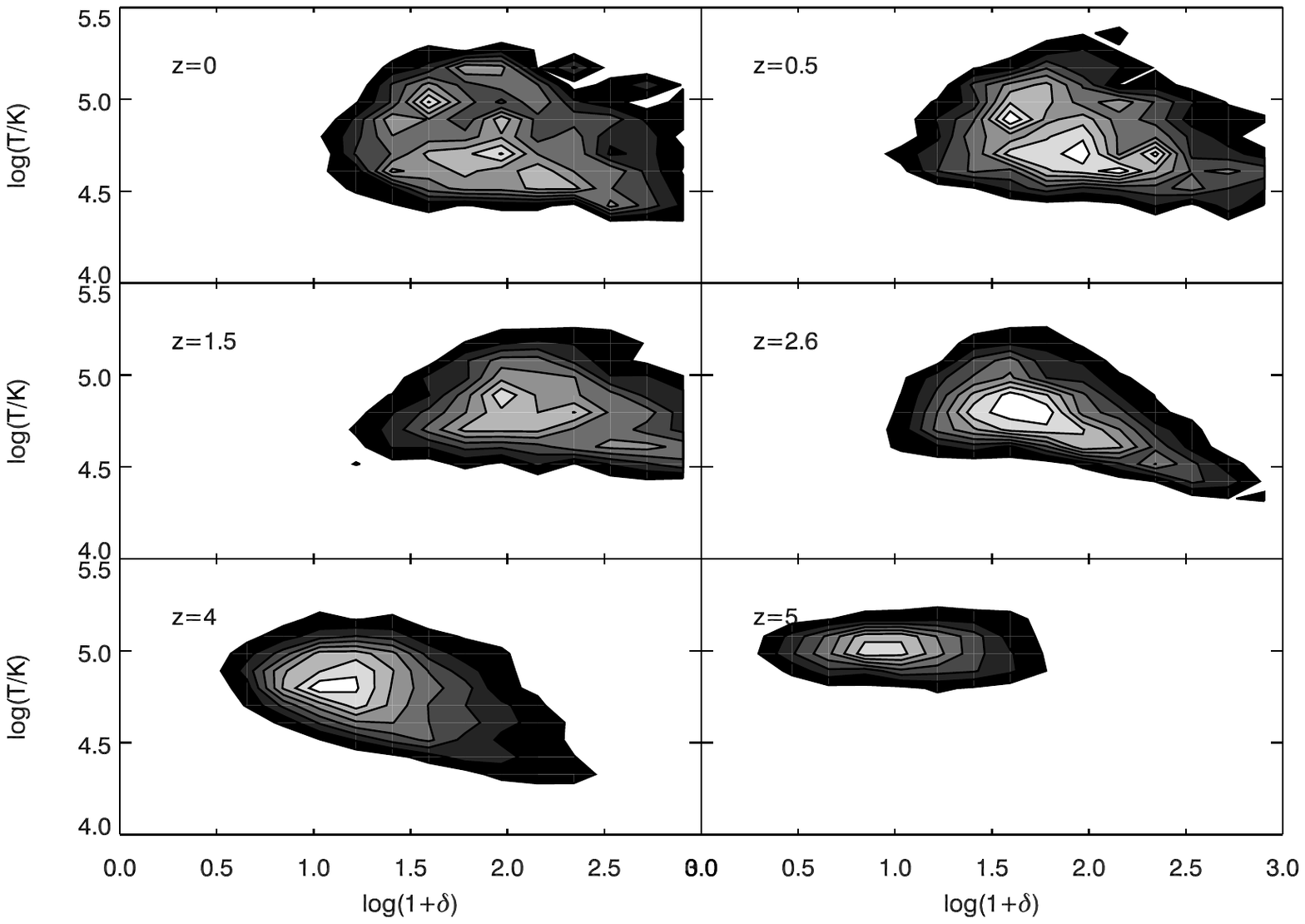,angle=0,width=4.1in} &
\hskip -1.1in
\epsfig{file=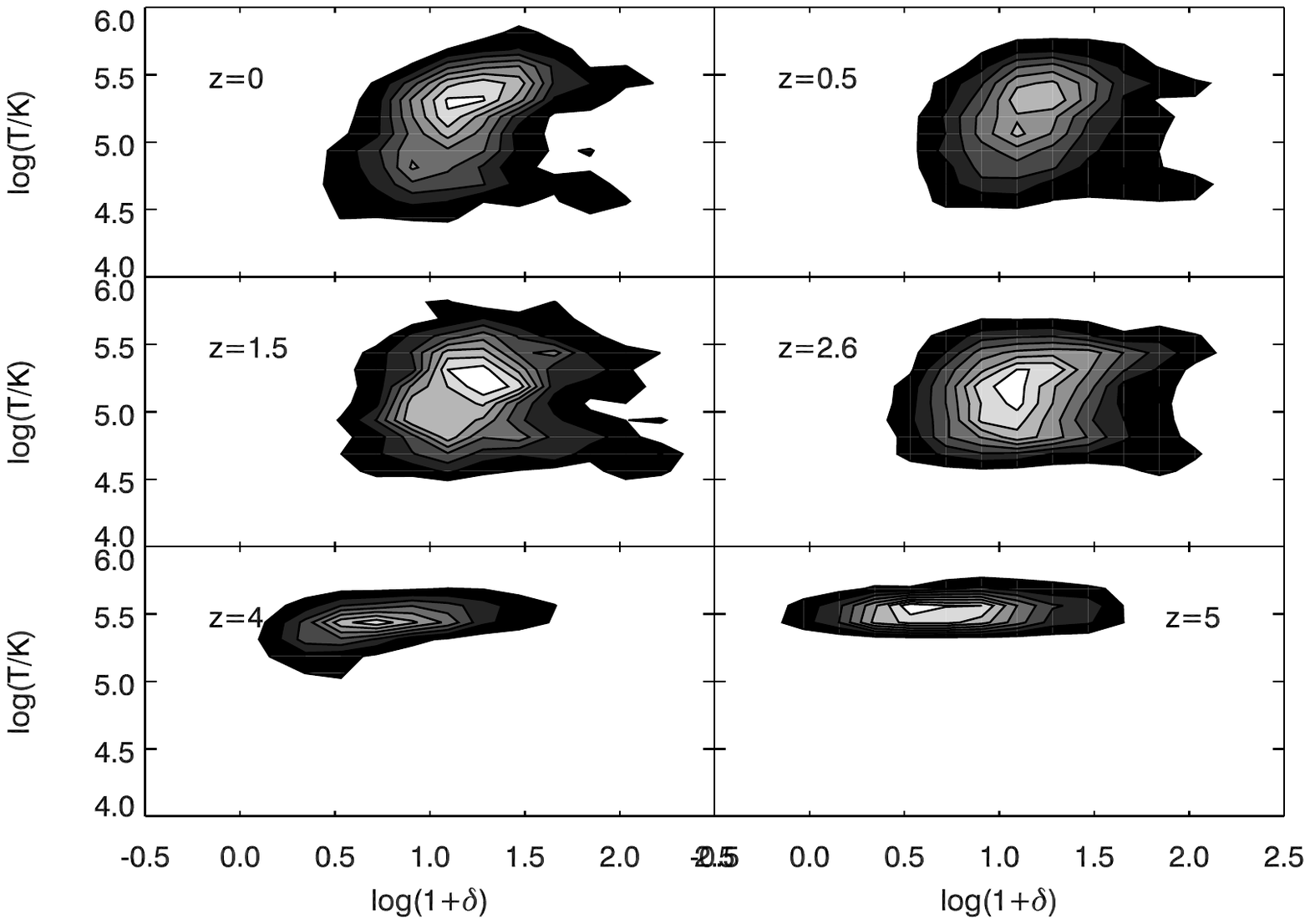,angle=0,width=4.1in}  \\
\end{tabular}
\vskip -2.4in
\caption{
The left panel shows the column density-weighted distribution of \civ lines
in the overdensity-temperature plane
at six different redshifts, $z=0,0.5,1.5,2.6,4,5$.
Right panel shows the counterpart for \ovi absorption lines.
}
\label{fig:CIVTdelta}
\end{figure*}
%****************************************************

Figure~\ref{fig:CIVTdelta} displays 
the distribution of \civ (left) and \ovi (right) absorbers in the overdensity-temperature plane
at redshift $z=0,0.5,1.5,3,4,5$.
We again see relatively narrow peaked temperature distribution at 
redshift $z=4-5$ for \ovi absorbers,
whereas at the same redshifts the \civ absorbers have a relatively broader temperature distribution.
Towards lower redshift there appears to be a multi-modal distribution in temperature for \civ absorbers, 
with the lower temperature peak at $T\sim 10^{4.2-4.5}$K
being progressively more important with decreasing redshift and becoming
dominant by $z=0$.
The lower temperature peak is photoionized. 
At redshift $z=1.5-2.6$ a higher temperature peak at $T\sim 10^{4.5-4.8}$K
is dominant, which is likely a mixture of collisional and photoionization.
It is interesting to note that at $z=2.6$ the radiation background is high enough
to allow for the existence of a small peak at ($\delta \ge 200$, $T=10^{4.2}$K) for \civ absorbers,
clearly arising from gas that is within virialized regions.
At redshift $z=0-0.5$ the peak at $T\sim 10^{4.5-4.8}$K is still prominent.
But, another peak at still higher temperature of $T\sim 10^{5.0-5.2}$K
emerges, which is likely dominated by collisional ionization.
Overall, the composition of \civ absorbers changes from being dominated by collisional ionization
at $z=4-5$, through a mixture of collisional and photoionization at $z=1.5-2.6$,
to being dominated by photoionization by $z=0$.
The distinct high temperature peak at $T\sim 10^{5}$K and density $\delta\sim 20$ at $z=0$
is rooted in the Warm-Hot Intergalactic Medium (WHIM; \citet[][]{1999Cen, 2001Dave, 2006Cen}),
where the intergalactic medium has been heated up by gravitational shocks due to the formation
of the large-scale structure.

A similar progression from mainly collisionally ionized to a mixture of collisional
and photoionization for \ovi absorbers is also seen.
However, for \ovi absorbers, the photoionization peak at 
$T\le 10^{5}$K never dominates at any redshift.
For both \civ and \ovi absorbers there is no visible correlation between overdensity and temperature for \ovi absorbers
at all redshifts. 
For example, there is no evidence of these regions obeying 
the so-called equation of state \citep[][]{1997HuiGnedin} that is applicable 
to low redshift $\lya$ forest clouds.
This just reinforces the statement that 
these regions are shock heated, in a dynamical state and perhaps transient,
and do not resemble photo-heated $\lya$ forest region.
We have also plotted (not shown)
the distribution of \civ and \ovi absorbers in the overdensity-metallicity plane
and find no visible correlation between them.

%**********************************************************
\begin{figure*}[h]
\centering
\vskip -0.5in
\begin{tabular}{cc}
\hskip -0.5in
\epsfig{file=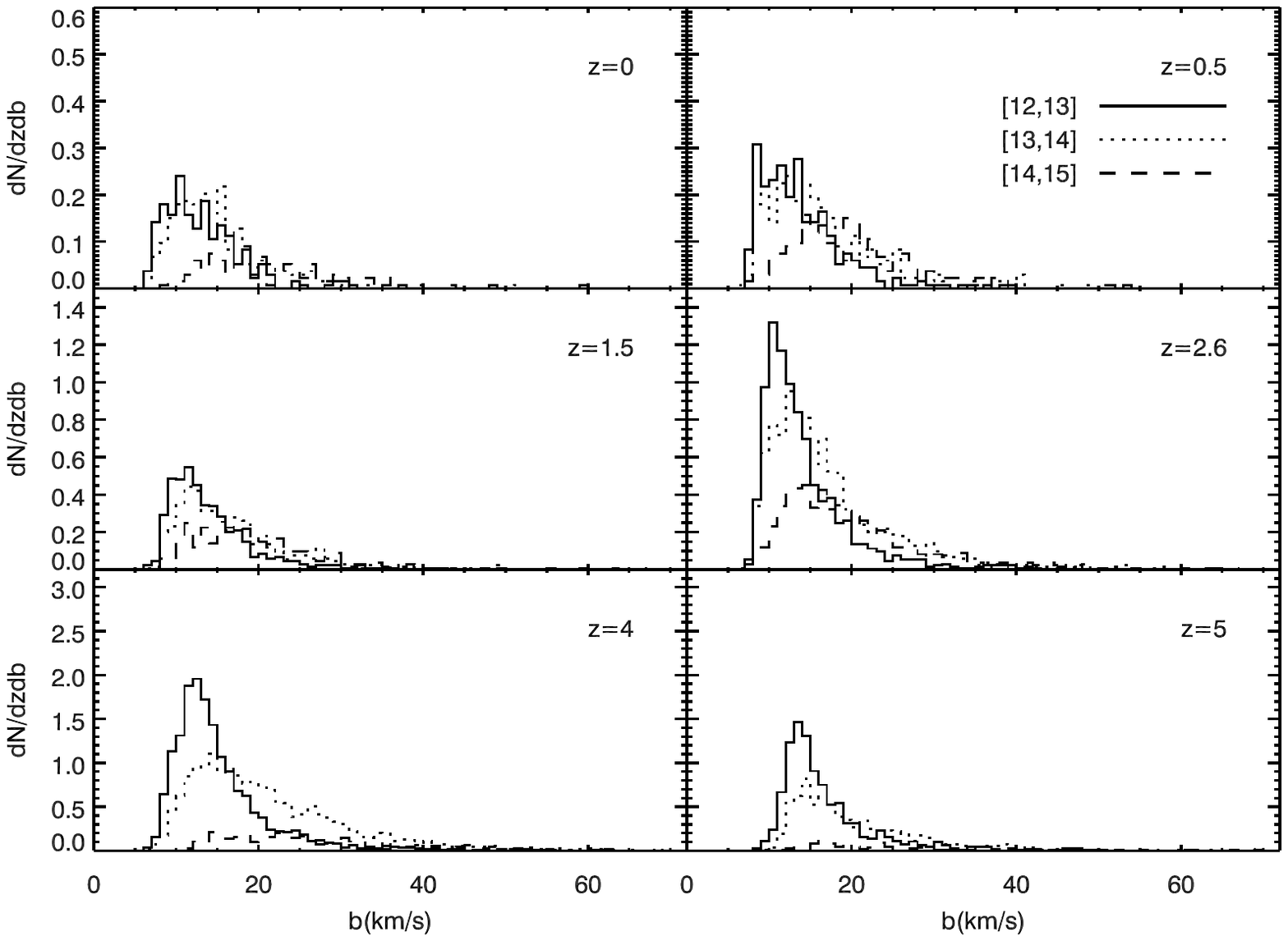,angle=0,width=4.1in} &
\hskip -1.1in
\epsfig{file=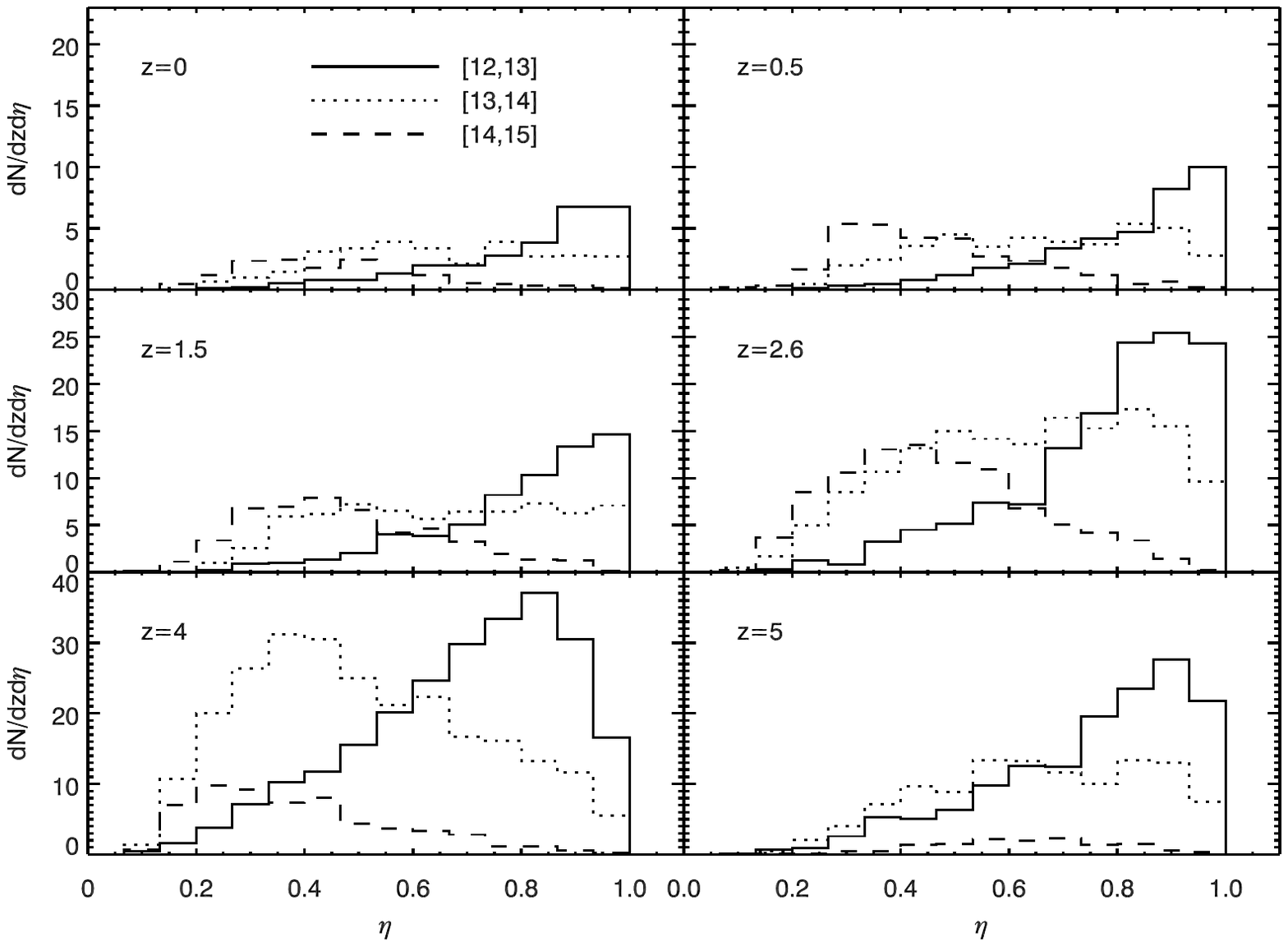,angle=0,width=4.1in} \\
\end{tabular}
\vskip -2.4in
\caption{
Left panel shows the distribution of Doppler width of computed CIV absorption lines
at six different redshifts, $z=0,0.5,1.5,2.6,4,5$,
separately for three subsets of lines of column density in the range
of $\log$N$_{\civ} {\rm cm}^2$=[12,13],[13,14],[14,15], respectively.
Right panel shows the distribution of the parameter $\eta$
 at four different redshifts, $z=1.5,2.6,4,5$,
separately for three subsets of lines of column density in the range
of $\log$N$_{\civ} {\rm cm}^2$=[12,13],[13,14],[14,15], respectively.
We note that $\eta=1$ corresponds to a Doppler width that is 100\% thermally
broadened, whereas $\eta=0$ corresponds to a Doppler width that has
no thermal contribution.
}
\label{fig:CIVdndb}
\end{figure*}
%*******************************************************************

The left panel of Figure~\ref{fig:CIVdndb}  
shows the distribution of Doppler width of computed \civ absorption lines.
The Doppler width distributions generally peak at $10-20$km/s at all redshifts.
Such a Doppler width peak is consistent with thermal broadening by gas temperature
$T\sim 10^{4.5}-10^5$K as seen Figure~\ref{fig:CIVdndT}.  
Because of the different definition of absorption lines we use compared to
Voigt profile fitting procedure for obtaining lines observationally,
a direct comparison is not possible.
%New paragraph
Nonetheless, our results are consistent with the Doppler widths of the CIV absorber sample in \citet[][]{2008Danforth}, the mean Doppler parameter at $\langle z \rangle=0.06$ is $\langle b_{\civ} \rangle=23\pm13$, while for our whole sample at $z=0$, the mean is $\langle b_{\civ} \rangle 15.6\pm7.1$ (1$\sigma$ interval). 
Comparisons to other samples, such as the one in \citet{2003Boksenberg}, are difficult. The reason for this is that it is common in observational investigations to fit a several number of components (with a Gaussian velocty distribution each) to each absorption line. This ``component'' vs. ``system'' definition makes comparisons between our work and observations subtle at least. Our definition of an absorber by establishing a flux threshold more closely resembles the standard definition of a ``system'', and in general we limit our comparisons to observational samples of ``systems''. Using this method, a large number of ``componentes'' might be fitted to one ``system''. In the sample of \citet[][]{2003Boksenberg}, this is as large as $32$ components for one given system at $z=2.438$; on average, there are $4.8$ ``components'' per ``system'' in this sample ranging between $1.6 < z < 4.4$.

The right panel of Figure~\ref{fig:CIVdndb} characterizes the nature of the Doppler width of computed \civ absorption lines using parameter $\eta$~$\equiv \sqrt{\frac{2kT}{m_{\rm ion}b^2}}$. It is indeed seen that most of the lower column density \civ absorbers with $\log($N$_{\civ}{\rm cm}^2)=[12,13]$ are dominated by thermal broadening. However, for higher column \civ absorbers, there appears to be roughly equal contributions to the Doppler width from thermal broadening and bulk velocity broadening. What this suggests is that lower column \civ absorbers tend to lie in quiescent regions, whereas high column ones typically reside in regions with significant velocity structures. This was seen earlier in Figures~\ref{fig:spectrum1},~\ref{fig:spectrum2},~\ref{fig:spectrum3}. 
Once again, it is important to stress that, even though the relative contribution to the line width from velocity structure is moderate for most \civ lines, the most likely physical explanation for the \civ producing regions is that they were shock heated by sweeping feedback shocks originating from nearby galaxies, have cooled to about $10^{4.5}-10^5$K and perhaps somewhat compressed in the process. Most \civ lines are far from shock fronts, whose velocity structures would otherwise make the lines significantly wider. \citet{1996Rauch} suggested that the quiescence of \civ lines may be due to the adiabatic compression of gas, which would not produce large velocity gradients. We show that this explanation may be incorrect given that most of the regions producing \civ lines at $z>2$ lie outside virialized regions. Rather, the quiescence is due to a combination of two things: the thermalization of previous shocks that reduces the random velocities and velocity gradients, and the narrow range of the region in physical space that produces the \civ line, which limits the velocity difference.

%**************************************************************
\begin{figure*} [h]
\centering
\vskip -0.5in
\begin{tabular}{cc}
\hskip -0.5in
\epsfig{file=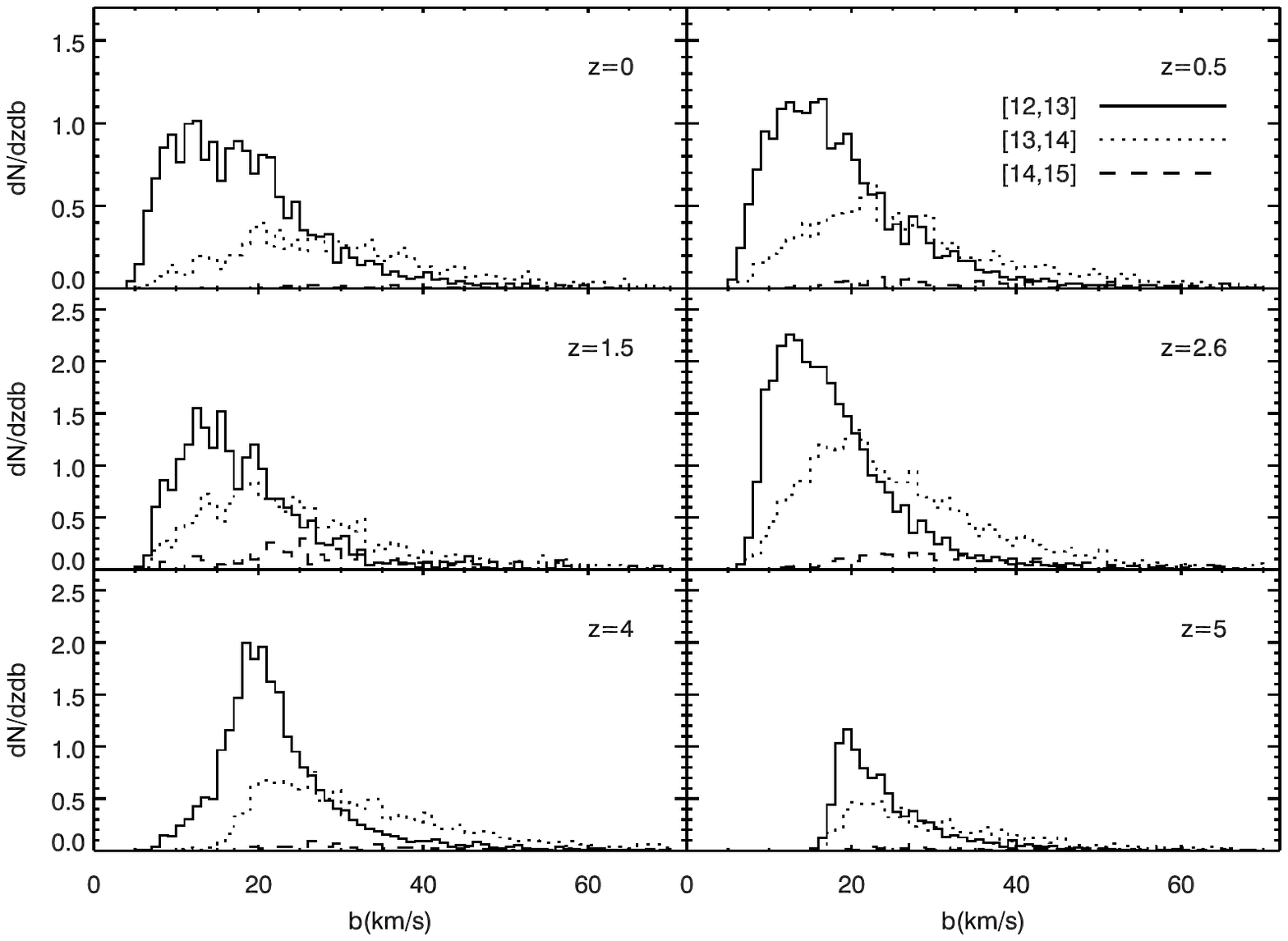,angle=0,width=4.1in} &
\hskip -1.1in
\epsfig{file=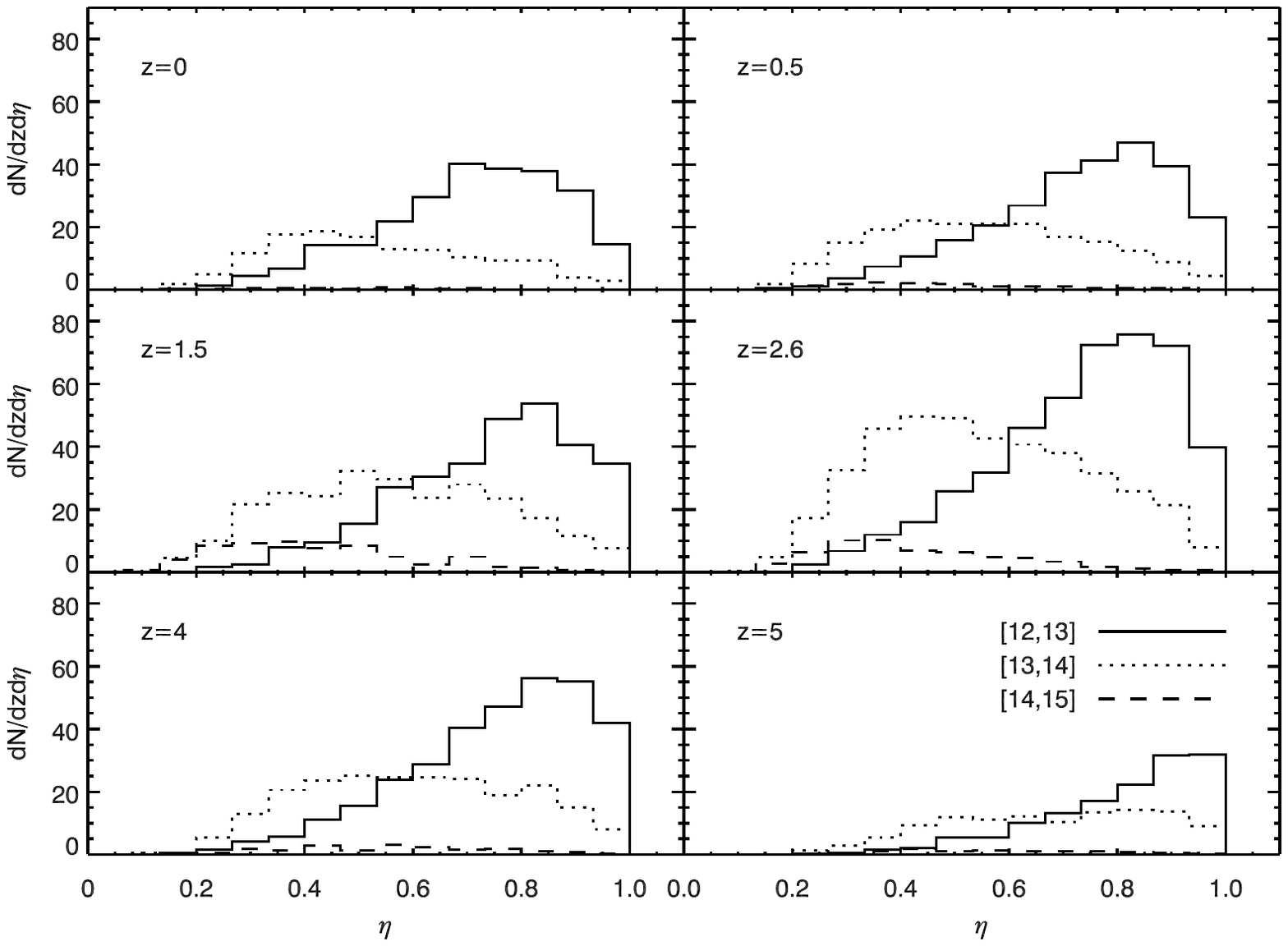,angle=0,width=4.1in} \\
\end{tabular}
\vskip -2.4in
\caption{Left panel shows the distribution of Doppler width of computed \ovi absorption lines
at four different redshifts, $z=1.5,2.6,4,5$,
separately for three subsets of lines of column density in the range
of $\log$N$_{\ovi} {\rm cm}^2$=[12,13],[13,14],[14,15], respectively.
Right panel shows the distribution of the parameter $\eta$ at four different redshifts, $z=1.5,2.6,4,5$,
separately for three subsets of lines of column density in the range
of $\log$N$_{\ovi} {\rm cm}^2$=[12,13],[13,14],[14,15], respectively.
We note that $\eta=1$ corresponds to a Doppler width that is 100\% thermally
broadened, whereas $\eta=0$ corresponds to a Doppler width that has
no thermal contribution.
}
\label{fig:OVIdndb}
\end{figure*}
%********************************************************************

The left panel of Figure~\ref{fig:OVIdndb} shows the distribution of Doppler width of computed \ovi absorption lines. For the OVI absorber sample of \citet[][]{2008Danforth}, the mean Doppler parameter at $\langle z \rangle=0.06$ is $\langle b_{\ovi} \rangle=30\pm16$, 
while for our whole sample at $z=0$, the mean is $\langle b_{\ovi} \rangle 22\pm13$ (1$\sigma$ interval is quoted in both cases).
In \citet[][]{2008Thoma}, Voigt profile fitting yields a mean number of $\sim1.4$ ``components'' in $27$ absorbers along 16 lines-of-sight towards QSOs, with a mean redshift of $\sim0.25$ and a corresponding Doppler width and $1\sigma$ of $\langle b_{\ovi}\rangle=27\pm17$.
Thus, within the errorbars our results agree with both observations.
A comparison with \civ lines shown Figure~\ref{fig:CIVdndb} is instructive. First, while the distributions for \civ and \ovi lines of $\log$N$_{\ovi}\,{\rm cm}^2$=[12,13] peak at comparable $b\sim 10\kms$ at $z=1.5$ and $z=2.6$, suggesting limited velocity contribution to the widths of both lines, the distribution for \ovi lines peaks at $b\sim 20\kms$ at $z=4-5$, significantly higher than that of \civ lines at the same redshifts.
%
%%%%%%%%%%%%%%%%I include the referee's comment: NEED TO REPHRASE
This is indeed to be expected: the ratios of \civ and C ($f_{\civ}$) and of \ovi and O ($f_{\ovi}$)
have a different dependence on density and temperature. At these densities and high temperatures, $f_{\civ}$ increases with increasing density, whereas $f_{\ovi}$ decreases with increasing density.  So If you are looking for broad lines, you will in \civ have an advantage going to high-z (where physical gas densities are higher), but not in \ovi.
Second, it is clear that a significant larger fraction of higher column \ovi lines of $\log$N$\,{\rm cm}^2$=[13,15] have larger Doppler width with $b\ge 40\kms$ at all redshifts than \civ lines, suggesting that there are significantly more \ovi lines that \civ lines that are in dynamically hot regions, such as around shocks where velocity gradients are high. Since these dynamically hot regions likely also have higher temperatures, collisional ionization would make a larger contribution to \ovi lines than \civ lines, consistent with our earlier statements.
Third, let us take a close look at $\eta$ distribution for $\log($N$_{\ovi}\,{\rm cm}^2)$=[13,14] \ovi lines and compare to that of \civ in Figure~\ref{fig:CIVdndb}: for \ovi lines it appears that the velocity contribution to the Doppler width is highest (i.e., lowest $\eta$) at $z=2.6$, whereas for \civ lines that occurs at $z=4$, suggesting that the fraction of \civ that are in dynamically hot regions peaks at a higher redshift than that for \ovi lines.
This is intriguing and likely due to a combination of several factors, including the evolution of the mixture of photoionized and collisionally ionized
absorbers, evolution of metal enrichment and feedback shock strengths as a function of redshift.
Potentially, useful and quantitative measures may be constructed to probe feedback processes using \civ, \ovi and other lines jointly.

%******************************************************
\begin{figure*}[h]
\centering
\vskip -0.5in
\begin{tabular}{cc}
\hskip -0.5in
\epsfig{file=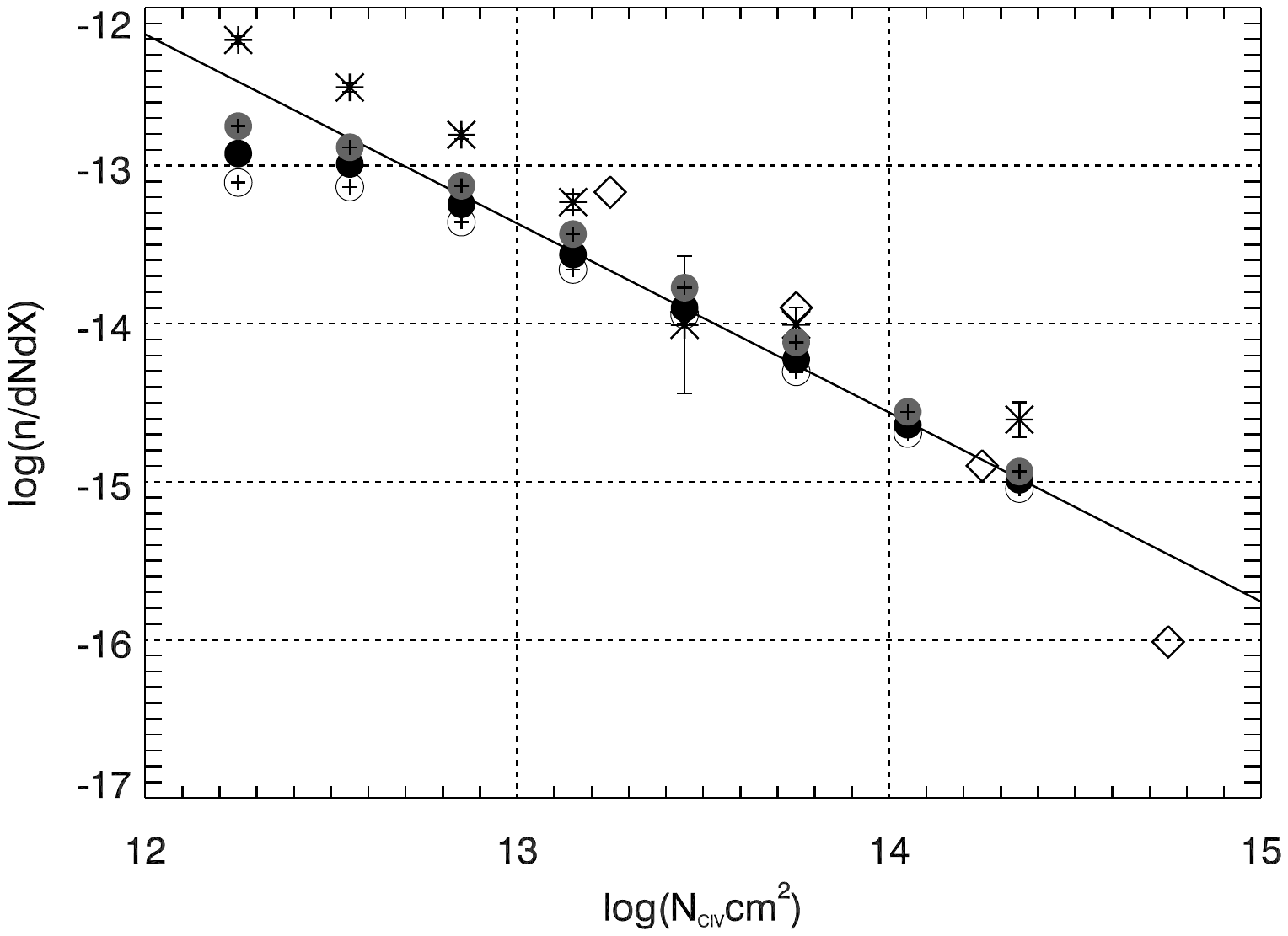,angle=0,width=4.1in} &
\hskip -1.1in
\epsfig{file=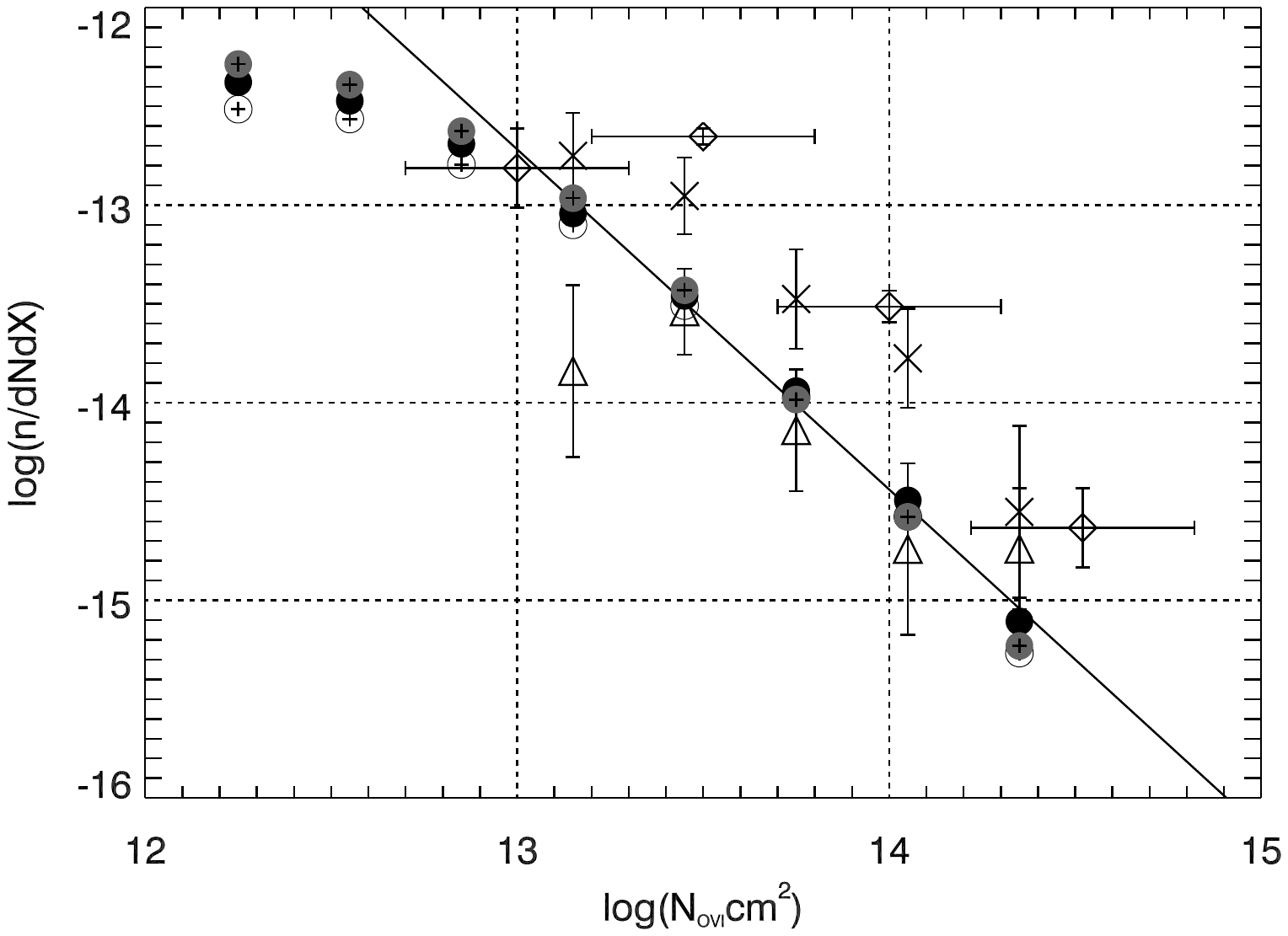,angle=0,width=4.1in} \\
\end{tabular}
\vskip -2.4in
\caption{
Left panel: the computed column density distribution for the C IV absorption line at $z=2.5$ for runs ``M'' (filled black circles), ``MR'' (open circles) and ``M2'' (filled grey circles). The solid line is the best power-law fit to our simulated results from run ``M'' performed for column densities in the range $[13,14.5]$. The slope of the fit is $-1.196\pm0.028$. Diamonds are observational data from \citet{2005Songaila} and \citet[][]{2003Boksenberg} at a mean redshift of $2.7$ and $2.6$, respectively, corrected for our cosmology. 
Right panel: the computed column density distribution for the \ovi absorption line at $z=2.5$ is shown as the solid line, which is the best power-law fit
to our simulated results, with slope $-1.723\pm0.075$. The circles have the same meaning as in the left panel. The observational data are drawn from \citet{2002Carswell} (squares), \citet{2005Bergeron} (diamonds), and \citet{2002Simcoe}(triangles) corrected for our cosmology.
}
\label{fig:CIVdndN}
\end{figure*}
%************************************************************

\subsection{\civ And \ovi Absorbers As Baryonic Matter Reservoirs}

Having gained a good understanding of the physical nature of \civ and \ovi lines,
we now turn to their overall column density distributions at $z=2.5$,
where observational data is most accurate, shown in Figure~\ref{fig:CIVdndN}. 
For both \civ and \ovi, the results obtained from runs ``M'', ``'MR' and ``M2'' 
show some small difference that is smaller than the magnitude of the difference between 
different observational studies and comparable to the difference between simulations
and observations.
This shows that our simulations are reasonably converged 
and not too sensitive to a factor $2$ or so variation in the strength of the UV backgrond. 
It is noted, however, that the convergence becomes much better
for clouds with column density greater than $10^{13}$, 
indicating that our current simulation resolution probably still
somewhat underestimates the abundance of clouds with columns smaller than that.
The error bars are not visible for the simulated values because they lie within the symbols plotted. 
Overall, we find the agreement of the computed distributions from the simulation to the observed ones 
is at the level that we could have hoped for.
We believe that differences may be contributable in part to cosmic variance,
in part to our resolution at the lower column density (as evidenced by the noticeable flattening)
and in part due to different methods of identifying clouds
(flux thresholding in our case versus Voigt profile fitting in the observed results, with 
the latter often producing multiple components for a single physical system).
Given the fact that our simulation has essentially only one free parameter ($e_{GSW}$) that has already been significantly constrained by the SFR history of the universe, it is really remarkable that we are able to match the observed column density distribution of both \civ and \ovi lines to within a factor of 2-3. 
Since the regions probed by \civ lines and \ovi lines are often physically different and to some extent reflect the different stages of the evolution of the feedback shocks, the fair agreement between our simulations and observations suggests that our treatment of the feedback process provides a good approximation to what happens in nature in terms of heating and
enriching the IGM, 
and it is indirect but strong evidence that feedback from star formation plays the central role in enriching
the IGM with its energy and metals. No additional, significantly energetic feedback from AGN seems required to account for the enrichment history
of the IGM. Therefore, it is very encouraging to note that the overall picture of the process of star formation feedback may be jointly probed by \civ, \ovi lines and other diagnostics. Detailed comparisons between simulations and observations in that regard would be the next logical step to further constrain theories of overall star formation in galaxies and feedback.

%****************************************************
\begin{figure*}[ht]
\centering
\vskip -0.5in
\begin{tabular}{cc}
\hskip -0.5in
\epsfig{file=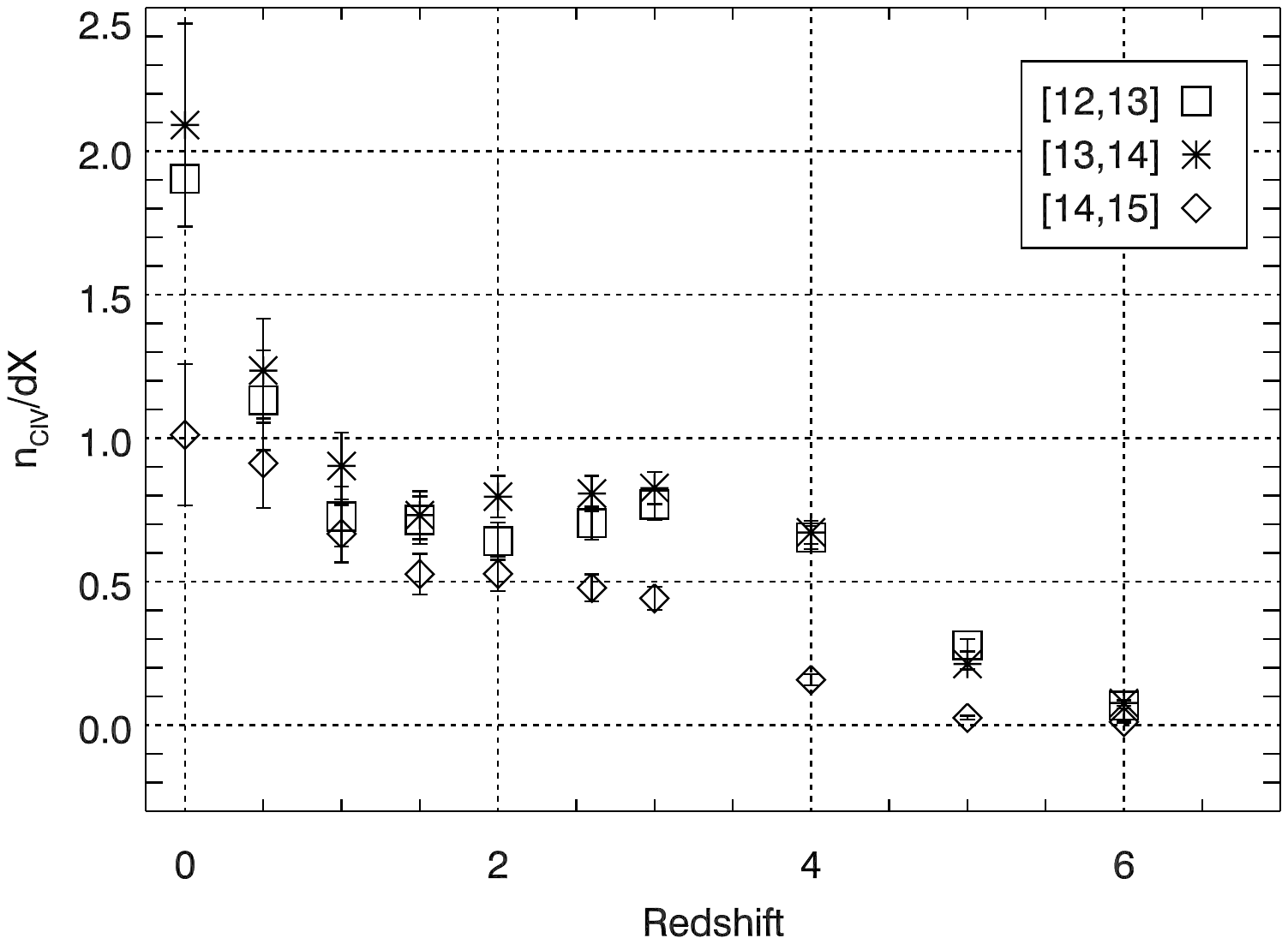,angle=0,width=2.6in} &
\hskip -0.5in
\epsfig{file=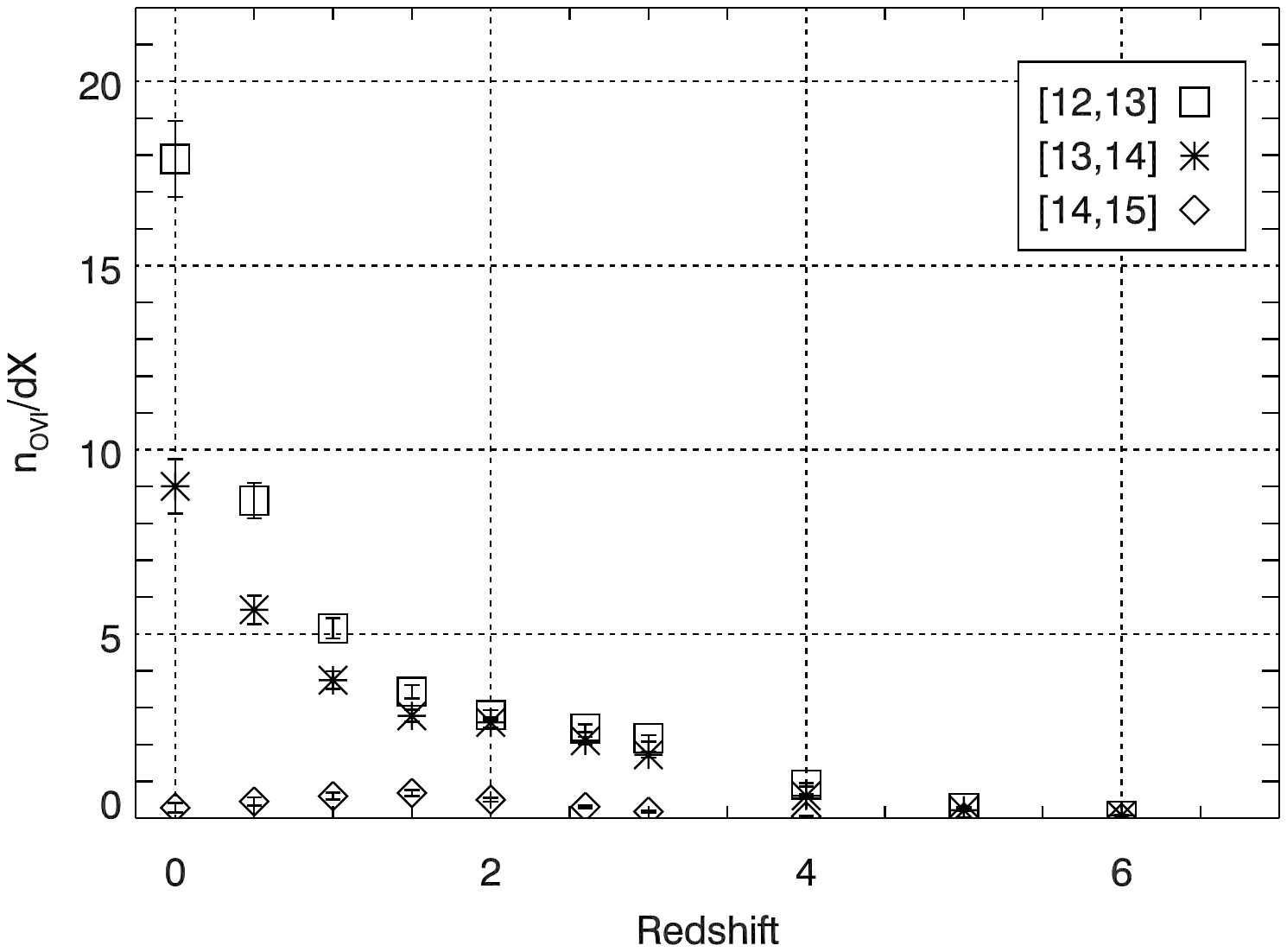,angle=0,width=2.6in} \\
\end{tabular}
\vskip -1.9in
\begin{tabular}{cc}
\hskip -0.25in
\epsfig{file=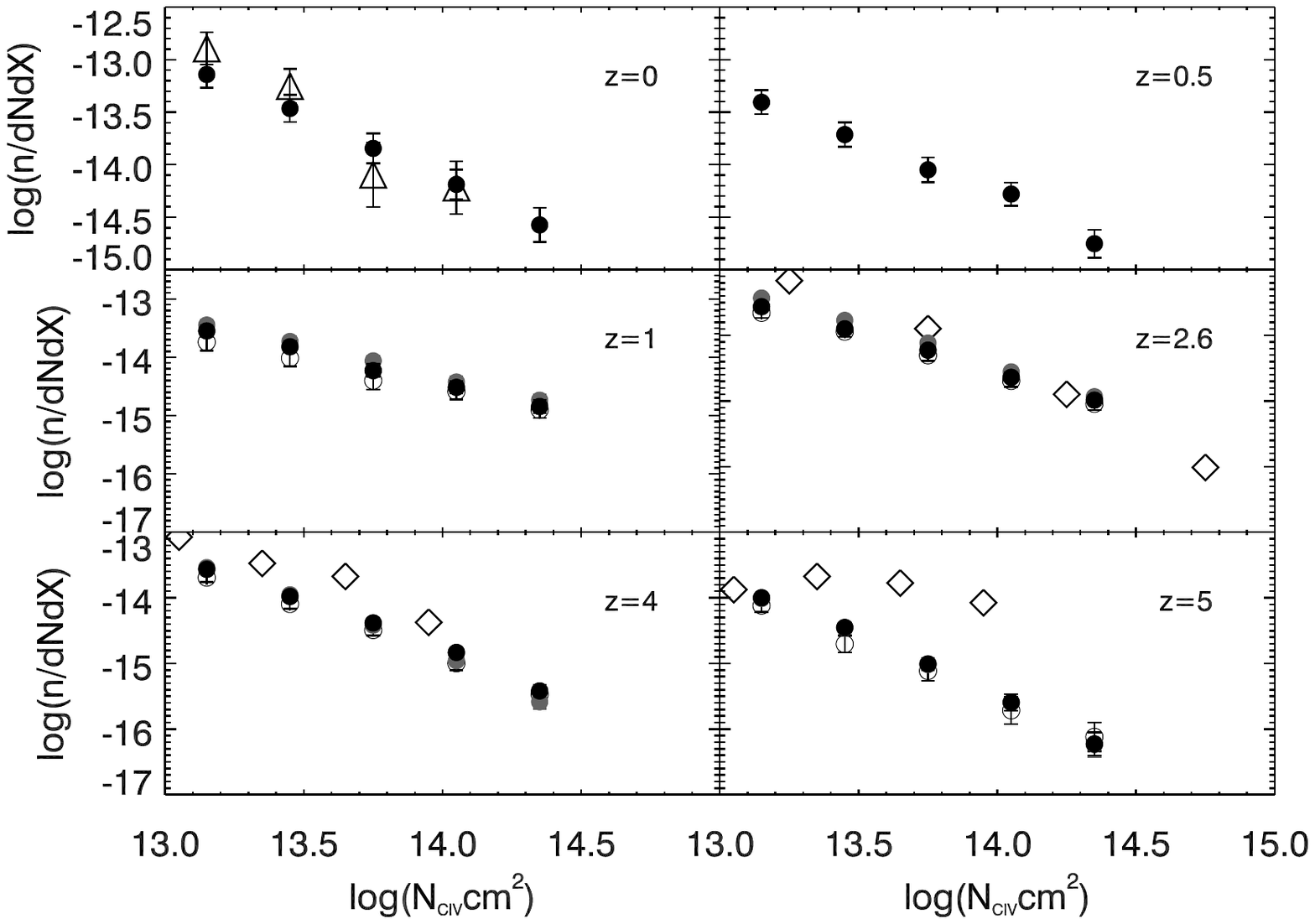,angle=0,width=2.6in} &
\hskip -0.5in
\epsfig{file=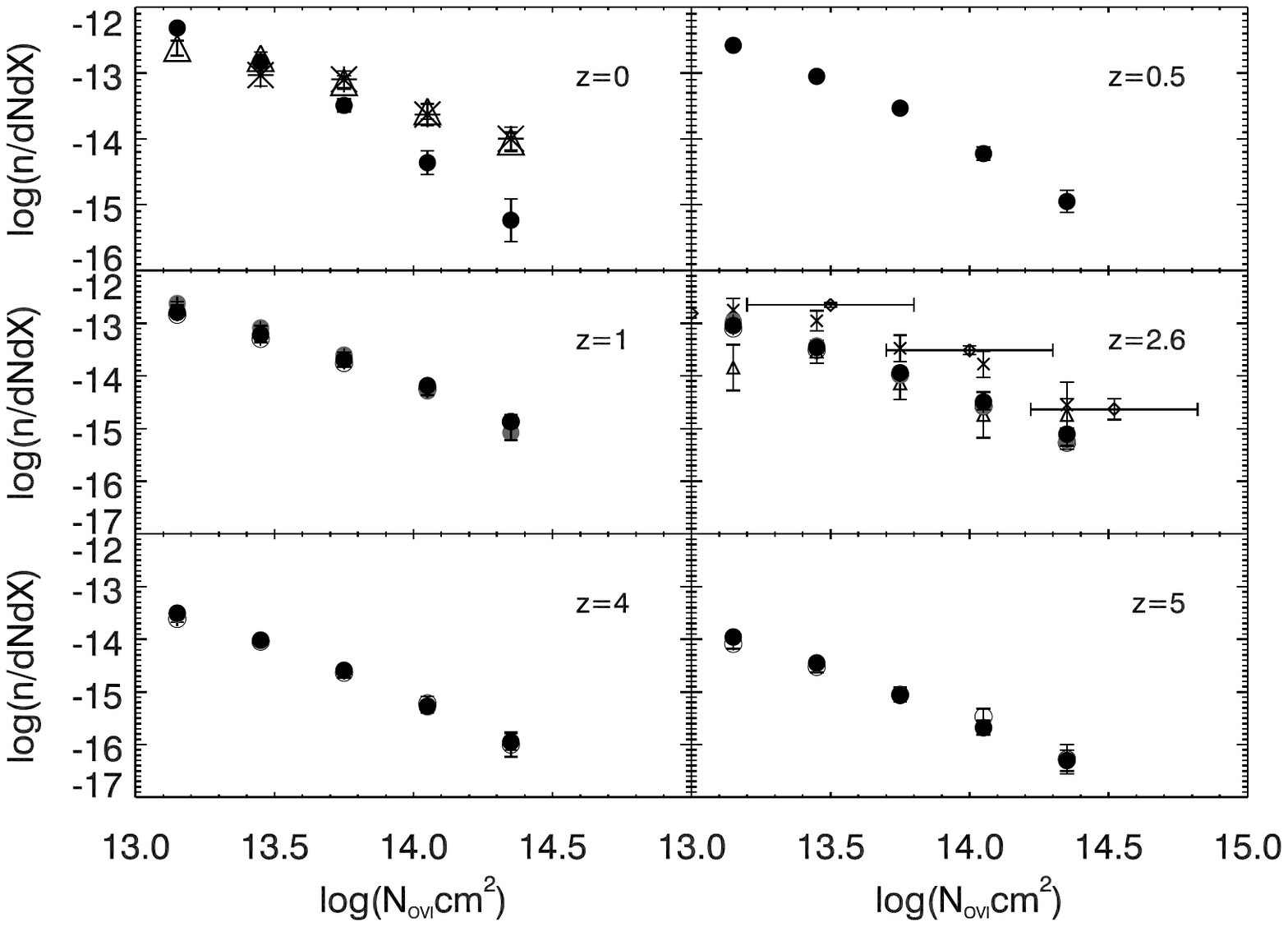,angle=0,width=2.6in} \\
\end{tabular}
\vskip -1.7in
\caption{
Top left panel: the evolution of the abundance of \civ absorbers separately for three subsets with column density in the range
$\log$N$_{\civ} {\rm cm}^2$=[12,13],[13,14],[14,15], respectively. Top right panel: the same for \ovi absorbers. %with references for the panel at $z=2.5$ the same as in Fig. \ref{fig:CIVdndN}.
Bottom left panel: column density distribution ($\log$f($N$)) for \civ absorbers at $z=0,0.5,1,2.6,4,5$. The results for our runs ``M'' (filled black circles), ``MR'' (open circles) and ``M2'' (filled grey circles) are shown.  Open diamonds correspond to observations \citep[][]{2001Songaila} corrected for our adopted cosmology, except at z=2.6, when they correspond to \citet[][]{2005Songaila}. At $z=2.6$ and $z=4$, asterisks correspond to \citet[][]{2003Boksenberg} for $3$ and $1$ sightline respectively in a $0.5$ redshift interval around the mean redshift. At $z=0$, the observational data correspond to \citet[][]{2008Danforth} (triangles) and \citet[][]{2008Thoma} (asterisks). 
Bottom right panel: $\log$f($N$) for \ovi absorbers at $z=0,0.5,1,2.6,4,5$. Observational data is available at redshift $z=2.5$: \citet{2002Carswell} (squares), \citet{2005Bergeron} (diamonds), and \citet{2002Simcoe}(triangles) corrected for our cosmology. At $z=0$ we compare our results to \citet[][]{2008Danforth} (triangles).
}
\label{fig:CIVdndNz}
\end{figure*}
%******************************************************************

In Figure~\ref{fig:CIVdndNz}, we show the evolution of the abundance of absorbers for different subsets of column densities (top panels) and the evolution of $\log(f($N$))$ (bottom panels). The number of \civ absorbers per unit redshift pathlength decreases with increasing redshift at both the low redshift interval $z\sim 0-2$ and the high redshift interval $z>4$ but stays roughly constant in the redshift interval $z\sim 2-4$. For \ovi absorbers, the number of absorbers per unit redshift pathlength decreases monotonically with increasing redshift for absorbers with column densities in the intervals 
$\log $N${\rm cm}^2=$[12,13],[13,14]. There are substantially fewer \ovi absorbers in the high column density range and their number peaks around $z \sim 1-2$. 
Comparing \civ and \ovi absorbers at each column density interval, we see that at $\log $N${\rm cm}^2=$[14,15] \civ and \ovi absorbers have comparable numbers at $z\ge 1$, but \civ absorbers outnumber \ovi absorbers by $z=0$ by a factor of a few, due to an upturn in \civ absorber number versus a downturn in \ovi absorber number from $z=1$ to $z=0$. This is probably caused by a combination of the rapidly diminished star formation activity and a lower radiation background towards $z=0$, which create an unfavorable condition for producing \ovi absorbers in denser environments either collisionally or by photoionization. At $\log $N${\rm cm}^2=[12,14]$ \ovi absorbers outnumber \civ absorbers at all redshifts.
From the lower panels we observe that the slope of $\log(f($N$))$ for \civ absorbers progressively becomes steeper at high redshifts. 
Our results seem to be consistent with observational results from \citet[][]{2005Songaila} and \citet[][]{2003Boksenberg} at redshift $z=2.6$ where observational data have the highest accuracy. 
We attribute the discrepancies at high column density between our simulations and observations
to cosmic variance: the size of our box is not large enough to host the higher column density structures. 
At $z=4-5$ the agreement is not as good, where we produce a steeper slope for $f($N$)$ than observed; 
this is likely in part due to cosmic variance and in part due to an underestimated UV background used.

What fraction of the metals in the IGM is directly seen in \civ and \ovi absorbers? From the column density distribution of the absorbers, we can estimate the ion baryon density of the IGM. Two different methods are typically used to do so. We can estimate $\Omega_{\rm ion}$ from
\begin{equation}
\label{sum}
\Omega_{\rm ion}=\frac{H_0 m_{\rm ion}}{c\rho_c}\frac{\sum_i N_{i,\rm ion}}{\sum \Delta X}
\end{equation}
where $H_0$ is Hubble's constant today, $m_{\rm ion}$ is the mass of the considered ion, $\rho_c$ is the critical density, $N_{\rm ion}$ is the absorber column density and $\sum \Delta X$ accounts for the total redshift pathlength covered by the sample of sightlines. In a flat Friedmann universe, this quantity is given by 
\begin{equation}
X(z)=\int_0^zdz'\frac{(1+z')^2}{[\Omega_M(1+z')^3+\Omega_\Lambda]^{1/2}}
\end{equation}
Another possibility is to construct the column density distribution per column density interval and unit $\Delta X$ 
\begin{equation}
f(N) = \frac{\sum_i N_{i,\rm ion}}{\Delta N \sum \Delta X}
\end{equation}
where the sum in the numerator is carried on the column densities of the absorbers present in bin $i$ and $\Delta \log $N$ = 0.3 $ in our case. 
The distribution $f(N)$ is typically fitted by a power-law $f(N) = K N^{\alpha}$. We can then obtain $\Omega_{\rm ion}$ from the fit by
\begin{equation}
\Omega_{\rm ion}=\frac{H_0 m_{\rm ion}}{c\rho_c}\int_{N_{\rm min}}^{N^{\rm max}}Nf(N)dN
\label{fit}
\end{equation}
\begin{equation}
\label{integral}
\Omega_{\rm ion}=\frac{8\pi G m_{\rm ion}}{3H_0c}K\frac{N^{\alpha+2}}{\alpha+2}|_{N{\rm min}}^{N_{\rm max}}
\end{equation}
Following \citet[][]{2009Becker} and  \citet[][]{2005Bergeron}, 
we will integrate $\Omega_{\rm ion}$ in the interval $\log $N$=[13,15]$, but the fit will be performed in the interval $\log $N$=[13,14.5]$ due to incompleteness of the sample at high values of $N$, as we have already mentioned. 
%%%%%%%%%%%%%%%%%%%%%%%IMPORTANT COMMENT

%*************************************************************
\begin{figure*}[h]
\centering
\vskip -0.9in
\begin{tabular}{cc}
\hskip -0.5in
\epsfig{file=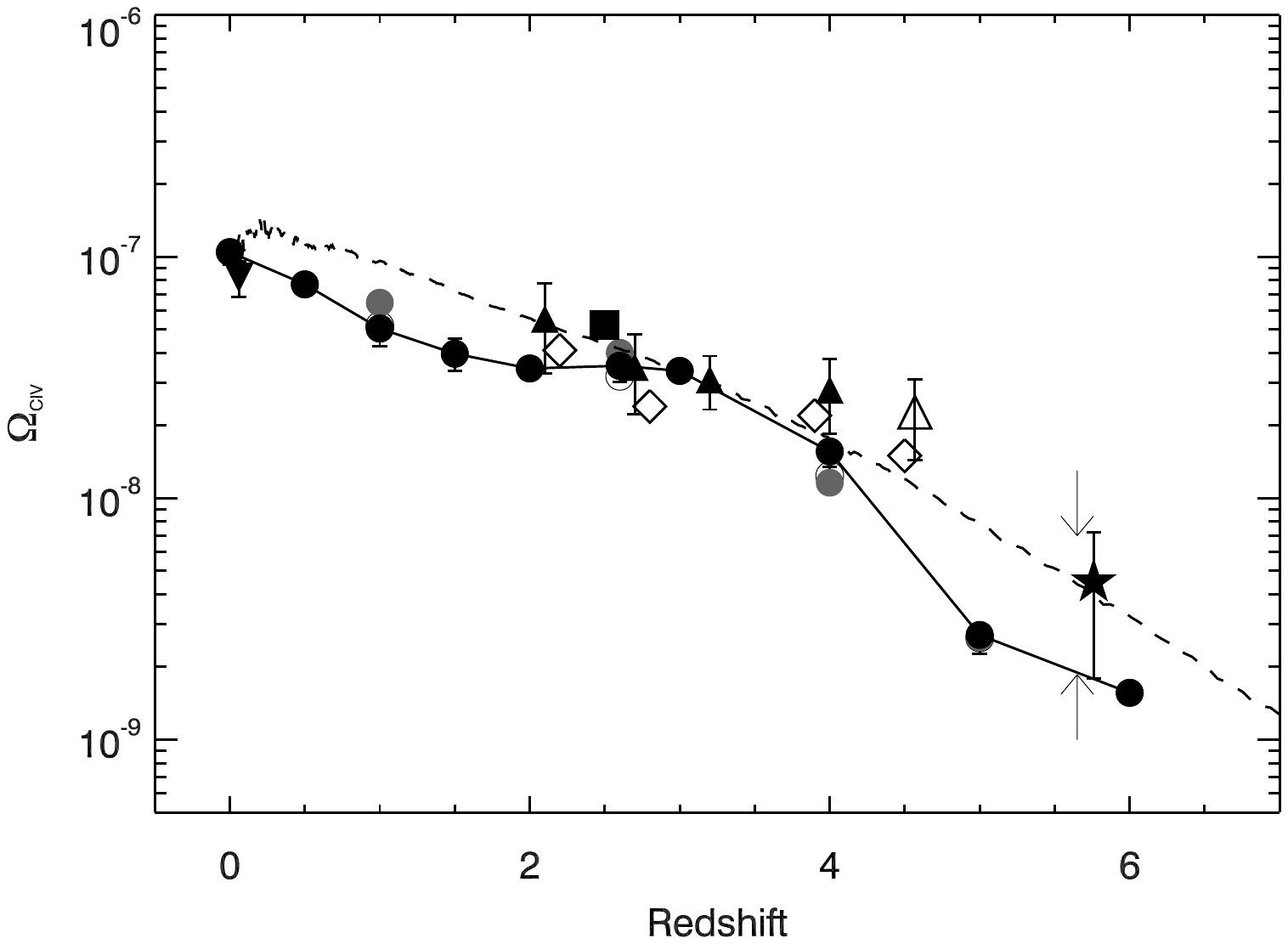,angle=0,width=2.6in} & 
\hskip -0.5in
\epsfig{file=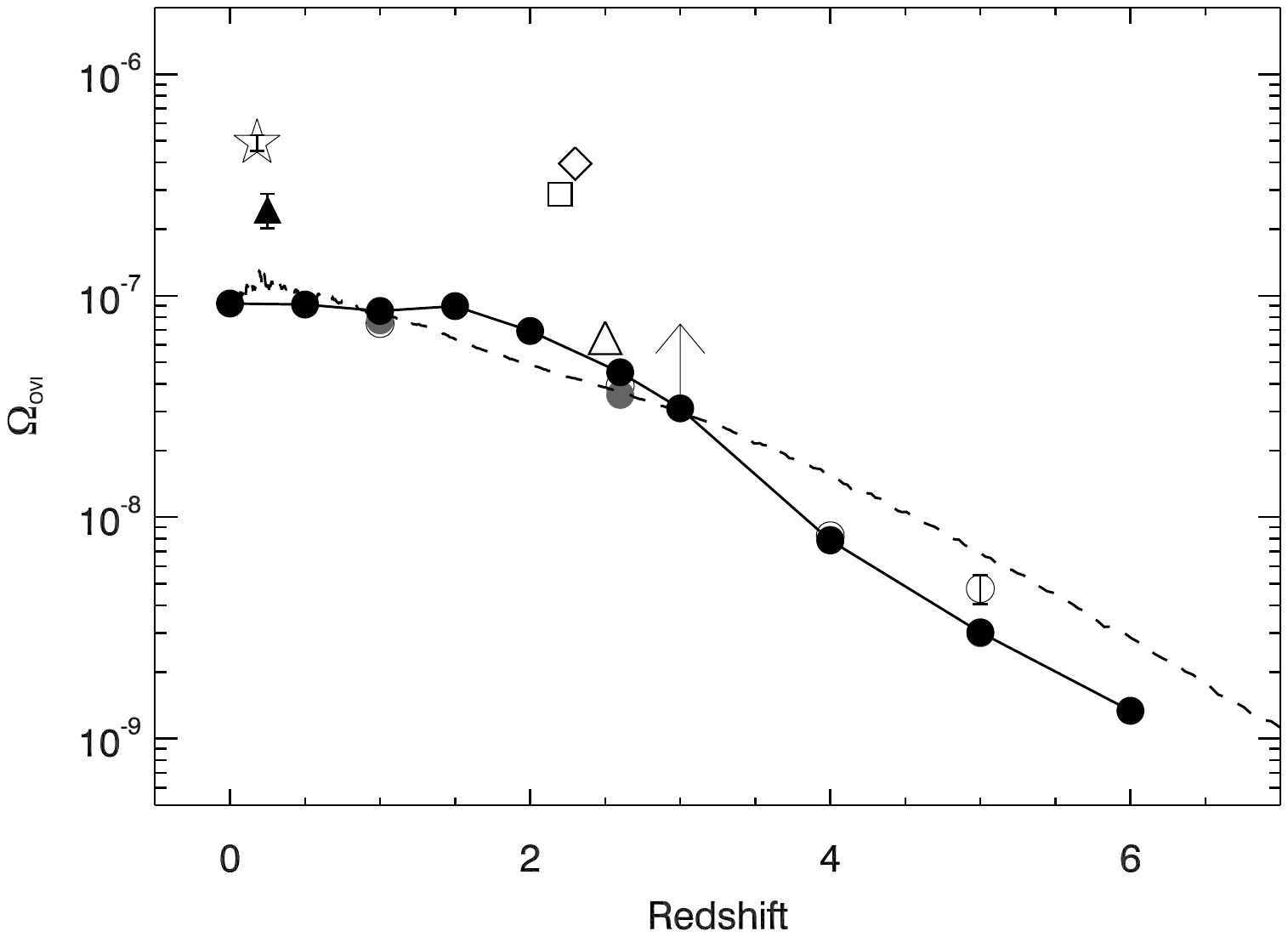,angle=0,width=2.6in} \\
\end{tabular}
\vskip -2.0in 
\begin{tabular}{cc}
\hskip -0.45in
\epsfig{file=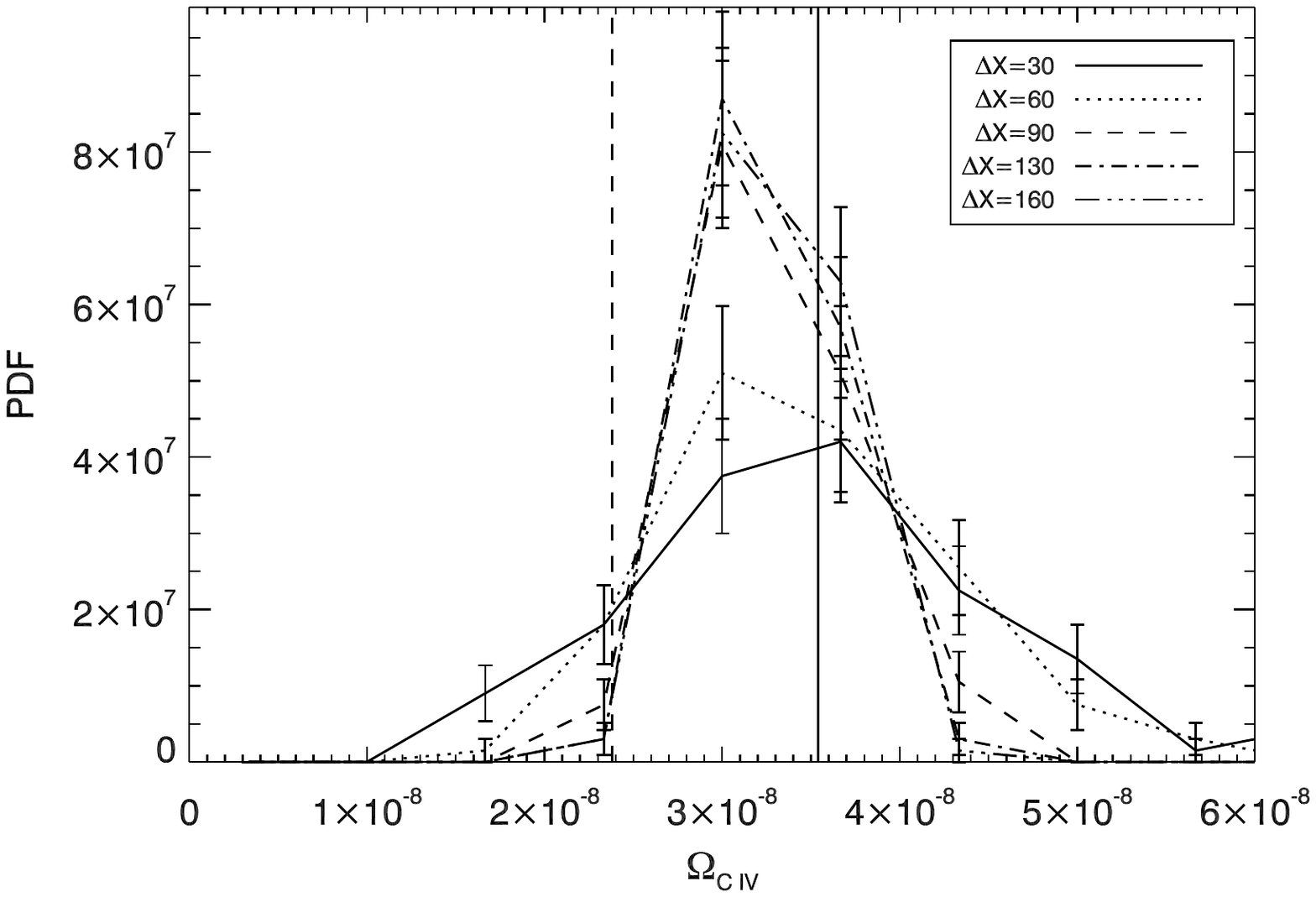,angle=0,width=2.6in} &
\hskip -0.5in
\epsfig{file=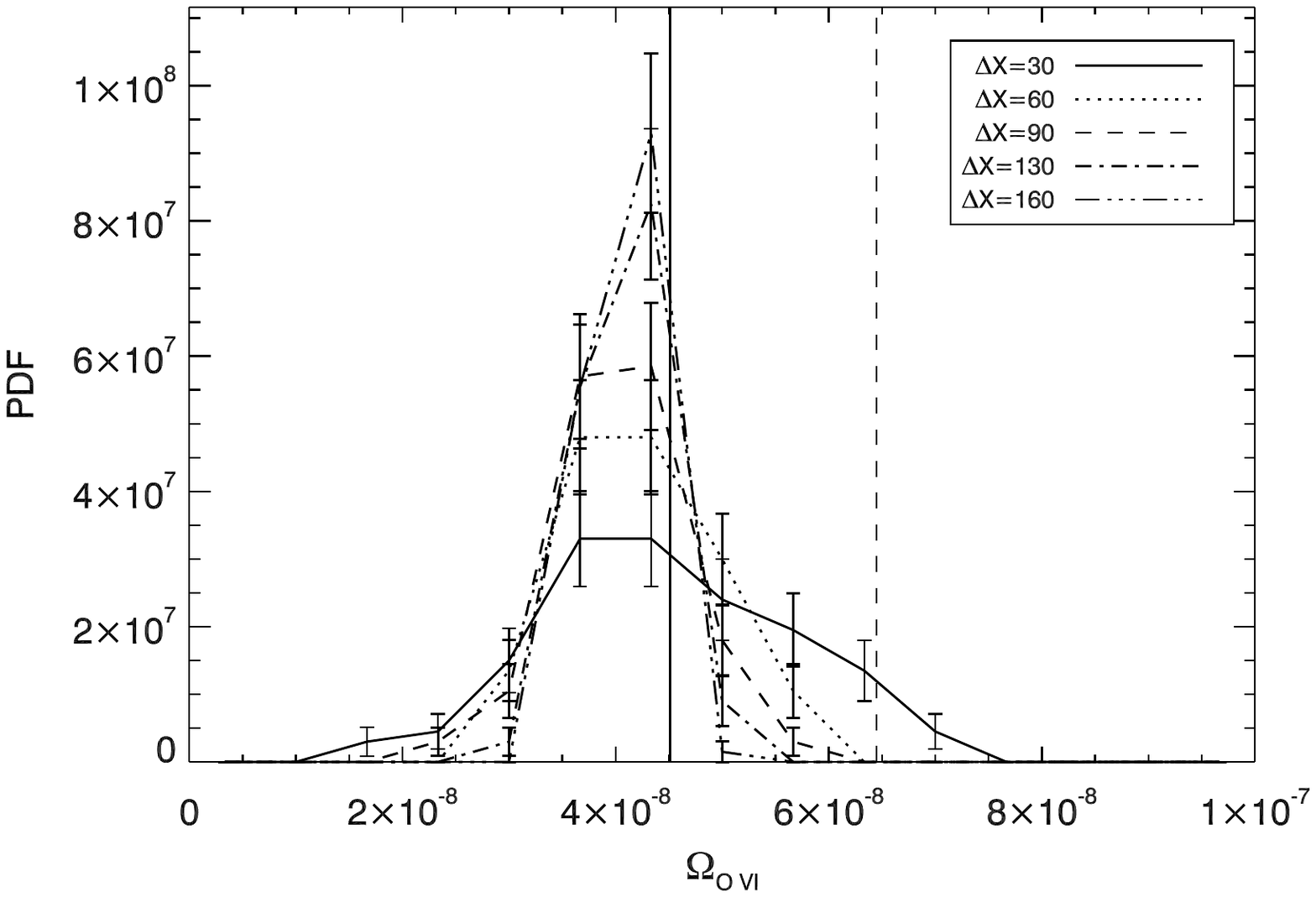,angle=0,width=2.6in} \\
\end{tabular}
\vskip -1.7in
\caption{
Top left: redshift evolution of $\omegaciv$ from simulations:
run ``M'' (filled black circles), run ``MR'' (open circles)
and run ``M2'' (filled grey circles). 
Observational data are from
\citet{2005Songaila} (open diamonds),
\citet{2009Becker} (arrows as limits),
\citet{2003Pettini}(open triangle),
\citet{2009Ryanweber} (filled star),
\citet{2003Boksenberg} (filled upright triangles),
\citet{2008Danforth} (filled downright triangle)
and  
\citet{2006Simcoe}(filled square).
The dashed curve is a simple physical model to explain 
the evolution of $\omegaciv$ (see text in \S 3.2).
Top right: redshift evolution of $\omegaovi$.
Observational data are from
\citet{2002Carswell} (open square),
\citet{2005Bergeron} (open diamond),
\citet{2002Simcoe}(open triangle),
\citet{2008Danforth}(open star),
\citet{2008Thomb}(filled triangle)
and 
\citet{2008Frank} (lower limit, arrow).
Bottom left: different curves are the expected PDFs for $\omegaciv$ at z=2.6,
based on our simulations, for observational samples of various sizes
(i.e., $\Delta X$ values). 
The solid vertical line indicates the median of the simulation results,
whereas the vertical dashed line is \citet{2005Songaila} value at z=2.5.
Bottom right: the expected PDFs for $\omegaovi$ at z=2.6.
The solid vertical line indicates the median of the simulation results,
whereas the vertical dashed line is \citet{2002Simcoe} value at z=2.5.
}
\label{fig:OmegaCIV}
\end{figure*}
%***************************************************************

Figure~\ref{fig:OmegaCIV} shows the evolution of the mass density contained in the \civ (left) and \ovi (right) absorption lines, respectively.
Considering the observational uncertainties and cosmic variance, it is very encouraging to see the excellent agreement between our simulated results and 
observations over the entire redshift range $z\sim 2-6$, where comparisons may be made.

%HERE COMMENT ON OBSERVATIONS
As a new finding from our simulation, we note that a significant dispersion, i.e., cosmic variance, in $\omegaciv$ is expected for available data samples with limited size (i.e., pathlength). In the bottom panels of Figure~\ref{fig:OmegaCIV} 
we show the expected distribution based on our simulations
for various sample sizes. We find that 
with $\Delta X=30$, 
the variance $\sigma=1.4\times10^{-8}$
for \civ and 
$1.2\times10^{-8}$
for \ovi;
for $\Delta X=60$, 
$\sigma=1.0\times10^{-8}$
for \civ and 
$8.5\times10^{-9}$
for \ovi;
with $\Delta X=160$, 
$\sigma=5.7\times 10^{-9}$
for \civ and 
$4.6\times 10^{-9}$
for \ovi.
Comparing \civ and \ovi lines it is seen that the total amount of mass contained in the \ovi line is 
comparable to that in the \civ line at all redshifts within a factor of 2 or so.
Note that the size for the observational sample of \citet{2005Songaila} is $\Delta X\le 20$.
Again, this suggests that some of the discrepancies between simulations and observations, and between observations may be accounted for by cosmic variance. For example, the slightly smaller value obtained from the observational sample of \citet{2005Songaila} for $\omegaciv$ is statistically consistent with simulations within $0.5\sigma$; the slightly larger value obtained from the observational sample of \citet{2002Simcoe} for $\omegaovi$ is statistically consistent with simulations within $1\sigma$.

%{\bf Note that the side of our simulated box, $50$Mpc$h^{-1}$, corresponds to the following values of 
%$\Delta X$: $\{1.7\times10^{-2},3.8\times10^{-2},6.7\times10^{-2},0.10,0.15,0.22,0.27,0.42,0.60,0.82\}$ 
%at redshifts $z=\{0,0.5,1,1.5,2,2.6,3,4,5,6\}$. In terms of the redshift interval spanned, these correspond to:
%$\Delta z=\{1.7\times10^{-2},2.2\times10^{-2},2.9\times10^{-2},3.8\times10^{-2},4.8\times10^{-2},6.2\times10^{-2},7.2\times10^{-2},0.10,0.13,0.16\}$. The $\Delta X$ value for a single sightline is two or three order of magnitudes smaller than the typical observational redshift path lengths. As a consequence, even if we add a large number
%of sightlines, we might be missing large column density absorbers in comparison typical with observational
%samples. }

In agreement with observations, the mass density contained in the \civ absorption line
is, within a factor of two, constant from $z=1$ to $z=4$ and subsequently drops
by a factor of $\sim (10, 20)$ by $z=(5,6)$.
Some rather subtle difference between \ovi and \civ lines may be noted.
While the metal density contained in the \civ absorption line
is nearly constant from $z=1$ to $z=4$,
that plateau for the \ovi line is attained only
for $z=0-2$. 
Because the total amount of metals in the IGM has increased significantly
in the redshift range $z=0-4$, it seems
that the near constancy of 
$\omegaciv$ at the redshift range $z=1-4$
and $\omegaovi$ at the redshift range $z=0-2$ does not reflect 
the amount of metals in the IGM, which has already been pointed out
earlier by \citet{2006Oppenheimer}. 
This probably reflects a ``selection effect" of \civ systems of the overall metals in the IGM,
which may be due to a combination of several different processes,
including the evolution of the mean gas density as $(1+z)^3$,
the evolution of the overdensity of the regions that produce \civ lines,
the density dependence of the IGM metallicity and its evolution,
the evolution of the radiation background and hierarchical build-up hence gravitational 
shock heating of the large-scale structure.

Our results contradict previous claims that observational data
point towards a near constancy of $\omegaciv$ with redshift
\citep[e.g.,][]{2001Songaila,2005Songaila, 2006Oppenheimer}.
However, more recent results have provided evidence of a downturn
in $\omegaciv$ towards $z \sim 6$. \citet[][]{2009Becker}
find no \civ absorbers in 4 sightlines towards $z \sim 6$ QSOs. They set 
limits on $\omegaciv$ and attribute the downturn to a decline
at least by a factor $\sim4.4$ (to 95\% confidence) in the number of \civ absorbers
at $z=5.3-6$ as compared to $z=2-4.5$. The decline shown in Figure \ref{fig:CIVdndNz}
is higher, at least a factor of $\sim 7$ for low column densities absorbers.   
\citet[][]{2009Ryanweber} perform the most extensive survey of intergalactic
metals at $z>5$, looking at the sightlines of 9 QSOs. They find evidence
of a drop by a factor $\sim 3.5$ in the mass density of \civ from redshift
$z=4.7$ to $z=5.7$. In comparison, we find a drop by a factor $\sim 1.7$
in $\omegaciv$ in the interval $z=5-6$. 

%***************************************************************
\begin{figure*}[h]
\centering
\vskip -0.5in
\begin{tabular}{cc}
\hskip -0.5in
\epsfig{file=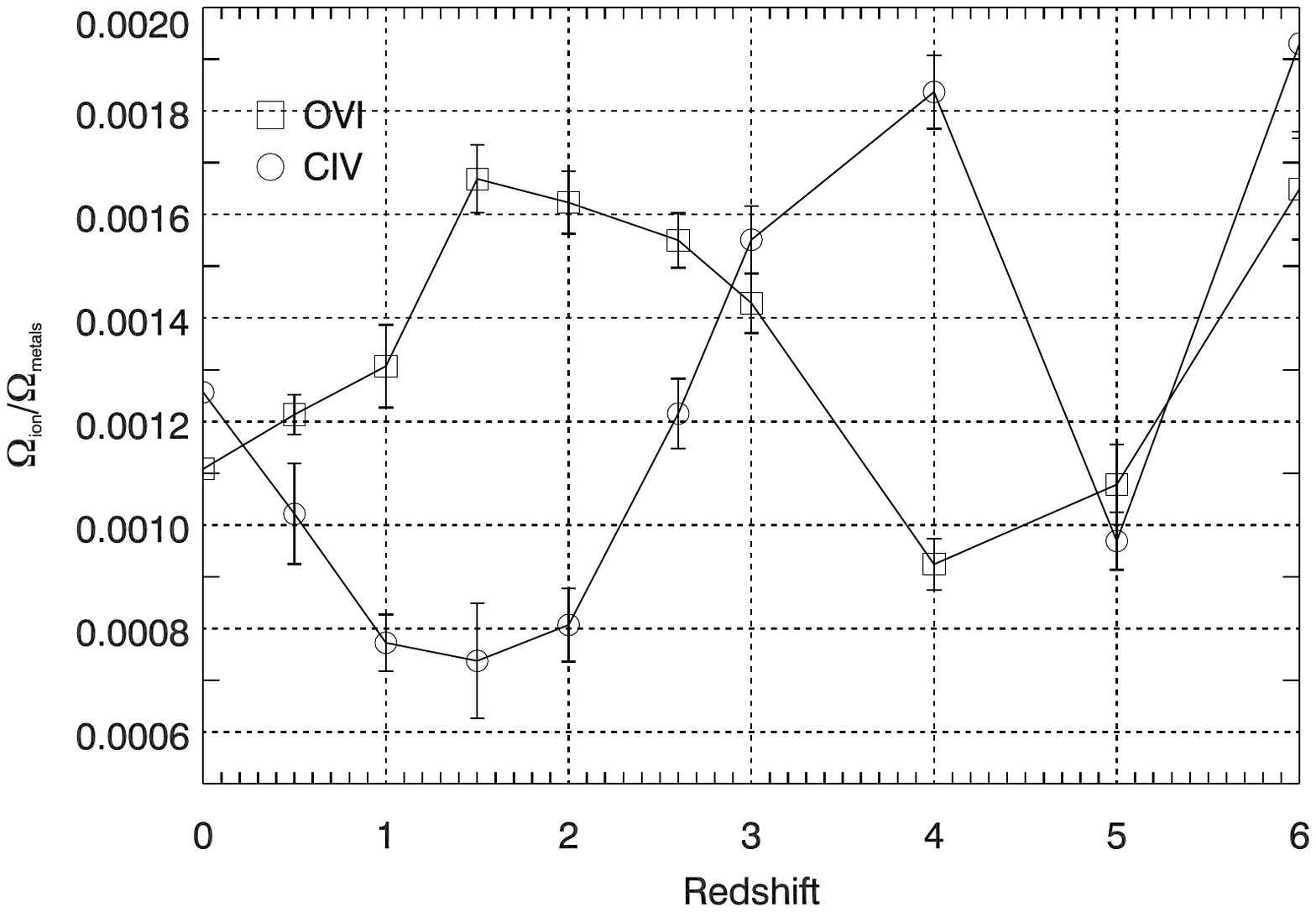,angle=0,width=4.1in} &
\hskip -0.9in
\epsfig{file=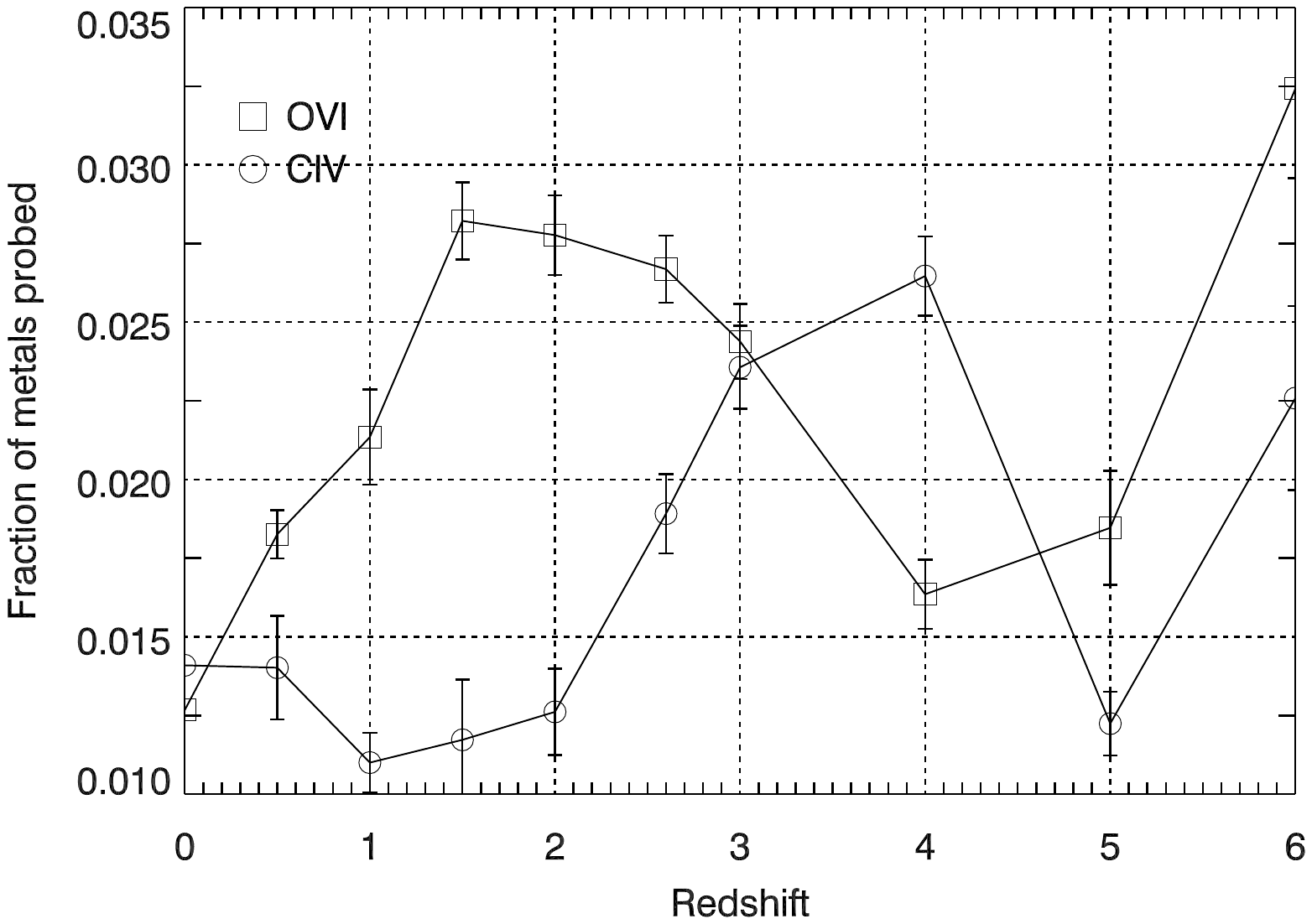,angle=0,width=4.1in} \\
\end{tabular}
\vskip -2.5in
\caption{
Left panel: the fraction of metals contained in \civ (circles) and \ovi (squares)
lines separately
in terms of the overall amount of metals in the IGM at each redshift.
Right panel: the fraction of metals contained in regions 
probed by \civ (circles) and \ovi (squares) lines, respectively,
in terms of the overall amount of metals in the IGM at each redshift.
}
\label{fig:omega_metals}
\end{figure*}
%*************************************************************

%***************************************************************
\begin{figure*}[h]
\centering
\vskip -0.5in
\begin{tabular}{c}
\hskip -0.5in
\epsfig{file=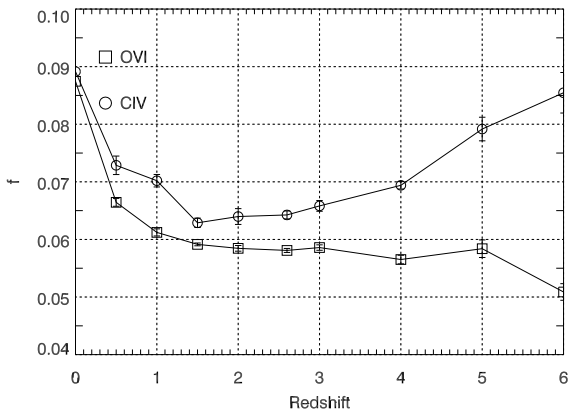,angle=0,width=4.1in} 
\end{tabular}
\vskip -2.5in
\caption{ shows the \civ ratio of $n_{\civ}/n_{\rm C,tot}$ (open circles) and the \ovi ratio of $n_{\ovi}/n_{\rm O,tot}$ (open squares) as a function of redshift. Note that at the optimal temperature with collisional ionization, $f_{max}$ for \civ and \ovi is $29\%$ at $\log T_{\rm max}=5.00$ and $22\%$ at $\log T_{\rm max}=5.45$, respectively \citep[][]{Sutherland93}.}
%Renyue could you check this phrase? Is it 29% or 39%?
\label{fig:fsw}
\end{figure*}
%*************************************************************

The second point, perhaps the most overlooked, is that the amount of metals contained in the \civ and \ovi absorption line
is a very small fraction of the overall metals. The left panel of Figure~\ref{fig:omega_metals} shows the ratios of mass density measured in the \civ (diamonds) and \ovi lines (triangles) over the total amount of metals in the IGM as a function of redshift. We see that the amount of mass contained in the \civ line remains at $\sim 0.13\%$ within a dispersion of $40\%$, and at $\sim 0.13\%$ within a dispersion of $25\%$ for the \ovi line.
In the right panel of Figure~\ref{fig:omega_metals} we show the amount of metals probed by each line as a function of redshift. Here is how we compute the metals probed by each line and use \civ as an example.
For each detected \civ line, a range of spatial locations (i.e., gas cells along the line of sight) contributes to its column density (see Figures~\ref{fig:spectrum1},\ref{fig:spectrum2},\ref{fig:spectrum3}).  
Roughly speaking, the amount of metals effectively probed by the \civ line will be larger than
the metals directly seen in the \civ line by a factor of $f=n_{C,tot}/n_{CIV}$ (a similar relation for the \ovi line).   
This ratio $f$ for the \civ and \ovi line is shown in Figure \ref{fig:fsw}.
One point worth noting is that for \civ there is an upturn of $f$ from $\sim 0.1$ at $z<2$ towards high redshift,
reaching again the same value at $z=6$.
This is caused from a transition from more collisionally dominated \civ absorber population
at $z>4$ to a more photoionization dominated one at intermediate redshifts.
The trend for the \ovi line is much less pronounced, indicative of a dominance of 
collisionally ionized \ovi absorbers over the entire redshift range $z=0-6$,
with a trend that it is more so at higher redshift.

From the right panel of Figure~\ref{fig:omega_metals} we see that, within a factor of $2$,
the amount of metals probed by either \civ or \ovi line 
is roughtly $2\%$.
Combining the fact that the majority of \civ-producing regions have not collapsed and virialized
(see Figure~\ref{fig:CIVdnddelta}) and 
a small fraction of all metals is probed by \civ and \ovi lines at all redshifts,
\civ and \ovi absorbers are ``transients";
in other words, only a small fraction of metals in the IGM
get ``lit up" as the \civ or \ovi line at any given time.
As we demonstrated earlier, these regions that produce \civ absorption lines 
have a set of properties that seem to be created by a combination
of physical processes including feedback shock heating and radiative cooling
(see Figures~\ref{fig:spectrum1},\ref{fig:spectrum2},\ref{fig:spectrum3}).  
These close observations suggest that only a fraction of metals at any given time
that has recently passed through shocks and cooled to an appropriate temperature
shows up as \civ absorption lines.
In this sense, \civ absorption lines trace the current feedback processes from star formation 
and how the current feedback energy and metal-enriched gas interact with the surrounding IGM.
Similar statements about the transient nature could be made for the \ovi line, except that 
the \ovi line corresponds to somewhat different physical states of the shocked regions:
they are slightly hotter in temperature and dynamically hotter.

Third, returning to Figure~\ref{fig:OmegaCIV},
we would like to emphasize that, 
consistent with recent observations
\citep[e.g.,][]{2009Becker,2009Ryanweber},
there is indeed a sharp drop in $\omegaciv$
from $z=3$ to $z=(4,5,6)$
by a factor of $\sim (2, 10, 20)$, respectively.
This has less to do with the evolution of 
the total amount of metals produced, rather it is tracing the phase of \civ gas
at any given time.
At redshift $z\ge 3$, \civ lines at different redshifts
appear to come from regions of comparable overdensity (see Figure~\ref{fig:CIVdnddelta})
and comparable metallicity (see Figure~\ref{fig:CIVdndZ}).
This allows us to test a very simple physical picture for the origin of \civ lines.
They are produced by regions that
were shock heated earlier by feedback shocks and have cooled to 
the temperature of $T\sim 10^{4.5}-10^5$K when they are seen,
and the duration of each \civ line in this ``\civ phase" would then be inversely proportional to the cooling time of the gas in
this phase,
which is proportional to $\Lambda^{-1}(T,Z) (1+z)^{-3} (1+\delta)^{-1}$, where 
$\Lambda(T,Z)$ is cooling function at temperature $T$ and metallicity $Z$
and the z-dependent term is due to density evolution with redshift.
Then, the total amount of metals in \civ lines, $\omegaciv$, will be proportional to
$\dot M_{star}(z) \Lambda(T,Z)^{-1} (1+z)^{-3} (1+\delta)^{-1}$,
where $\dot M_{star}(z)$ is the star formation rate at $z$.
Taking $\delta$, $Z$ and $T$ as roughly being constant
(see Figures ~\ref{fig:CIVdnddelta}, \ref{fig:CIVdndZ}, \ref{fig:CIVdndT}),
we have $\omegaciv\propto \dot M_{star}(z) (1+z)^{-3}$,
which is shown as the dashed curve 
on the left panel of Figure~\ref{fig:OmegaCIV}.
It provides a reasonably good fit for the actual computed evolution of $\omegaciv$.

\subsection{Global Metal Enrichment of the IGM and Missing Metals}

We now turn to present a global metal enrichment history of the IGM
to supplement what is captured by the \civ and \ovi absorption lines.
As in \citet{1999Cen}, 
in our analysis
we divide the IGM into three components by temperature:
(1) $T<10^5~$K cold-warm gas, which is in low density regions or cooling, star forming gas,
(2) WHIM at $10^7~$K$>T>10^5~$K,
(3) Hot X-ray emitting gas at $T > 10^7$K.
One additional component (4) is the baryons that have left the IGM and been condensed
into stellar objects, which we designate as ``stars".

%************************************************************
\begin{figure}[h]
\vskip -1in
\centerline{
\epsfig{file=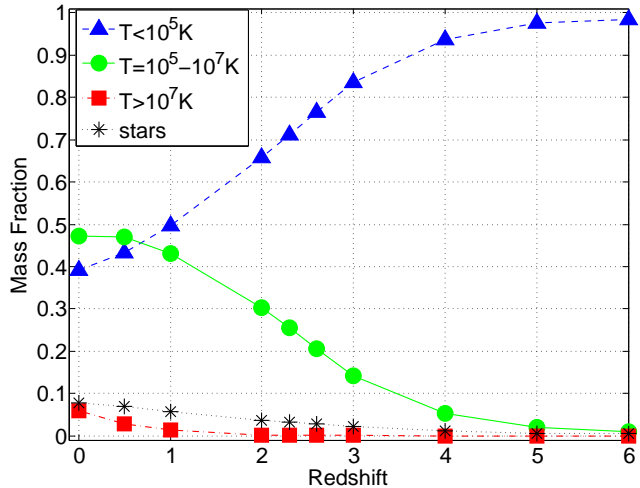,angle=0,width=3.5in}}
\vskip -1in
\caption{
shows the evolution of baryons for the four mutually exclusive components:
(1) $T<10^5~$K cold-warm gas, 
(2) WHIM at $10^7~$K$>T>10^5~$K,
(3) Hot X-ray emitting gas at $T > 10^7$K
and (4) ``stars".
}
\label{fig:compz}
\end{figure}

\begin{figure*}[h]
\vskip -1.6in
\centerline{
\epsfig{file=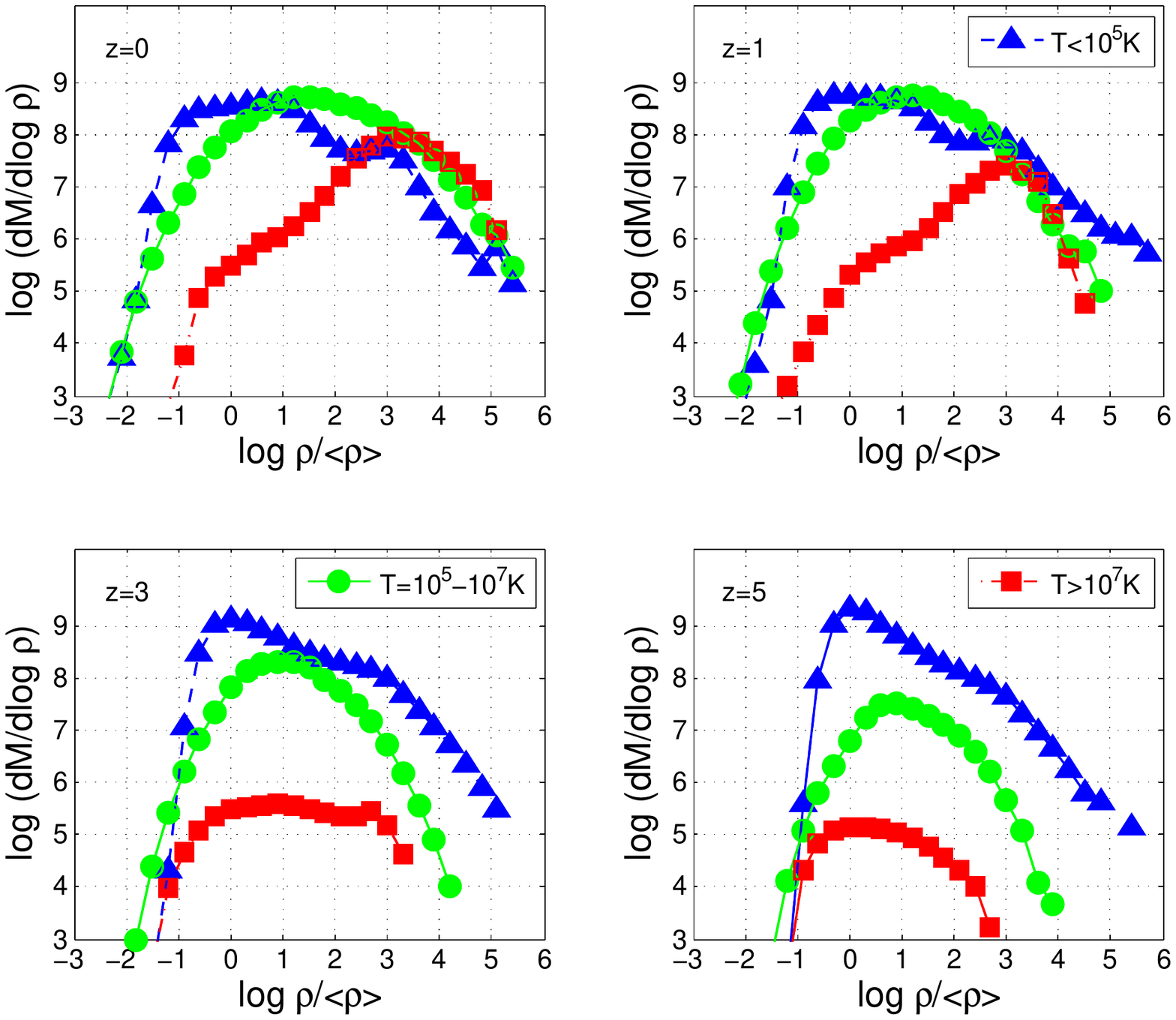,angle=0,width=6.0in}}
\vskip -1.4in
\caption{
shows the mass distribution of the three IGM components - 
(1) cold-warm gas at $T<10^5~$K, 
(2) WHIM at $10^7~$K$>T>10^5~$K,
(3) Hot X-ray emitting gas at $T > 10^7$K - 
as a function of overdensity at four different redshifts
$z=0,1,3,5$.
Note that the area under each curve is proportional to the 
mass contained.
}
\label{fig:hisrho}
\end{figure*}
%**************************************************************

Figure~\ref{fig:compz} shows the evolution of these four components.
The overall evolution of the four components are in good agreement 
with earlier findings \citep[][]{1999Cen, 2001Dave, 2006Cen} and relevant observations
\citep[e.g.,][]{1998Fukugita}.
In particular, we see that $40-50\%$ of all baryons 
are in WHIM by the time $z=0$,
which is in excellent agreement with our previous findings
\citep[][]{1999Cen, 2001Dave, 2006Cen}. 
%A comparison of Run N and Run H at $z=0$
%indicates that, while galactic superwinds are subdominant to
%gravitational heating caused by the collapse of large-scale density waves,
%they make, nevertheless, about a $20\%$ contribution to the overall WHIM mass by $z=0$.
It is also noted that $\sim 40\%$ of the baryons at $z=0$ reside
in a relatively cool but diffuse component with $T<10^5$K (the triangles in 
Figure~\ref{fig:compz}).
It is likely that a significant portion of this cool component at $z=0$, in the form of
$\lya$ forest, is already seen by 
UV observations \citep[e.g.,][]{2004Penton}. 
As we noted earlier, the strength of feedback from star formation is chosen
to match the observed overall star formation history.
%The gas in the hot component reached 7\% in this simulation by $z=0$,
%which is somewhat smaller than the $19\%$ found in our previous work (CO).
%The difference is partly due to difference in the model parameters
%and partly due to cosmic variance.
%This component is not greatly affected by GSW (cf. Figure 13c).

Each of the IGM components is composed of different regions that have gone
through distinct evolutionary paths and thus spans a wide range in density,
shown in Figure~\ref{fig:hisrho}.
The distribution of the cold-warm component (triangles) is always peaked
at the mean density at all redshifts, reflecting the initial gaussian distribution
of gas around the cosmic mean and indicating that the bulk 
of the IGM at mean density or lower has never been shock heated 
by either strong gravitational shocks or feedback shocks.
The cold-warm gas extends to very high densities ($\ge 10^5$).
It is interesting to note that the amount of cold-warm gas 
that could potentially feed the star formation, i.e., the cold-warm gas
at density $\log\rho/\langle\rho\rangle\ge 2-3$,
remains constant, within a factor of $\sim 2$, over the range redshift shown $z=0-5$.
This is consistent with observations of the nearly non-evolving amount of gas probed
by DLAs 
\citep[e.g.,][]{2003Peroux,2005Zwaan,2006Rao,2009Prochaska,2009Noterdaeme}. 
The physical relation between this apparently non-evolving gas 
and the precipitous drop of star formation rate at $z<1$ is currently unclear.

\begin{figure*}[h]
\centerline{
\epsfig{file=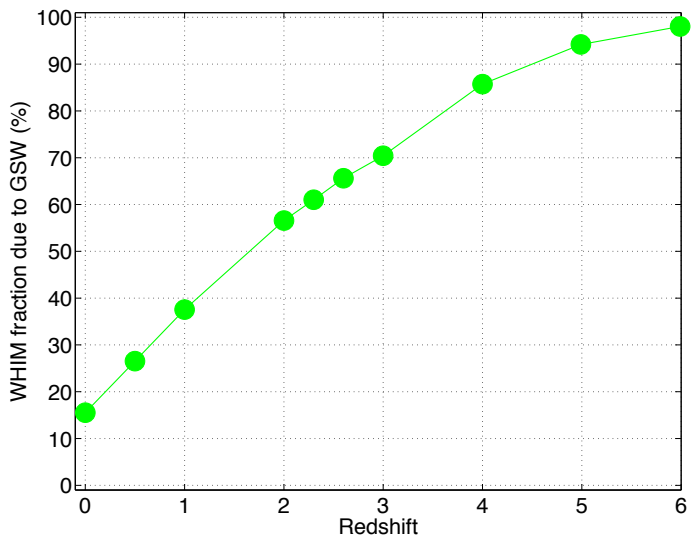,angle=0,width=4.5in}}
\caption{
shows the (cumulative) fraction of WHIM that is produced by 
GSW as a function of redshift.
}
\label{fig:GSWz}
\end{figure*}

The distribution of the WHIM also appears to peak at a constant overdensity
of about 10 times the mean density.
This is rather intriguing.
In order to properly interpret this interesting phenomenon
it is useful to understand the heating sources of WHIM.
There are two primary heating sources for WHIM: 
shocks due to the collapse of large-scale structure
and GSW produced shocks.
Earlier works have already 
shown that gravitational shock heating due to the formation of 
large-scale structure dominates the energy input for heating up
and thus turning about $50\%$ of the IGM into WHIM by $z=0$
\citep[][]{1999Cen, 2001Dave, 2006Cen}. 
It is, however, expected that heating due to 
hydrodynamic shocks emanating from galactic superwinds 
become increasingly more important at higher redshifts.
This is because the amount of energy from gravitational collapse of large-scale structure
as well as the resulting shock velocity decreases steeply towards higher redshift. 
The reason for this is simple: in the standard cosmological model the amount of power
is peaked at a wavelength of $\sim 300$Mpc/h and drops steeply towards small scales.
To quantify the relative contribution of GSW in turning the IGM into WHIM,
we compare the simulation with GSW feedback to that without GSW feedback
(run N in Table 1).
Then, we make the simple assertion that the difference in the amount of WHIM 
between the two simulations is due to GSW.
Figure~\ref{fig:GSWz} shows the fraction of WHIM that is produced (cumulatively) by 
GSW as a function of redshift.
Consistent with previous results \citep[][]{1999Cen, 2006Cen},
the contribution from GSW to heating up WHIM by $z=0$ 
is subdominant at $10-20\%$.
This relatively small contribution to WHIM from GSW can be
understood based on simple energetics estimates.
But we see the GSW fraction increases rapidly with increasing redshift.
At redshift $z=1.5$ the GSW fraction 
is about $50\%$, then reaching $70\%$ at $z=3$ and $95\%$ at $z=5$.
Thus, we see the primary heating source of WHIM at $z>1.5$ is GSW,
whereas gravitational shocks due to structure formation 
are mostly responsible for heating the WHIM at $z<1.5$.

%*******************************************************
%*********************************************************

From Figure~\ref{fig:hisrho} 
it seems clear that WHIM does not distinguish between gravitational
shocks and feedback shocks.
In both cases shocks have largely stopped at overdensity of about $10$.
Let us try to understand why that happened.
First we note that the shocks originate approximately from the central
regions of filaments, where pancakes collapse and shock for the case
of gravitational shocks and galaxies are generally located for the case of GSW shocks.
For gas shock heated to $10^5$K the shock velocity
is roughly $70\kms$. With that velocity the shock will be able to
travel roughly $700(1+z)^{-1}$kpc comoving over the Hubble time
at any redshift. 
Therefore, one should expect to see shocks have reached a few hundred
kpc comoving at any redshift, which are about one to a few times the virial radius
of typical large galaxies, which in turn correspond an overdensity in the vicinity of $10$
and are thus in good agreement with simulation results.
Some shocks penetrate deeper into the IGM, especially along directions
with lower densities and steeper density gradients,
as seen in Figure~\ref{fig:CIVslice};
but the amount of mass effected in these low density regions
is small, corresponding to the sharp drop
of WHIM mass at the low density end (Figure~\ref{fig:hisrho}). 
This last point is best corroborated by the distribution
of the hot gas at high redshift ($z=3,5$), in the bottom two panels
of Figure~\ref{fig:hisrho}.
There we see a small amount of hot gas heated up by GSW shocks
is indeed produced in regions of density lower than the mean density
and traces a larger amount of WHIM gas that is also produced there.
At $z\le 1$, some comparable, small amount of hot gas is still
produced at low density regions.
But the vast majority of hot X-ray emitting gas is now residing
in the deep potential wells of X-ray clusters of galaxies,
when the cluster scale turns nonlinear and collapses.

%**********************************************************
\begin{figure*}[h]
\vskip -1in
\centerline{
\epsfig{file=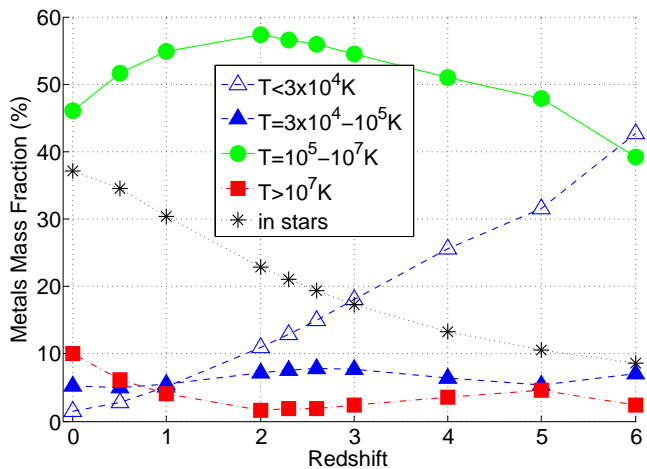,angle=0,width=3.5in}}
\vskip -1in
\caption{
shows the evolution of fractions of all metals produced that are contained
in each of the five components, as a function of redshift:
(1C) $T<3\times 10^4~$K cold gas,
(1W) $T=3\times 10^4-10^5~$K warm gas,
(2) WHIM at $10^7~$K$>T>10^5~$K,
(3) Hot X-ray emitting gas at $T > 10^7$K,
(4) ``stars".
}
\label{fig:mtlcompz}
\end{figure*}

Having obtained an overview of the thermal history of the IGM,
we now turn to the metal story.
We will first focus our attention on the WHIM here, 
because that is where most of the energy and metal exchanges
between galaxies and the IGM take place,
as shown in Figure~\ref{fig:hisrho}.
Observationally, integrating the observed star formation rate history from high redshift
down to $z=2.5$
suggests that the vast majority (possibly $\ge 80\%$) of cosmic metals %produced in stars
at $z\sim 2.5$ appear to be missing \citep[e.g.,][]{1999Pagel, 1999bPettini}. % 2005Ferrara}. 
Note that this conclusion is insensitive to the choice of IMF, since
both UV light and metals are, to zeroth order, produced by the same massive stars.
Metals that have been accounted for in the estimates
include those in stars of Lyman break galaxies (LBG), 
damped Lyman alpha systems (DLAs) and $\lya$ forest,
i.e., cold-warm gas and stars.
Given the dominant heating of WHIM by GSW,
one may immediately ask:
Could a significant fraction of metals that accompanies
the GSW energy be heated up and in a phase like WHIM that is different from those
where metals have been inventoried?
To better address this open question, 
we further break down the IGM component (1) ($T<10^5~$K cold-warm gas)
into two sub-components with (1C) ($T<3\times 10^4$K cold gas)
and (1W) ($T=3\times 10^4-10^5$K warm gas).
The purpose of this finer division is to separate out the cold gas (1C), which
can be more appropriately identified with $\lya$ forest clouds and DLAs.
The results are shown in Figure~\ref{fig:mtlcompz}.
We see that about one third of all metals produced by $z=0$ is locked up in stars,
decreasing monotonically towards high redshift, dropping to about $10\%$ by $z=5$.
The fraction of metals in the hot X-ray emitting component is at about $10\%$ level
at $z=0$, plummeting to about $2\%$ at $z=2$ and slowly rising back to about $6\%$ at $z=6$.
It is likely that the metal fraction in the hot X-ray component at $z<1$ be somewhat underestimated
given the relatively moderate simulation boxsize.
The remaining metals are in the general photoionized $\lya$ forest and the WHIM.
At $z=6$ the $\lya$ forest ($T<3\times 10^4$K, open triangles) contains about
$43\%$ of all metals,
while WHIM ($T=10^5-10^7$K, open circles), 
and warm IGM ($T=3\times 10^4-10^5$K, solid triangles) contain $39\%$ and $7\%$, respectively.
But the fraction of metals in the $\lya$ forest decreases steadily with time 
and becomes a minor component by $z=0$ at $<3\%$.
Most of the metals is seen to be contained in the WHIM at all times below redshift five
at $50-60\%$, peaking at $\sim 60\%$ at redshift $z\sim 2$.
In total, the amount of metals contained in the IGM with temperature $T>3\times 10^4$
constitutes about 2/3 of all metals produced by $z=2.5$.
Metals in this temperature range were not accounted for in the quoted observational inventory
at $z=2.5$.
Thus, it seems probable that 
the missing metals problem at $z=2-3$ can be largely rectified,
if one counts the metals in the IGM at $T>3\times 10^4$K.

\begin{figure*}[h]
\vskip -1.3in
\centerline{
\epsfig{file=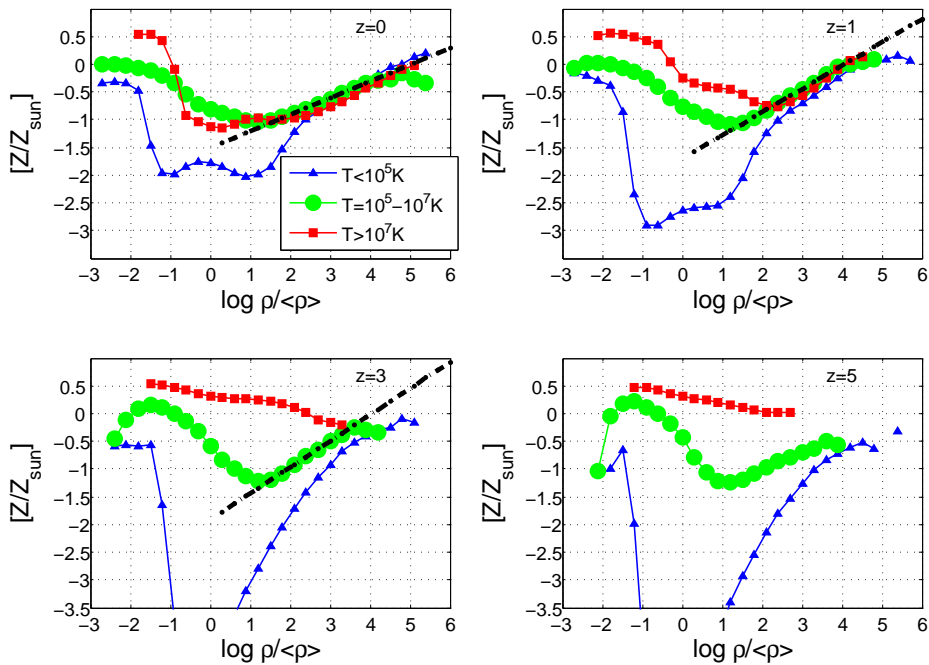,angle=0,width=5.0in}}
\vskip -1.4in
\caption{
shows the metallicity of the three IGM components - 
(1) cold-warm gas at $T<10^5~$K,
(2) WHIM at $10^7~$K$>T>10^5~$K, (3) Hot X-ray emitting gas at $T > 10^7$K -
as a function of overdensity at four different redshifts
$z=0,1,3,5$.
}
\label{fig:Zrho}
\end{figure*}

By now we have learned that a large amount of metals could be hidden
in the WHIM of temperature $10^5-10^7$K spanning a wide range in density.
Since the metallicity is a strong function of density,
it is still unclear the location of the WHIM that dominates 
the missing metals.
Figure~\ref{fig:Zrho} shows the mean metallicity of the three IGM components 
as a function of overdensity at four different redshifts.
It is evident that within each IGM component there is
a wide range in metallicity that is a non-trivial function of overdensity.
Let us examine their behaviors in detail.
For all three IGM components
there is a strong correlation between the mean metallicity and overdensity
at overdensity $\delta \ge 10$ and they converge at the highest density.
While the metallicity of the cold-warm gas at the high density end 
remains at about solar at high density,
its mean metallicity at the low overdensity drops rapidly with increasing
redshift.  For example, at $\delta=10$, the mean metallicity is 
(-2, -2.5, -3, -4) in solar units at $z=(0,1,3,5)$. 
%This is indicative that GSW, perhaps assisted with gravitational interactions,
%are continuously spreading a small amount of metals to very low density regions
%with time.

%********************************************************************

One may notice that all three distributions exhibit a minimum metallicity
at some intermediate density range, $\delta=0.1-10$ for the cold warm-gas,
$\delta=10$ for the WHIM and $\delta=1-100$ for hot gas (only at $z=0-1$).
This is entirely in agreement with the physical picture that we described 
earlier for the GSW shock propagation through the IGM.
Figure~\ref{fig:Zrho} confirms that the transformation of cold gas to WHIM
roughly stops at $\delta=10$. 
Additional metal-enriched gas is further transported along some directions,
such as those perpendicular to the filaments,
to very low density regions and enrich these regions to higher metallicity
(due to a negligible amount of pre-existing gas there). 
The behavior of cold and hot components at the low density end can be understood 
in the same way as the WHIM.
The metallicity of hot gas at the centers of clusters of galaxies
(at overdensity $\rho/\langle \rho \rangle \ge 500$) appear
to stay in narrow range around $[Z/\zsun]\sim -0.5$ over the redshift range
$z=0-1$, consistent with observations
\citep[e.g.,][]{1994Arnaud, 1996Mushotzky, 1996Tamura, 1997Mushotzky}. 
There is some indication of a still higher metallicity towards
higher density regions, which may be in agreement with observations
\citep[e.g.,][]{2001Iwasawa}. 
The metallicity of the WHIM at the peak of its mass distribution 
($\rho/\langle\rho\rangle\sim 10$) at $z=0$ is $[Z/\zsun]\sim -1$,
in good agreement with observations \citep[e.g.,][]{2005Danforth}. 
We find that the following formula fits well the metallicity 
of the WHIM as a function of overdensity $\rho/\langle\rho\rangle$ at the redshift range $z=0-3$:
\begin{equation}
[Z/Z_\odot]_{WHIM}=-1.2 - 0.08z + (0.3+0.12z^{1/3})(\log\rho/\langle\rho\rangle -1),
\end{equation}
which are shown as the straight lines in 
the three panels of Figure~\ref{fig:Zrho}.

%*********************************************************
\begin{figure*}[h]
\vskip -1.6in
\centerline{
\epsfig{file=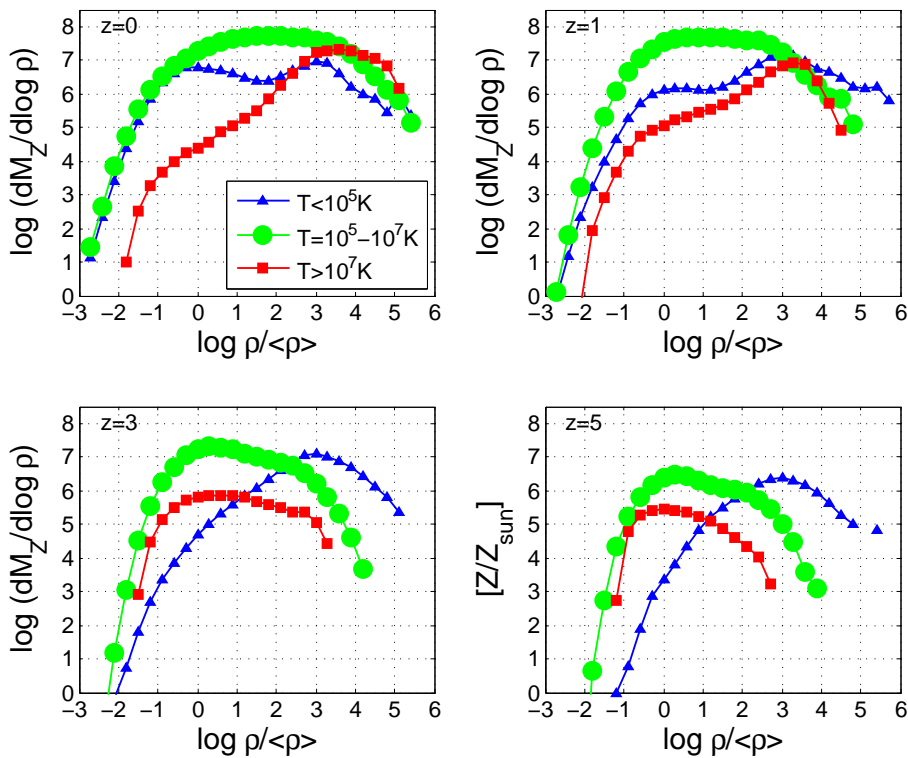,angle=0,width=6.0in}}
\vskip -1.6in
\caption{
shows the distributions of metals mass for the three IGM components - 
(1) cold-warm gas at $T<10^5~$K,
(2) WHIM at $10^7~$K$>T>10^5~$K, (3) Hot X-ray emitting gas at $T > 10^7$K -
as a function of overdensity at four different redshifts
$z=0,1,3,5$.
Note that the area under each curve is proportional to the 
metals mass contained.
}
\label{fig:mtlrho}
\end{figure*}
%**********************************************************

We now examine directly the distribution of metal mass as a function of density
for each IGM component, shown in Figure~\ref{fig:mtlrho}.
A very interesting result is that
at high redshift ($z=3,5$)
the metals mass in the WHIM tends to peak at
a somewhat lower overdensity than that
for the overall WHIM mass, thanks to the
upturn of metallicity of the WHIM at low overdensity end.
Specifically, at $z=3-5$ it appears that
the metals mass peaks at $\delta\sim 2$,
whereas the total WHIM mass peaks at $\delta\sim 10$.
This trend is reversed at lower redshift; for example,
at $z=0$ the metals in WHIM is now broadly peaked at $\delta\sim 100$,
while the WHIM mass peaks at $\delta\sim 10$.
This reversal is likely due to accretion of metal-enriched gas onto high density regions
during recent formation of large-scale structures.
Quantitatively, we find that, at $z=2.5$, only
about $15\%$ of the metals in warm and WHIM gas is located within virialized regions.
About $73\%$ of the metals in warm and WHIM gas 
resides in the IGM with $\delta=1-100$,
with the remaining $12\%$ in underdense regions.
This confirms an earlier expectation that some of the missing
metals may be in the hot halos of galaxies \citep[e.g.,][]{1999bPettini,2005Ferrara};
but that accounts for only a small fraction of the total missing metals.
Combining with our earlier statements on missing metals at $z=2-3$,
our finding on missing metals is that most of the missing metals are in the warm and WHIM gas
with moderate overdensity broadly distributed between $\delta\sim 1-10$.

%***********************************************************************
\begin{figure*}[h]
\vskip -1.0in
\centerline{
\epsfig{file=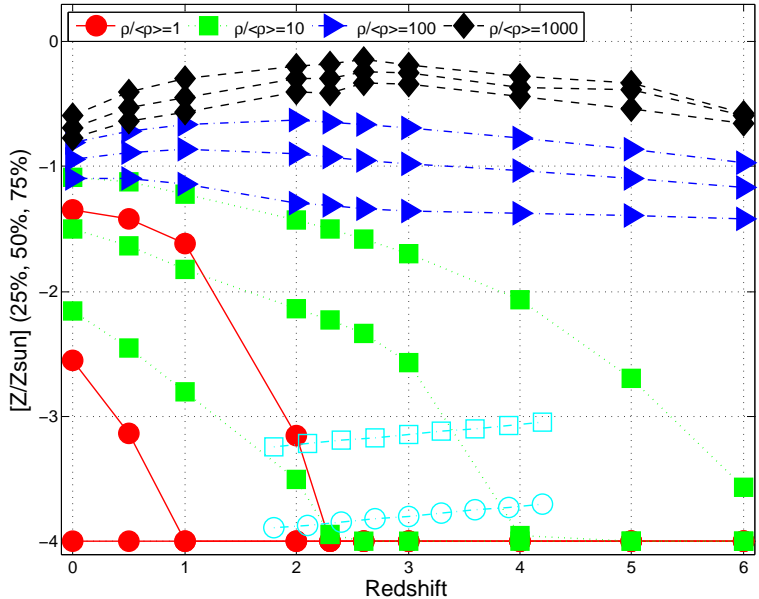,angle=0,width=4.5in}}
\vskip -1.4in
\caption{
shows metallicity evolution as a function of redshift
at four fixed densities, $\rho/\langle\rho\rangle=1,10,100,1000$.
For each density there are three curves, corresponding to
($25\%$, $50\%$, $75\%$) percentiles.
The open squares and open circles are
the observed median metallicity evolution at overdensity
equal to $10$ and $1$, respectively \citep[][]{2003Schaye}.
Note that the metallicity $[Z/\zsun]=-4$ is a floor value.
}
\label{fig:Zdelta}
\end{figure*}
%*************************************************************************

Finally, our attention is turned to the cold-warm component, which
displays a dramatic trough at the mean density.
Physically, it suggests that GSW does not affect bulk of the IGM.
Comparisons with observations are useful here to shed light on this
dramatic behavior.
We note that the mean metallicity at overdensity $\rho/\langle\rho\rangle < 10$
drops quickly below $[Z/\zsun]=-3$.
The typical $\lya$ forest clouds of column density 
$10^{13}-10^{14}$~cm$^{-2}$ arise in these moderate
density regions.
Our simulations suggest that most of these clouds are not expected to be
enriched to a level higher than $[Z/\zsun]=-3$,
which appear to be in agreement with direct metallicity measurements of 
$\lya$ forest clouds
\citep[e.g.,][]{1994Tytler, 1998Lu}. 
However, our results are at variance with recent measurements
of metallicity in these moderate density regions using POD method 
in the sense that the observed metallicity
seems to far exceed what we obtain in our simulations.
To illustrate the disagreement we cast the information presented in
Figures~\ref{fig:hisrho} and \ref{fig:Zrho} into
a different form in Figure~\ref{fig:Zdelta},
where we show the evolution of metallicity 
as a function of redshift at four fixed densities, $\rho/ \langle \rho \rangle=1,10,100,1000$,
for the ease of comparison.
If one compares the middle solid square curve 
(the median metallicity at overdensity $10$ from our simulations)
and the open squares curve
(the median metallicity at overdensity $10$ from observations, \citet{2003Schaye}),
the middle solid dots curve 
(the median metallicity at overdensity $1$ from our simulations)
and the open circles curve 
(the median metallicity at overdensity $1$ from observations, \citet{2003Schaye}),
the disagreement is clear and dramatic.
We predict that the metallicity in regions with overdensity less than about $10$
generally increases quite rapidly with decreasing redshift,
whereas the observationally inferred trend goes in the opposite direction with
a mild rate of change.

Is our simulation incomplete or are the observations misinterpreted? Recall from Figure~\ref{fig:CIVdnddelta} 
that the typical overdensity for low column \civ lines is about $10$, comparable to that of $\lya$ forest clouds.
But that is a mere coincidence: the two types of absorbers are generally not co-located in physical space.
If we go back to Figures~\ref{fig:spectrum1},\ref{fig:spectrum2},\ref{fig:spectrum3} and study the temperature (second rows) and metallicity (third row), we see there is a strong spatial correlation between temperature and metallicity;
regions where a significant amount of \civ reside tend to have an elevated temperature that exceeds $2\times 10^4$K,
whereas the metallicity in lower temperature regions, where HI reside in abundance to give rise 
$\lya$ forest clouds, seems extremely low. As we noted earlier, the regions with elevated temperature and CIV lines
have a width that corresponds to one to several hundred km/s. Interestingly, these regions also typically have peculiar velocities of several hundred km/s (fourth row from top of Figures~\ref{fig:spectrum1},\ref{fig:spectrum2},\ref{fig:spectrum3}). As a result, there should be some overlap in velocity space between some \civ lines and $\lya$ forest lines, 
even when they are significantly displaced in physical space.
This overlap may ``diffuse", in velocity space, some of the metals in regions that producing \civ lines into the $\lya$ forest lines, causing an apparent, moderate metallicity level in $\lya$ forest,
as inferred by \citet[][]{2003Schaye}), when a method such as POD is employed.

A closer look at the left panel of Figure~\ref{fig:CIVdndZ} indicates
that typical \civ absorbers show a decrease of metallicity 
with decreasing redshift in the range $2-5$:
roughly $[Z/\zsun]=[-2.0,-1.5],[-2.3,1.4],[-2.6,-1.5]$
at $z=5,4,2.6$.
This is in accord with the observed weak trend 
of increasing metallicity with increasing redshift,
which otherwise is extremely difficult to understand in
the context of the standard cosmological model.
Needless to say, the \ovi lines located in regions that are spatially
close to \civ lines will also ``diffuse" into the $\lya$ forest in velocity space.
The fact that \ovi lines tend to have a higher metallicity,
about $[Z/\zsun]=0.2$ to $0.4$, than the \civ lines 
over the redshift range of $z\sim 2-4$
(comparing the left and right panels of Figure~\ref{fig:CIVdndZ})
and there are more \ovi lines than \civ lines
(comparing the left and right panels of Figure~\ref{fig:CIVdndZ})
would suggest that one may expect that
the apparent oxygen abundance in the $\lya$ forest 
inferred from POD should be higher than that of \civ lines. 
This is indeed the case: \citet[][]{2008Aguirre}
found that $[O/C]=0.66^{+0.06}_{-0.2}$.
We argue that this provides independent, supporting evidence
for our explanation that is self-consistent and physically plausible.
Alternatively, the IGM may be enriched to the observed level by first generation, Pop III galaxies
that are not properly captured in our simulations.

\begin{figure*}[h]
\vskip -0.7in
\centerline{
\epsfig{file=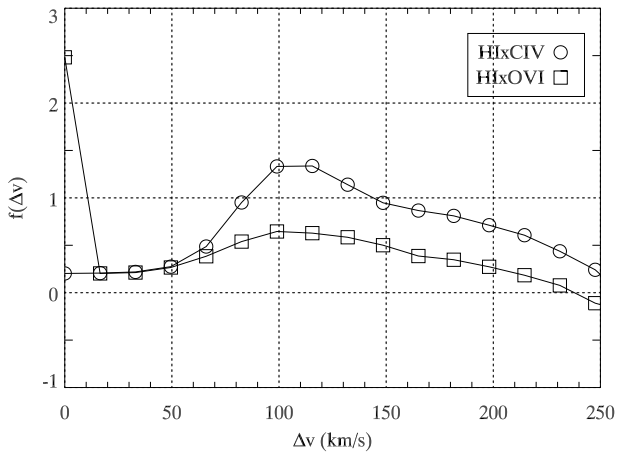,angle=0,width=4.5in}}
\vskip -2.8in
\caption{Comparison of the cross-correlation functions of \civ and \ovi with and without peculiar velocities. The function plotted is $f(\Delta v)$, defined in the text. Values greater than $0$ imply there is a stronger correlation between the ion and the $\lya$ spectrum in the case where peculiar velocities are taken into account. }
\label{podcheck}
\end{figure*}

To further test our ``diffusion'' hypothesis, we have computed the cross-correlation between $\lya$, \civ and \ovi spectra and taken the mean along all lines of sight at $z=2.6$ for two cases: run ``M'' of our simulations with and without the effect of peculiar velocities taken into account. We present in Figure~\ref{podcheck} the following function:
\begin{equation}
f(\Delta v)\equiv\frac{\xi_{p,{\rm HI\,x\,ion}}(\Delta v)}{\xi_{0,{\rm HI\,x\,ion}}(\Delta v)}-1
\end{equation}
where $\xi_{p,{\rm HI\,x\,ion}}(v)$ is the cross-correlation function for the spectrum of HI and the corresponding ion averaged over all lines of sight and symmetrized for positive and negative velocity lags at $z=2.6$. $\xi_{0,{\rm HI\,x\,ion}}(v)$ is the same function computed in the case where there are no peculiar velocities.
Figure~\ref{podcheck} shows that in the case of no peculiar velocity, the cross-correlation between $\lya$ and \civ, and $\lya$ and \ovi, is weaker than in the case where peculiar velocities are considered. 
This is compelling evidence that peculiar velocities effects could artificially diffuse metals into the $\lya$ forest.
%that actually lie in a different physical environment.

%***********************************************************************
%*************************************************************************

\section{Conclusions}

We have carried out the state-of-the-art cosmological hydrodynamic simulations of the standard cold dark matter model to investigate the process of metal enrichment of the intergalactic medium. Our simulations have substantially higher resolution than our previous simulations to address this problem. More importantly, we can now constrain 
the strength of the feedback process by matching the star formation history in our current simulation to the observed one in the range $z=0-6$.
We find that our model reproduces the observed 
mean flux of the $\lya$ forest and the mass density of \civ and \ovi absorbers.
It is also in general consistent with observed physical properties of absorbers. This indicates that we can explain the metal enrichment of the IGM by considering star formation to be the main feedback mechanism, with no apparent need of significant contribution from AGN in terms of additional energy. We conclude from our results that:

(1) The overall star formation history depends rather sensitively on the feedback strength. 
This is likely due to GSW significantly reducing the concentration of cold gas around halos. Nevertheless, GSW do not significantly alter the overall large-scale filamentary baryonic structure that follows the cosmic web of dark matter distribution. While GSW could travel far into low density regions sometimes, the amount of energy and metals that are deposited in underdense regions is very small. 
Most of the GSW energy and metals remain in regions of overdensity $\delta \ge 10$,
with the distance of influence of GSW from galaxies limited to about $\le 0.5$Mpc. 
Metal bubbles blown by GSW coincide with temperatures bubbles, suggesting a tight coupling of energy and metal deposition,
and they are terminated by shock fronts. 

(2) Both \civ and \ovi absorbers are located in regions that have been swept by feedback shocks, of elevated temperature ($T\ge 2\times 10^4$K), density ($\delta \ge 10$) and metallicity ($[Z/\zsun]=[-2.5,-0.5]$), demarcated by a double shock propagating outwards, with \ovi absorbers typically having a higher metallicity than \civ absorbers. Within these shocked regions, most of \civ absorbers tend to arise from moderate density peaks that are troughs in temperature and are thus relatively quiescent. The \ovi absorbers are from regions that are dynamically hotter near shock fronts.
There is a trend for the population of \civ and \ovi absorbers to be more collisionally ionized at higher redshift; for \ovi collisional ionization dominates over the entire redshift range $z=0-6$, whereas for \civ the transition occurs at moderate redshift $z\sim 3$ from collisionally dominated to photoionization dominated.

(3) The evolution of the mass density contained in \civ and \ovi lines, $\omegaciv$ is in good agreement with observations, with both the latest observations and simulations of $\omegaciv$ exhibiting an exponential drop beyond redshift $z=4$;
$\omegaciv$ drop exponentially beyond redshift $z=3$; the near constancy of $\omegaciv$ at redshift $z=1-3$ does not reflect
the evolution of the overall metal content in the IGM. 
In the case of $\omegaovi$, we find a less good agreement between observations and out results. 
This might be in part due to cosmic variance. 

(4) Most of \civ and \ovi absorbers, while clustered around galaxies, are transient and intergalactic in origin,
produced by galactic superwinds in the process of transporting both energy and metals from galaxies into the IGM; 
the metal mass densities contained in \civ and \ovi lines in the range $\log N {\rm cm}^2=12-15$ each constitutes $\sim 0.1\%$ of total metal density at all redshifts; the amount of metals probed by \civ and \ovi lines in the range $\log N {\rm cm}^2=12-15$ is $\sim 1\%$ 
%(with a factor of 2) 
of the total metal density at all redshifts.

(5) While gravitational shocks from large-scale structure formation dominate the energy budget ($80-90\%$) for
turning about $50\%$ of IGM to the warm-hot intergalactic medium (WHIM) by $z=0$, galactic superwind feedback shocks are energetically dominant over gravitational shocks at $z\ge 1-2$.

(6) Most of the so-called ``missing metals" at $z=2-3$ are hidden in a warm-hot gaseous phase ($T>3\times 10^4$K) that is heated up by star formation feedback shocks. Their mass distribution is broadly peaked at overdensity $1-10$ in the IGM, 
outside virialized halos.
Approximately $(37,46,10,7)\%$ of the total metals at $z=0$ are in (stars, WHIM, X-ray gas, cold gas); the distribution stands at $(23,57,2,18)\%$ and $(14,51,4,31)\%$ at $z=2$ and $z=4$, respectively.

(7) The metallicity of the IGM with moderate overdensities ($1-10$) that are probed by the $\lya$ forest 
shows a rapid increase with decreasing redshift. 
We show that velocity ``diffusion'' effect that arises from the peculiar velocities 
could enhance the ``apparent" metallicity of the $\lya$ forest clouds,
as supported by our cross-correlation analysis.
Tentatively, we suggest that this may reconcile, at least in part, the discrepancy between our simulations
and observations at $z=2-4$ based on pixel optical depth (POD) method.

\acknowledgments

We are thankful to Ben Oppenheimer for useful conversations on the subject
and an anonymous referee for a demanding but constructive report that
helps significantly improve the paper.
Computing resources were in part provided by the NASA High-
End Computing (HEC) Program through the NASA Advanced
Supercomputing (NAS) Division at Ames Research Center.
This work is supported in part by grants NNX08AH31G and NAS8-03060.

%\bibliographystyle{apj}
%\bibliography{astro}

\begin{thebibliography}{104}
\expandafter\ifx\csname natexlab\endcsname\relax\def\natexlab#1{#1}\fi

\bibitem[{{Adelberger} {et~al.}(2003){Adelberger}, {Steidel}, {Shapley}, \&
  {Pettini}}]{2003Adelberger}
{Adelberger}, K.~L., {Steidel}, C.~C., {Shapley}, A.~E., \& {Pettini}, M. 2003,
  \apj, 584, 45

\bibitem[{{Aguirre} {et~al.}(2008){Aguirre}, {Dow-Hygelund}, {Schaye}, \&
  {Theuns}}]{2008Aguirre}
{Aguirre}, A., {Dow-Hygelund}, C., {Schaye}, J., \& {Theuns}, T. 2008, \apj,
  689, 851

\bibitem[{{Aguirre} {et~al.}(2001){Aguirre}, {Hernquist}, {Schaye}, {Katz},
  {Weinberg}, \& {Gardner}}]{2001Aguirre}
{Aguirre}, A., {Hernquist}, L., {Schaye}, J., {Katz}, N., {Weinberg}, D.~H., \&
  {Gardner}, J. 2001, \apj, 561, 521

\bibitem[{{Aguirre} {et~al.}(2005){Aguirre}, {Schaye}, {Hernquist}, {Kay},
  {Springel}, \& {Theuns}}]{2005Aguirre}
{Aguirre}, A., {Schaye}, J., {Hernquist}, L., {Kay}, S., {Springel}, V., \&
  {Theuns}, T. 2005, \apjl, 620, L13

\bibitem[{{Arnaud} {et~al.}(1994){Arnaud}, {Mushotzky}, {Ezawa}, {Fukazawa},
  {Ohashi}, {Bautz}, {Crewe}, {Gendreau}, {Yamashita}, {Kamata}, \&
  {Akimoto}}]{1994Arnaud}
{Arnaud}, K.~A., {Mushotzky}, R.~F., {Ezawa}, H., {Fukazawa}, Y., {Ohashi}, T.,
  {Bautz}, M.~W., {Crewe}, G.~B., {Gendreau}, K.~C., {Yamashita}, K., {Kamata},
  Y., \& {Akimoto}, F. 1994, \apjl, 436, L67

\bibitem[{{Bahcall} \& {Spitzer}(1969)}]{1969Bahcall}
{Bahcall}, J.~N. \& {Spitzer}, Jr., L. 1969, \apjl, 156, L63+

\bibitem[{Barger {et~al.}(2000)Barger, Cowie, \& Richards}]{Barger00}
Barger, A.~J., Cowie, L.~L., \& Richards, E.~A. 2000, AJ, 119, 2092

\bibitem[{{Becker} {et~al.}(2009){Becker}, {Rauch}, \& {Sargent}}]{2009Becker}
{Becker}, G.~D., {Rauch}, M., \& {Sargent}, W.~L.~W. 2009, \apj, 698, 1010

\bibitem[{{Bergeron} {et~al.}(2002){Bergeron}, {Aracil}, {Petitjean}, \&
  {Pichon}}]{2002Bergeron}
{Bergeron}, J., {Aracil}, B., {Petitjean}, P., \& {Pichon}, C. 2002, \aap, 396,
  L11

\bibitem[{{Bergeron} \& {Herbert-Fort}(2005)}]{2005Bergeron}
{Bergeron}, J. \& {Herbert-Fort}, S. 2005, ArXiv Astrophysics e-prints

\bibitem[{{Blanchard} {et~al.}(1992){Blanchard}, {Valls-Gabaud}, \&
  {Mamon}}]{1992Blanchard}
{Blanchard}, A., {Valls-Gabaud}, D., \& {Mamon}, G.~A. 1992, \aap, 264, 365

\bibitem[{{Boksenberg} {et~al.}(2003){Boksenberg}, {Sargent}, \&
  {Rauch}}]{2003Boksenberg}
{Boksenberg}, A., {Sargent}, W.~L.~W., \& {Rauch}, M. 2003, ArXiv Astrophysics
  e-prints

\bibitem[{{Bouwens} {et~al.}(2005){Bouwens}, {Illingworth}, {Thompson}, \&
  {Franx}}]{Bouwens05}
{Bouwens}, R.~J., {Illingworth}, G.~D., {Thompson}, R.~I., \& {Franx}, M. 2005,
  ApJL, 624, L5

\bibitem[{{Burles} {et~al.}(2001){Burles}, {Nollett}, \& {Turner}}]{2001Burles}
{Burles}, S., {Nollett}, K.~M., \& {Turner}, M.~S. 2001, \apjl, 552, L1

\bibitem[{{Carswell} {et~al.}(2002){Carswell}, {Schaye}, \&
  {Kim}}]{2002Carswell}
{Carswell}, B., {Schaye}, J., \& {Kim}, T. 2002, \apj, 578, 43

\bibitem[{{Cen} {et~al.}(1995){Cen}, {Kang}, {Ostriker}, \& {Ryu}}]{1995Cen}
{Cen}, R., {Kang}, H., {Ostriker}, J.~P., \& {Ryu}, D. 1995, \apj, 451, 436

\bibitem[{{Cen} {et~al.}(1994){Cen}, {Miralda-Escude}, {Ostriker}, \&
  {Rauch}}]{1994Cen}
{Cen}, R., {Miralda-Escude}, J., {Ostriker}, J.~P., \& {Rauch}, M. 1994, \apjl,
  437, L9

\bibitem[{{Cen} {et~al.}(2005){Cen}, {Nagamine}, \& {Ostriker}}]{2005Cen}
{Cen}, R., {Nagamine}, K., \& {Ostriker}, J.~P. 2005, \apj, 635, 86

\bibitem[{{Cen} \& {Ostriker}(1999{\natexlab{a}})}]{1999bCen}
{Cen}, R. \& {Ostriker}, J.~P. 1999{\natexlab{a}}, \apjl, 519, L109

\bibitem[{{Cen} \& {Ostriker}(1999{\natexlab{b}})}]{1999Cen}
---. 1999{\natexlab{b}}, \apj, 514, 1

\bibitem[{{Cen} \& {Ostriker}(2006)}]{2006Cen}
---. 2006, \apj, 650, 560

\bibitem[{{Cen} {et~al.}(2001){Cen}, {Tripp}, {Ostriker}, \&
  {Jenkins}}]{2001Cen}
{Cen}, R., {Tripp}, T.~M., {Ostriker}, J.~P., \& {Jenkins}, E.~B. 2001, \apjl,
  559, L5

\bibitem[{{Chaffee} {et~al.}(1986){Chaffee}, {Foltz}, {Bechtold}, \&
  {Weymann}}]{1986Chaffee}
{Chaffee}, Jr., F.~H., {Foltz}, C.~B., {Bechtold}, J., \& {Weymann}, R.~J.
  1986, \apj, 301, 116

\bibitem[{{Cole}(1991)}]{1991Cole}
{Cole}, S. 1991, \apj, 367, 45

\bibitem[{{Cole} {et~al.}(2001){Cole}, {Norberg}, {Baugh}, {Frenk},
  {Bland-Hawthorn}, {Bridges}, {Cannon}, {Colless}, {Collins}, {Couch},
  {Cross}, {Dalton}, {De Propris}, {Driver}, {Efstathiou}, {Ellis},
  {Glazebrook}, {Jackson}, {Lahav}, {Lewis}, {Lumsden}, {Maddox}, {Madgwick},
  {Peacock}, {Peterson}, {Sutherland}, \& {Taylor}}]{Cole01}
{Cole}, S., {Norberg}, P., {Baugh}, C.~M., {Frenk}, C.~S., {Bland-Hawthorn},
  J., {Bridges}, T., {Cannon}, R., {Colless}, M., {Collins}, C., {Couch}, W.,
  {Cross}, N., {Dalton}, G., {De Propris}, R., {Driver}, S.~P., {Efstathiou},
  G., {Ellis}, R.~S., {Glazebrook}, K., {Jackson}, C., {Lahav}, O., {Lewis},
  I., {Lumsden}, S., {Maddox}, S., {Madgwick}, D., {Peacock}, J.~A.,
  {Peterson}, B.~A., {Sutherland}, W., \& {Taylor}, K. 2001, \mnras, 326, 255

\bibitem[{{Cooksey} {et~al.}(2009){Cooksey}, {Thom}, {Prochaska}, \&
  {Chen}}]{2009Cooksey}
{Cooksey}, K.~L., {Thom}, C., {Prochaska}, J.~X., \& {Chen}, H. 2009, ArXiv
  e-prints

\bibitem[{Cowie {et~al.}(1999)Cowie, Songaila, \& Barger}]{Cowie99}
Cowie, L.~L., Songaila, A., \& Barger, A.~J. 1999, AJ, 118, 603

\bibitem[{{Dalla Vecchia} \& {Schaye}(2008)}]{2008DallaVecchia}
{Dalla Vecchia}, C. \& {Schaye}, J. 2008, \mnras, 387, 1431

\bibitem[{{Danforth} \& {Shull}(2005)}]{2005Danforth}
{Danforth}, C.~W. \& {Shull}, J.~M. 2005, \apj, 624, 555

\bibitem[{{Danforth} \& {Shull}(2008)}]{2008Danforth}
---. 2008, \apj, 679, 194

\bibitem[{{Dav{\'e}}(2008)}]{2008Dave}
{Dav{\'e}}, R. 2008, \mnras, 385, 147

\bibitem[{{Dav{\'e}} {et~al.}(2001){Dav{\'e}}, {Cen}, {Ostriker}, {Bryan},
  {Hernquist}, {Katz}, {Weinberg}, {Norman}, \& {O'Shea}}]{2001Dave}
{Dav{\'e}}, R., {Cen}, R., {Ostriker}, J.~P., {Bryan}, G.~L., {Hernquist}, L.,
  {Katz}, N., {Weinberg}, D.~H., {Norman}, M.~L., \& {O'Shea}, B. 2001, \apj,
  552, 473

\bibitem[{{D'Odorico} {et~al.}(2009){D'Odorico}, {Calura}, {Cristiani}, \&
  {Viel}}]{2009Dodorico}
{D'Odorico}, V., {Calura}, F., {Cristiani}, S., \& {Viel}, M. 2009, ArXiv
  e-prints

\bibitem[{{Ferland} {et~al.}(1998){Ferland}, {Korista}, {Verner}, {Ferguson},
  {Kingdon}, \& {Verner}}]{1998Ferland}
{Ferland}, G.~J., {Korista}, K.~T., {Verner}, D.~A., {Ferguson}, J.~W.,
  {Kingdon}, J.~B., \& {Verner}, E.~M. 1998, \pasp, 110, 761

\bibitem[{{Ferrara} {et~al.}(2005){Ferrara}, {Scannapieco}, \&
  {Bergeron}}]{2005Ferrara}
{Ferrara}, A., {Scannapieco}, E., \& {Bergeron}, J. 2005, \apjl, 634, L37

\bibitem[{{Frank} {et~al.}(2008){Frank}, {Mathur}, \& {York}}]{2008Frank}
{Frank}, S., {Mathur}, S., \& {York}, D.~G. 2008, ArXiv e-prints

\bibitem[{{Fukugita} {et~al.}(1998){Fukugita}, {Hogan}, \&
  {Peebles}}]{1998Fukugita}
{Fukugita}, M., {Hogan}, C.~J., \& {Peebles}, P.~J.~E. 1998, \apj, 503, 518

\bibitem[{Gabasch {et~al.}(2004)Gabasch, Salvato, Saglia, Bender,
  {et~al.}}]{Gabasch04}
Gabasch, A., Salvato, M., Saglia, R.~P., Bender, R., {et~al.} 2004, ApJ, 616,
  L83

\bibitem[{Giavalisco {et~al.}(2004)Giavalisco, Dickinson, Ferguson,
  Ravindranath, Kretchmer, Moustakas, Madau, Fall, {et~al.}}]{Giavalisco04}
Giavalisco, M., Dickinson, M., Ferguson, H.~C., Ravindranath, S., Kretchmer,
  C., Moustakas, L.~A., Madau, P., Fall, M., {et~al.} 2004, ApJ, 600, L103

\bibitem[{{Gnedin} \& {Ostriker}(1997)}]{1997Gnedin}
{Gnedin}, N.~Y. \& {Ostriker}, J.~P. 1997, \apj, 486, 581

\bibitem[{{Haardt} \& {Madau}(1996)}]{1996Haardt}
{Haardt}, F. \& {Madau}, P. 1996, \apj, 461, 20

\bibitem[{Heavens {et~al.}(2004)Heavens, Panter, Jimenez, \&
  Dunlop}]{Heavens04}
Heavens, A.~F., Panter, B., Jimenez, R., \& Dunlop, J. 2004, Nature, 428, 625

\bibitem[{{Heckman}(2001)}]{2001Heckman}
{Heckman}, T.~M. 2001, in Astronomical Society of the Pacific Conference
  Series, Vol. 240, Gas and Galaxy Evolution, ed. J.~E. {Hibbard}, M.~{Rupen},
  \& J.~H. {van Gorkom}, 345

\bibitem[{{Hopkins} {et~al.}(2006){Hopkins}, {Hernquist}, {Cox}, {Robertson},
  {Di Matteo}, \& {Springel}}]{Hopk06a}
{Hopkins}, P.~F., {Hernquist}, L., {Cox}, T.~J., {Robertson}, B., {Di Matteo},
  T., \& {Springel}, V. 2006, \apj, 639, 700

\bibitem[{{Hui} \& {Gnedin}(1997)}]{1997HuiGnedin}
{Hui}, L. \& {Gnedin}, N.~Y. 1997, \mnras, 292, 27

\bibitem[{{Iwasawa} {et~al.}(2001){Iwasawa}, {Fabian}, {Allen}, \&
  {Ettori}}]{2001Iwasawa}
{Iwasawa}, K., {Fabian}, A.~C., {Allen}, S.~W., \& {Ettori}, S. 2001, \mnras,
  328, L5

\bibitem[{{Katz} {et~al.}(1996){Katz}, {Weinberg}, \& {Hernquist}}]{1996Katz}
{Katz}, N., {Weinberg}, D.~H., \& {Hernquist}, L. 1996, \apjs, 105, 19

\bibitem[{{Komatsu} {et~al.}(2009){Komatsu}, {Dunkley}, {Nolta}, {Bennett},
  {Gold}, {Hinshaw}, {Jarosik}, {Larson}, {Limon}, {Page}, {Spergel},
  {Halpern}, {Hill}, {Kogut}, {Meyer}, {Tucker}, {Weiland}, {Wollack}, \&
  {Wright}}]{2009Komatsu}
{Komatsu}, E., {Dunkley}, J., {Nolta}, M.~R., {Bennett}, C.~L., {Gold}, B.,
  {Hinshaw}, G., {Jarosik}, N., {Larson}, D., {Limon}, M., {Page}, L.,
  {Spergel}, D.~N., {Halpern}, M., {Hill}, R.~S., {Kogut}, A., {Meyer}, S.~S.,
  {Tucker}, G.~S., {Weiland}, J.~L., {Wollack}, E., \& {Wright}, E.~L. 2009,
  \apjs, 180, 330

\bibitem[{{Krumholz} \& {Tan}(2007)}]{2007Krumholz}
{Krumholz}, M.~R. \& {Tan}, J.~C. 2007, \apj, 654, 304

\bibitem[{{Li} {et~al.}(2008){Li}, {Li}, \& {Cen}}]{2008Li}
{Li}, S., {Li}, H., \& {Cen}, R. 2008, \apjs, 174, 1

\bibitem[{Lilly {et~al.}(1996)Lilly, F\`{e}vre, Hammer, \& Crampton}]{Lilly96}
Lilly, S.~J., F\`{e}vre, O.~L., Hammer, F., \& Crampton, D. 1996, ApJ, 460, L1

\bibitem[{{Lu} {et~al.}(1998){Lu}, {Sargent}, {Barlow}, \& {Rauch}}]{1998Lu}
{Lu}, L., {Sargent}, W.~L.~W., {Barlow}, T.~A., \& {Rauch}, M. 1998, ArXiv
  Astrophysics e-prints

\bibitem[{{Mac Low} \& {Ferrara}(1999)}]{1999MacLow}
{Mac Low}, M. \& {Ferrara}, A. 1999, \apj, 513, 142

\bibitem[{{Mathur} {et~al.}(2003){Mathur}, {Weinberg}, \& {Chen}}]{2003Mathur}
{Mathur}, S., {Weinberg}, D.~H., \& {Chen}, X. 2003, \apj, 582, 82

\bibitem[{{McDonald} {et~al.}(2000){McDonald}, {Miralda-Escud{\' e}}, {Rauch},
  {Sargent}, {Barlow}, {Cen}, \& {Ostriker}}]{2000McDonald}
{McDonald}, P., {Miralda-Escud{\' e}}, J., {Rauch}, M., {Sargent}, W.~L.~W.,
  {Barlow}, T.~A., {Cen}, R., \& {Ostriker}, J.~P. 2000, \apj, 543, 1

\bibitem[{{Mushotzky} {et~al.}(1996){Mushotzky}, {Loewenstein}, {Arnaud},
  {Tamura}, {Fukazawa}, {Matsushita}, {Kikuchi}, \&
  {Hatsukade}}]{1996Mushotzky}
{Mushotzky}, R., {Loewenstein}, M., {Arnaud}, K.~A., {Tamura}, T., {Fukazawa},
  Y., {Matsushita}, K., {Kikuchi}, K., \& {Hatsukade}, I. 1996, \apj, 466, 686

\bibitem[{{Mushotzky} \& {Loewenstein}(1997)}]{1997Mushotzky}
{Mushotzky}, R.~F. \& {Loewenstein}, M. 1997, \apjl, 481, L63+

\bibitem[{{Nakamura} {et~al.}(2004){Nakamura}, {Fukugita}, {Brinkmann}, \&
  {Schneider}}]{Nakamura04}
{Nakamura}, O., {Fukugita}, M., {Brinkmann}, J., \& {Schneider}, D.~P. 2004,
  \aj, 127, 2511

\bibitem[{{Nicastro} {et~al.}(2009){Nicastro}, {Krongold}, {Fields},
  {Conciatore}, {Zappacosta}, {Elvis}, {Mathur}, \& {Papadakis}}]{2009Nicastro}
{Nicastro}, F., {Krongold}, Y., {Fields}, D., {Conciatore}, M.~L.,
  {Zappacosta}, L., {Elvis}, M., {Mathur}, S., \& {Papadakis}, I. 2009, ArXiv
  e-prints

\bibitem[{Norman {et~al.}(2004)Norman, Ptak, Hornschemeier, Hasinger, Bergeron,
  Comastri, Giacconi, Gilli, {et~al.}}]{Norman04}
Norman, C., Ptak, A., Hornschemeier, A., Hasinger, G., Bergeron, J., Comastri,
  A., Giacconi, R., Gilli, R., {et~al.} 2004, ApJ, 607, 721

\bibitem[{{Norris} {et~al.}(1983){Norris}, {Peterson}, \&
  {Hartwick}}]{1983Norris}
{Norris}, J., {Peterson}, B.~A., \& {Hartwick}, F.~D.~A. 1983, \apj, 273, 450

\bibitem[{{Noterdaeme} {et~al.}(2009){Noterdaeme}, {Petitjean}, {Ledoux}, \&
  {Srianand}}]{2009Noterdaeme}
{Noterdaeme}, P., {Petitjean}, P., {Ledoux}, C., \& {Srianand}, R. 2009, \aap,
  505, 1087

\bibitem[{{Oppenheimer} \& {Dav{\'e}}(2006)}]{2006Oppenheimer}
{Oppenheimer}, B.~D. \& {Dav{\'e}}, R. 2006, \mnras, 373, 1265

\bibitem[{{Ouchi} {et~al.}(2004){Ouchi}, {Shimasaku}, {Okamura}, {Furusawa},
  {Kashikawa}, {Ota}, {Doi}, {Hamabe}, {Kimura}, {Komiyama}, {Miyazaki},
  {Miyazaki}, {Nakata}, {Sekiguchi}, {Yagi}, \& {Yasuda}}]{Ouchi04a}
{Ouchi}, M., {Shimasaku}, K., {Okamura}, S., {Furusawa}, H., {Kashikawa}, N.,
  {Ota}, K., {Doi}, M., {Hamabe}, M., {Kimura}, M., {Komiyama}, Y., {Miyazaki},
  M., {Miyazaki}, S., {Nakata}, F., {Sekiguchi}, M., {Yagi}, M., \& {Yasuda},
  N. 2004, \apj, 611, 660

\bibitem[{{Pagel}(1999)}]{1999Pagel}
{Pagel}, B.~E.~J. 1999, ArXiv Astrophysics e-prints

\bibitem[{{Pen}(1999)}]{1999Pen}
{Pen}, U. 1999, \apjl, 510, L1

\bibitem[{{Penton} {et~al.}(2004){Penton}, {Stocke}, \& {Shull}}]{2004Penton}
{Penton}, S.~V., {Stocke}, J.~T., \& {Shull}, J.~M. 2004, \apjs, 152, 29

\bibitem[{{P{\'e}roux} {et~al.}(2003){P{\'e}roux}, {McMahon},
  {Storrie-Lombardi}, \& {Irwin}}]{2003Peroux}
{P{\'e}roux}, C., {McMahon}, R.~G., {Storrie-Lombardi}, L.~J., \& {Irwin},
  M.~J. 2003, \mnras, 346, 1103

\bibitem[{{Pettini}(1999)}]{1999bPettini}
{Pettini}, M. 1999, in Chemical Evolution from Zero to High Redshift, ed. J.~R.
  {Walsh} \& M.~R. {Rosa}, 233--+

\bibitem[{{Pettini} {et~al.}(2003){Pettini}, {Madau}, {Bolte}, {Prochaska},
  {Ellison}, \& {Fan}}]{2003Pettini}
{Pettini}, M., {Madau}, P., {Bolte}, M., {Prochaska}, J.~X., {Ellison}, S.~L.,
  \& {Fan}, X. 2003, \apj, 594, 695

\bibitem[{{Pettini} {et~al.}(1997){Pettini}, {Smith}, {King}, \&
  {Hunstead}}]{1997Pettini}
{Pettini}, M., {Smith}, L.~J., {King}, D.~L., \& {Hunstead}, R.~W. 1997, \apj,
  486, 665

\bibitem[{Prochaska {et~al.}(2003)Prochaska, Gawiser, Wolfe, Castro, \&
  G.}]{2003Prochaska}
Prochaska, J.~X., Gawiser, E., Wolfe, A.~M., Castro, S., \& G., G.~D. 2003,
  ApJ, 595, L9

\bibitem[{{Prochaska} \& {Wolfe}(2009)}]{2009Prochaska}
{Prochaska}, J.~X. \& {Wolfe}, A.~M. 2009, \apj, 696, 1543

\bibitem[{{Rao} {et~al.}(2006){Rao}, {Turnshek}, \& {Nestor}}]{2006Rao}
{Rao}, S.~M., {Turnshek}, D.~A., \& {Nestor}, D.~B. 2006, \apj, 636, 610

\bibitem[{{Rauch} {et~al.}(1996){Rauch}, {Sargent}, {Womble}, \&
  {Barlow}}]{1996Rauch}
{Rauch}, M., {Sargent}, W.~L.~W., {Womble}, D.~S., \& {Barlow}, T.~A. 1996,
  \apjl, 467, L5+

\bibitem[{{Reddy} {et~al.}(2005){Reddy}, {Erb}, {Steidel}, {Shapley},
  {Adelberger}, \& {Pettini}}]{Reddy05}
{Reddy}, N.~A., {Erb}, D.~K., {Steidel}, C.~C., {Shapley}, A.~E., {Adelberger},
  K.~L., \& {Pettini}, M. 2005, ApJ, 633, 748

\bibitem[{{Ryan-Weber} {et~al.}(2006){Ryan-Weber}, {Pettini}, \&
  {Madau}}]{2006Ryanweber}
{Ryan-Weber}, E.~V., {Pettini}, M., \& {Madau}, P. 2006, \mnras, 371, L78

\bibitem[{{Ryan-Weber} {et~al.}(2009){Ryan-Weber}, {Pettini}, {Madau}, \&
  {Zych}}]{2009Ryanweber}
{Ryan-Weber}, E.~V., {Pettini}, M., {Madau}, P., \& {Zych}, B.~J. 2009, \mnras,
  395, 1476

\bibitem[{{Schaye} {et~al.}(2003){Schaye}, {Aguirre}, {Kim}, {Theuns}, {Rauch},
  \& {Sargent}}]{2003Schaye}
{Schaye}, J., {Aguirre}, A., {Kim}, T., {Theuns}, T., {Rauch}, M., \&
  {Sargent}, W.~L.~W. 2003, \apj, 596, 768

\bibitem[{{Schramm} \& {Turner}(1998)}]{1998Schramm}
{Schramm}, D.~N. \& {Turner}, M.~S. 1998, Reviews of Modern Physics, 70, 303

\bibitem[{{Shen} {et~al.}(2010){Shen}, {Wadsley}, \& {Stinson}}]{2010Shen}
{Shen}, S., {Wadsley}, J., \& {Stinson}, G. 2010, \mnras, 407, 1581

\bibitem[{{Simcoe}(2006)}]{2006Simcoe}
{Simcoe}, R.~A. 2006, \apj, 653, 977

\bibitem[{{Simcoe} {et~al.}(2002){Simcoe}, {Sargent}, \& {Rauch}}]{2002Simcoe}
{Simcoe}, R.~A., {Sargent}, W.~L.~W., \& {Rauch}, M. 2002, \apj, 578, 737

\bibitem[{{Simcoe} {et~al.}(2004){Simcoe}, {Sargent}, \& {Rauch}}]{2004Simcoe}
---. 2004, \apj, 606, 92

\bibitem[{{Songaila}(2001)}]{2001Songaila}
{Songaila}, A. 2001, \apjl, 561, L153

\bibitem[{{Songaila}(2005)}]{2005Songaila}
---. 2005, \aj, 130, 1996

\bibitem[{{Springel} \& {Hernquist}(2003)}]{2003Springel}
{Springel}, V. \& {Hernquist}, L. 2003, \mnras, 339, 289

\bibitem[{Steidel {et~al.}(1999)Steidel, Adelberger, Giavalisco, Dickinson, \&
  Pettini}]{Steidel99}
Steidel, C.~C., Adelberger, K.~L., Giavalisco, M., Dickinson, M., \& Pettini,
  M. 1999, ApJ, 519, 1

\bibitem[{{Steidel} \& {Sargent}(1992)}]{1992Steidel}
{Steidel}, C.~C. \& {Sargent}, W.~L.~W. 1992, \apjs, 80, 1

\bibitem[{{Steinmetz}(1996)}]{1996Steinmetz}
{Steinmetz}, M. 1996, \mnras, 278, 1005

\bibitem[{{Sutherland} \& {Dopita}(1993)}]{Sutherland93}
{Sutherland}, R.~S. \& {Dopita}, M.~A. 1993, \apjs, 88, 253

\bibitem[{{Tamura} {et~al.}(1996){Tamura}, {Day}, {Fukazawa}, {Hatsukade},
  {Ikebe}, {Makishima}, {Mushotzky}, {Ohashi}, {Takenaka}, \&
  {Yamashita}}]{1996Tamura}
{Tamura}, T., {Day}, C.~S., {Fukazawa}, Y., {Hatsukade}, I., {Ikebe}, Y.,
  {Makishima}, K., {Mushotzky}, R.~F., {Ohashi}, T., {Takenaka}, K., \&
  {Yamashita}, K. 1996, \pasj, 48, 671

\bibitem[{{Theuns} {et~al.}(2002{\natexlab{a}}){Theuns}, {Schaye}, {Zaroubi},
  {Kim}, {Tzanavaris}, \& {Carswell}}]{2002Theuns}
{Theuns}, T., {Schaye}, J., {Zaroubi}, S., {Kim}, T.-S., {Tzanavaris}, P., \&
  {Carswell}, B. 2002{\natexlab{a}}, \apjl, 567, L103

\bibitem[{{Theuns} {et~al.}(2002{\natexlab{b}}){Theuns}, {Viel}, {Kay},
  {Schaye}, {Carswell}, \& {Tzanavaris}}]{2002bTheuns}
{Theuns}, T., {Viel}, M., {Kay}, S., {Schaye}, J., {Carswell}, R.~F., \&
  {Tzanavaris}, P. 2002{\natexlab{b}}, \apjl, 578, L5

\bibitem[{{Thom} \& {Chen}(2008{\natexlab{a}})}]{2008Thoma}
{Thom}, C. \& {Chen}, H. 2008{\natexlab{a}}, \apjs, 179, 37

\bibitem[{{Thom} \& {Chen}(2008{\natexlab{b}})}]{2008Thomb}
---. 2008{\natexlab{b}}, \apj, 683, 22

\bibitem[{{Tripp} {et~al.}(2008){Tripp}, {Sembach}, {Bowen}, {Savage},
  {Jenkins}, {Lehner}, \& {Richter}}]{2008Tripp}
{Tripp}, T.~M., {Sembach}, K.~R., {Bowen}, D.~V., {Savage}, B.~D., {Jenkins},
  E.~B., {Lehner}, N., \& {Richter}, P. 2008, \apjs, 177, 39

\bibitem[{{Tytler} \& {Fan}(1994)}]{1994Tytler}
{Tytler}, D. \& {Fan}, X. 1994, \apjl, 424, L87

\bibitem[{{White} \& {Frenk}(1991)}]{1991White}
{White}, S.~D.~M. \& {Frenk}, C.~S. 1991, \apj, 379, 52

\bibitem[{{Wu} {et~al.}(2001){Wu}, {Fabian}, \& {Nulsen}}]{2001Wu}
{Wu}, K.~K.~S., {Fabian}, A.~C., \& {Nulsen}, P.~E.~J. 2001, \mnras, 324, 95

\bibitem[{{Yepes} {et~al.}(1997){Yepes}, {Kates}, {Khokhlov}, \&
  {Klypin}}]{1997Yepes}
{Yepes}, G., {Kates}, R., {Khokhlov}, A., \& {Klypin}, A. 1997, \mnras, 284,
  235

\bibitem[{{Young} {et~al.}(1982){Young}, {Sargent}, \&
  {Boksenberg}}]{1982Young}
{Young}, P., {Sargent}, W.~L.~W., \& {Boksenberg}, A. 1982, \apjs, 48, 455

\bibitem[{{Zhang} {et~al.}(2009){Zhang}, {Li}, {Kauffmann}, {Zou}, {Catinella},
  {Shen}, {Guo}, \& {Chang}}]{2009Zhang}
{Zhang}, W., {Li}, C., {Kauffmann}, G., {Zou}, H., {Catinella}, B., {Shen}, S.,
  {Guo}, Q., \& {Chang}, R. 2009, \mnras, 397, 1243

\bibitem[{{Zwaan} {et~al.}(2005){Zwaan}, {Meyer}, {Staveley-Smith}, \&
  {Webster}}]{2005Zwaan}
{Zwaan}, M.~A., {Meyer}, M.~J., {Staveley-Smith}, L., \& {Webster}, R.~L. 2005,
  \mnras, 359, L30

\end{thebibliography}

\end{document}